\documentclass[prx,twocolumn,english,longbibliography,superscriptaddress]{revtex4-2}

\usepackage{ifthen}
\usepackage{xcolor}
\usepackage{graphics,graphicx,epsfig}
\usepackage{epsf,epstopdf,wrapfig}
\usepackage{amssymb,amsfonts,amsmath}
\usepackage[export]{adjustbox}
\usepackage[english]{babel}
\usepackage{ifthen}
\usepackage{rotating}

\usepackage{booktabs}
\usepackage{multirow}

\definecolor{ocre}{RGB}{10,100,185}
\usepackage[colorlinks = true,
            linkcolor = ocre,
            urlcolor  = ocre,
            citecolor = ocre,
            anchorcolor = ocre]{hyperref}

\usepackage[normalem]{ulem}

\usepackage{hyperref}

\newcommand{\EQ}{\begin{equation}}
\newcommand{\EE}{\end{equation}}
\newcommand{\EQA}{\begin{eqnarray}}
\newcommand{\EEA}{\end{eqnarray}}
\newcommand{\Ag}{\text{Ag}}
\newcommand{\infc}{\text{inf}}
\newcommand{\res}{\text{res}}

\newcommand{\DiffSchematic}{S1}
\newcommand{\DesignResponseClearance}{S2}
\newcommand{\ClusterInput}{S3}
\newcommand{\BaselineOnHarm}{S4}
\newcommand{\PsiOnHarm}{S5}
\newcommand{\ParetoArchOne}{S6}
\newcommand{\BaselineOnMacroDyn}{S7} 
\newcommand{\PsieOnMacroDyn}{S8} 
\newcommand{\CancerAntigenicity}{S9}
\newcommand{\paramsensitivityFigureone}{S10} 
\newcommand{\paramsensitivityFiguretwo}{S11}
\newcommand{\noisesensitivityFigureone}{S12} 

\newcommand{\ParameterTable}{S1}

\newcommand{\corres}{Correspondence should be addressed to: Armita Nourmohammad: {armita.nourmohammad@yale.edu}.}

\begin{document}

\title{Design principles of the cytotoxic CD8$^+$ T-cell response}
\author{Obinna A. Ukogu}
\affiliation{Department of Applied Mathematics, University of Washington, Seattle, WA, USA}
\author{Zachary Montague}
\affiliation{Department of Physics, University of Washington, Seattle, WA, USA}
\author{Gr\'egoire Altan-Bonnet}
\affiliation{Immunodynamics Group, Laboratory of Integrative Cancer Immunology, Center for Cancer Research, National Cancer Institute, Bethesda, MD, USA}
\author{Armita Nourmohammad}
\thanks{\corres}
\affiliation{Department of Applied Mathematics, University of Washington, Seattle, WA, USA}
\affiliation{Department of Physics, University of Washington, Seattle, WA, USA}
\affiliation{Allen School of Computer Science and Engineering, University of Washington, Seattle, WA, USA}
\affiliation{Fred Hutchinson Cancer Research Center, Seattle, WA, USA}
\affiliation{Center for Systems and Engineering Immunology, and the Departments of Immunobiology, Biomedical Engineering, and Physics,  Yale University, New Haven, CT, USA}

\begin{abstract}
\noindent Cytotoxic T lymphocytes eliminate infected or malignant cells, safeguarding surrounding tissues. Although experimental and systems-immunology studies have cataloged many molecular and cellular actors involved in an immune response, the design principles governing how the speed and magnitude of T-cell responses emerge from cellular decision-making remain elusive. Here, we recast T-cell response as a feedback-controlled program, wherein the rates of activation, proliferation, differentiation and death are regulated through antigenic, pro- and anti-inflammatory cues. {By exploring a broad class of feedback-controller designs as potential immune programs, }we demonstrate how the speed and magnitude of T-cell responses emerge from optimizing signal-feedback to protect against diverse infection settings. We recover an inherent trade-off: infection clearance at the cost of immunopathology. We show how this trade-off is encoded into the logic of T-cell responses by hierarchical sensitivity to different immune signals. {Notably, we find the designs that balance the harm from acute infections and autoimmunity produce immune responses consistent with the experimentally observed patterns of T-cell effector expansion in mice.} Extending our model to immune-based T-cell therapies for cancer tumors, we {quantify the} trade-off between the affinity for tumor antigens (``quality") and the abundance (``quantity") of infused T-cells necessary for effective treatment. Finally, we show how therapeutic efficacy can be improved by targeted genetic perturbations to T-cells. Our findings offer a unified control-logic for cytotoxic T-cell responses and point to specific regulatory programs that can be engineered for more robust T-cell therapies.
\end{abstract}

\maketitle

\section{Introduction}

Cellular immunity hinges upon the regulated activation, expansion, and contraction of lymphocytes that clear an infection while limiting collateral tissue damage. We focus on CD8$^+$ T-cells---also known as killer or cytotoxic T-cells---which are crucial for immune surveillance and defense against infections and cancer. {Pathogen recognition is mediated by T-cell receptors (TCRs) that bind short pathogen-derived peptides (or antigens) presented by major histocompatibility complexes (MHCs) on antigen-presenting cells (APCs) and other target cells. To recognize a multitude of pathogens, the immune system generates a diverse repertoire of TCRs through a stochastic process known as V(D)J recombination, producing hundreds of billions of distinct T-cell clonotypes (i.e., T-cells with unique receptors)~\cite{Qi2014-pz, Lythe2016-xv, Mora2019-zn}.}

\begin{figure*}[t!]
\centering
\includegraphics[width=0.9\linewidth]{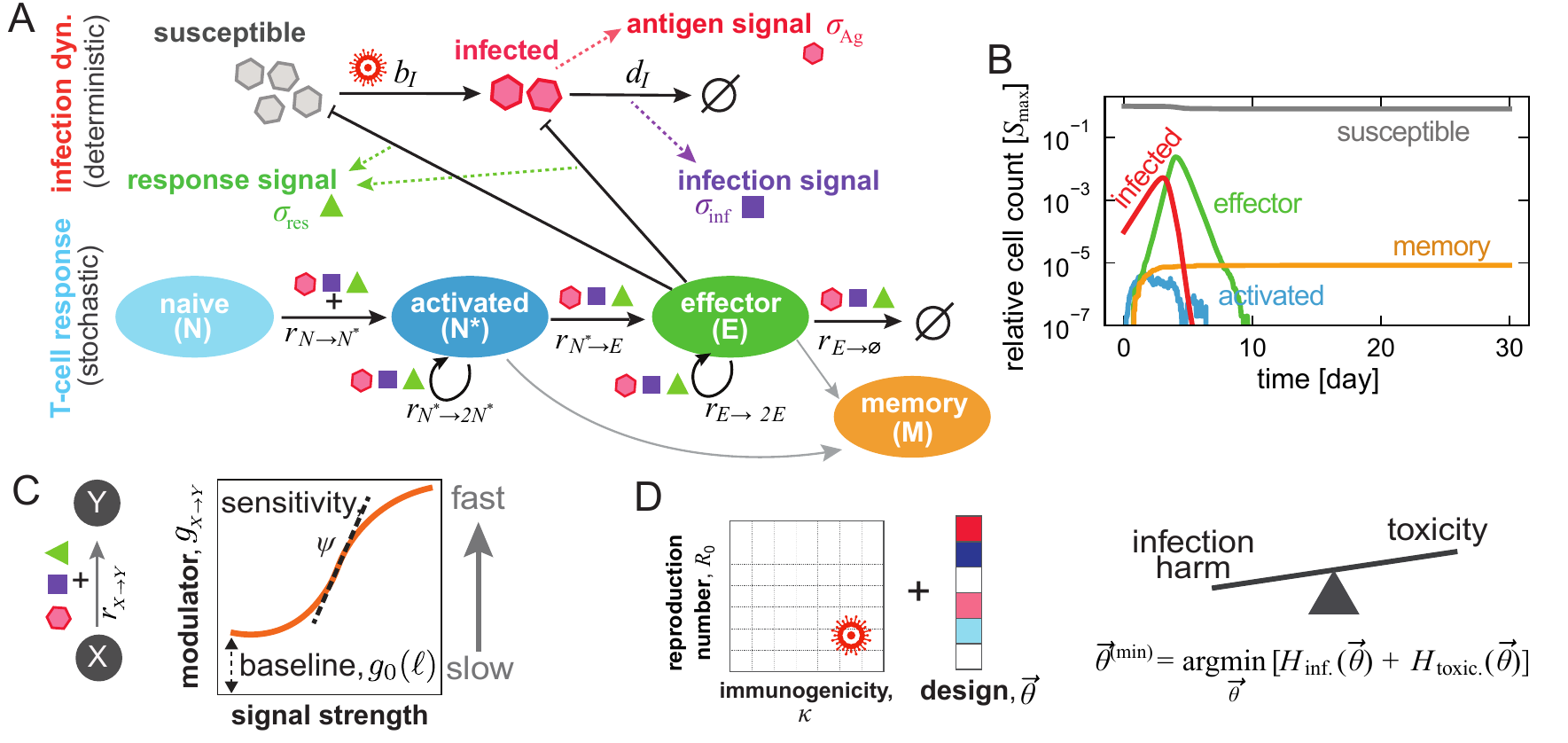}
\caption{{\bf CD8$^+$ T-cell response to infections.} {\bf (A)} The schematic shows how upon infection, susceptible cells $S$ become infected $I$ at rate $b_I$. Infected cells provide an antigen signal $\sigma_\Ag \propto I$, and die at rate $d_I$, which triggers an infection harm signal $\sigma_\infc$. In response to infection, na\"ive CD8$^+$ T-cells (with initial clone size of $N_\text{lin.}$) are activated $N\to N^*$, and can then differentiate into effector or memory fates ({Fig.~\DiffSchematic}). Effector cells can proliferate and kill infected and bystander cells, generating a response-harm signal $\sigma_\res$, and finally become terminal or differentiate into memory. Consequently, T-cell transitions are modulated by the threes signals ($\sigma_\Ag,\sigma_\infc,\sigma_\res$). Infection dynamics $(S,I)$ are modeled by ODEs coupled {to} stochastic dynamics of the T-cell response. Memory transitions (gray arrows) are neglected in our model.
{\bf (B)} Representative trajectories of susceptible, infected, effector, and memory CD8$^+$ T-cells are shown over time.
Cell counts are reported in units of the maximum number of susceptible cells $S_\text{max}$.
 {\bf (C)} Each T-cell transition has a rate $r_i = r^\text{max} g_i (\sigma_\Ag, \sigma_\infc, \sigma_\res; \vec\psi,\ell)$, modulated by a sigmoidal function with baseline $\ell$ and signal sensitivities $\vec\psi$ (eqs.~\ref{eq:rate_regulation},~\ref{eq:monod}). The regulatory parameters define the design of the effector program $\vec \theta$ (eq.~\ref{eq:theta}). 
{\bf (D)} We assess design performance over diverse infections sampled from a broad range of reproduction numbers $R_0$ (eq.~\ref{eq:R_0}) and immunogenicities $\kappa$ (eq.~\ref{eq:kappa}). 
For each infection, we can find an optimal design $\vec\theta^{(\mathrm{min})}$ that minimizes total cell loss (harm). Harm decomposes into infection harm $H_\text{inf.}$ (cells killed by the pathogen plus infected cells eliminated by effectors), and toxicity $H_\text{toxic.}$ (healthy bystanders killed by effectors), highlighting an intrinsic trade-off (eq.~\ref{eq: clear_tox}). 
Simulation parameters in {(B)}: $S_\text{max} = 10^7, N_\text{lin.} = 100, K_I = 10^{-3}S_\text{max}, K_S = S_\text{max}, K_H = 10^{-3}S_\text{max}, b_I = 2.0\times10^{-7} \text{day}^{-1}~\text{per cell},d_I = 0.5~\text{day}^{-1}, I_0 = 10^{-4}S_\text{max}, T = 30~\text{days}, \vec\theta =(\psi_{Ag} = 2, \psi_{I} = 2, \psi_{E} = -0.25, \ell_{N \to N^*} = -0.75,\ell_{N^*\to E} = -0.75,\ell_{E\to \emptyset} = 0.75$).}
\label{fig1}
\end{figure*}

Pathogen recognition triggers a response program in na\"ive T-cells, tuned by antigen signal, pro- and anti-inflammatory signaling molecules (cytokines), and other environmental signals generated by the infection and the host's response (Fig.~\ref{fig1}A)~\cite{Van_Stipdonk2003-ui,Joshi2007-fz, Marchingo2014-dh, Marchingo2016-ao, Eizenberg-Magar2017-cp}. These signals determine the extent of burst-like proliferation of short-lived effector T-cells versus differentiation into long-lived, slowly dividing memory subsets. Effector T-cells have the capacity to rapidly divide---up to 19 cell divisions~\cite{Badovinac2007-ii,Zhang2011-dv,Buchholz2013-ab,Marchingo2016-ao}---and secrete cytolytic chemical molecules that can degrade infected or unwanted cells~\cite{Sarkar2008-ob, Jenkins2008-if}. While most activated na\"ive cells differentiate into terminal effectors that undergo apoptosis once infection is cleared, a small fraction survive as long-lived effector or central memory cells~\cite{Seder2008-xh, Sarkar2008-ob, Youngblood2017-my, Abadie2024-rd}.

The logic of signal integration for immune decision-making throughout a response remains the central open problem. {Simple additive models of antigen and cytokine inputs can reproduce experimental patterns of clonal expansion in some settings~\cite{Marchingo2014-dh,Marchingo2016-ao,Eizenberg-Magar2017-cp}. However, genetic and single-cell analyses reveal far richer control circuits governing activation, differentiation, and contraction of T-cells that involve intricate genetic programs and probabilistic regulatory rules~\cite{Heinzel2017-qp, Cho2017-ul, Abadie2024-rd, Plambeck2022-dt, Xin2016-ut, Youngblood2017-my, Buchholz2013-ab, Kretschmer2020-ad}.}

Pathogens vary in their immunogenicity---the ease with which they are recognized as ``non-self"---and in the degree of harm they impose via secreted toxins, inflammation, and disruption of cellular function. The CD8$^+$ T-cell response must have the potential to curb ``harm" across diverse infection scenarios~\cite{Downie2021-xe, Mayer2016-mw}, yet the evolutionary objectives and biophysical constraints shaping their decision-making program remain unclear. Here, we formalize one plausible objective for T-cell decision-making: to minimize cumulative host healthy-cell loss due to infection and collateral death of healthy cells due to the immune response (immunopathology or toxicity).

We model the T-cell decision-making program as a flexible, coarse-grained signal-integration protocol. Specifically, we cast each CD8$^+$ T-cell as a feedback controller whose intra-cellular regulatory program modulates activation, differentiation, proliferation, and death in response to intra-host cues, such as antigen levels and cytokine signals, which themselves are dynamic in the course of an infection~\cite{Van_den_Berg2004-pn, Marchingo2016-ao}.

We then ask how alternative immunological contexts require distinct effector programs. First, we parameterize a broad class of feedback-controller designs, with varied signal sensitivities and integration rules. Next, by systematically sampling designs, we optimize for rapid pathogen clearance while minimizing immunopathology. {We find that optimal designs lie on a Pareto front: further reductions in infection-induced harm entail increased immunopathology. Accounting for the imperative to separate infection---microbial or malignant---from transient inflammatory events (e.g., tissue damage)~\cite{Green2009-el, Labbe2008-px,Butler2013-li,Cicchese2018-po, Medzhitov2021-to}, we show how a trade-off between pathogen-clearance and immunopathology (toxicity) may be encoded into the logic of T-cell responses. Optimizing these competing objectives yields effector programs consistent with experiments with vertebrate organisms.} Finally, applying our framework to a minimal tumor-growth model, we demonstrate how modifying the endogenously encoded effector program can improve the effectiveness of T-cell cancer therapies, pointing to signal-processing adjustments through genetic engineering that could enhance T-cell-based cancer immunotherapy.

The immune system's true performance objective, and how evolution has fine-tuned it, may ultimately be unknowable. Formulating and testing principled heuristics provides a baseline to gauge the performance of real immune systems in responding to infections. This perspective has yielded deep insights in other biological contexts, revealing the architectural logic of genetic and regulatory networks~\cite{Francois2004-mc, Shinar2007-mz, Lalanne2013-rr, Hart2014-ms, Alon2019-ep, Sokolowski2023-um}. In immunology, high-level mathematical models have demonstrated how a finite lymphocyte repertoire can be distributed to best confront a diverse pathogenic landscape~\cite{Perelson1979-kq,Mayer2015-ii, Bradde2020-lt} and how immune memory should be allocated to counter future threats~\cite{Mayer2019-tx, Schnaack2021-wh, Chardes2021-pf, Schnaack2022-jf, Qin2023-is}. By extending this perspective to the dynamics of CD8$^+$ T-cell signal integration and response, our work provides a quantitative framework that links observed decision-making to operating principles and trade-offs. 

\section{Model}

To characterize the immune response to an infection, we model the large-population dynamics of susceptible and infected cells in a host as deterministic processes, coupled to a stochastic, time-inhomogeneous birth-death process for the CD8$^+$ T-cell response (Fig.~\ref{fig1}A, B and SI Appendix).

\paragraph{Intra-host infection dynamics.} We model a population of  healthy, susceptible cells $S(t)$. 
Pathogen exposure seeds an infected subpopulation $I(t)$, which grows over time $t$ via encounters with an otherwise homeostatically stable susceptible cell population, and declines through infected-cell death,  triggering an immune response (Figs.~\ref{fig1}A, B~\DiffSchematic). Activated effector T-cells $E(t)$ can kill both infected and susceptible cells, with clearance governed by {\em recognition thresholds} $K_I$ and $K_S$, respectively, which quantify how readily effector cells recognize each cellular subpopulation; higher thresholds indicate that more effector cells are needed for recognition and killing. We model this dynamics as a deterministic birth-death process with  competitive interaction of T-cells with infected and healthy cells; see Materials and Methods  and the SI Appendix.

We quantify within-host reproductive success of a pathogen by the \emph{intra-host basic reproduction number},
\begin{equation}
\label{eq:R_0}
R_0 = \frac{b_I S_\text{max}}{d_I},
\end{equation}
where $b_I/d_I$ is the ratio of birth to death rates for infected cells, and $S_\text{max}$ is the maximum number of susceptible cells in a tissue (Materials and Methods).  $R_0$ measures the expected number of secondary infected cells generated by a single infected cell in an otherwise fully susceptible cell population. This definition mirrors the standard epidemiological reproduction number \cite{Milligan2015-yu, Abuin2020-tf}. 

We model the T-cell repertoire as initially consisting of a number of identical na\"ive T-cells, and assume that the T-cell recognition threshold for  infected cells is lower than that for susceptible cells  ($K_I \leq K_S$), such that T-cells more readily respond to infected cells. Accordingly, we define the {\em immunogenicity} of an infection as,
\begin{equation}
\label{eq:kappa}
\text{immunogenicity:}~~~~\kappa = K_I^{-1}K_S.
\end{equation}
This dimensionless quantity reflects how readily cognate T-cells recognize antigens presented by infected cells as ``non-self" relative to those presented by healthy cells (self). We consider infection scenarios with broad ranges of immunogenicity ($\kappa$) and basic reproduction numbers ($R_0$) (SI Appendix).

\paragraph{Signal cues to modulate T-cell decision-making.} We express damage during an infection in terms of cell deaths,
\begin{align}
\begin{split}
    h(t) &=
    h_{\infc}(t) + \underbrace{ h_{E,I}(t) + h_{E,S}(t)}_{=h_\res(t)},\label{eq:harm}
    \end{split}
\end{align}
where $h_\infc(t)$ represents the number of cells killed directly by the pathogen (infection), and $h_{E,I}(t)$ and $h_{E,S}(t)$ denote the number of infected and healthy cells, respectively, killed by the effector T-cell response, amounting to the total damage {from} the response $h_\res(t)$.

Cell deaths can produce different types of signals. Unprogrammed cell death (e.g., pyroptosis triggered by viruses and bacteria) can lead to the  release of pro-inflammatory danger-associated molecular patterns (DAMPs) \cite{Green2009-el}. On the other hand, cell death initiated by CD8$^+$ T-cells generates a heterogeneous signal that generates pro- and anti-inflammatory cytokines {(e.g., IFN$\gamma$, TNF$\alpha$, IL-1ra, IL-4, IL-10, TGF-$\beta$)}~\cite{Green2009-el, Medzhitov2021-to, Labbe2008-px}.

We simplify this picture by partitioning non-antigen cues into (i) infection-induced and (ii) response-induced harms, whose accumulated effects trigger infection and response harm-induced signals, $\sigma_{\text{inf}}(t)$ and $\sigma_{\text{res}}(t)$, respectively. Presented antigens on APCs serve as an additional  cue. We take the infected-cell burden $I(t)$ as a proxy for the instantaneous antigen signal $\sigma_{\text{Ag}}(t)$, which together with the other two signals, modulate the cellular decision-making processes in Fig.~\ref{fig1}A; see Materials and Methods and SI Appendix for details. By disentangling these sources of (potentially redundant) signals, we probe how CD8$^+$ T-cells use these cues to make fate decisions. We do not pre-assign any signals to a pro- or anti-inflammatory role but allow optimization goals to determine their {influence} across diverse infection scenarios.

We should emphasize that CD8$^+$ responses do not occur in isolation: the innate immune system, regulatory T-cells (Tregs),  helper T-cells, and B-cells can all directly or indirectly shape CD8$^+$ dynamics. While these interactions are not modeled explicitly, we interpret the harm and antigen signals as coarse-grained proxies for inputs from other immune compartments, capturing their net pro- and anti-inflammatory effects. In the same vein, our immunogenicity parameter $\kappa$ (eq.~\ref{eq:kappa}) may also subsume regulatory constraints, e.g. through  Treg-mediated suppression that limits effector responses to self-like antigens, which in our framework correspond to low-immunogenicity targets.  Incorporating these additional components explicitly would be an important direction for future work, enabling the study of richer feedbacks and more granular immune regulation and dynamics.

\paragraph{Design of T-cell programs for infection response.} Na\"ive T-cells cycle through lymphoid tissues, becoming activated when they durably engage cognate peptide-MHC complexes on activated APCs~\cite{Van_Stipdonk2001-nm,Francois2013-yl}. Activated T-cells can differentiate into an effector phenotype, which participates in infection clearance, or into a long-lived memory state ({Figs.~\ref{fig1}A,~\DiffSchematic}). Here, we examine the course of CD8$^+$ T-cells response during an infection, dividing the process into four canonical response modules (Fig.~\ref{fig1}A):
\begin{enumerate}
    \item {\bf Activation:} priming of na\"ive T-cells, $N \to N^*$,
    \item {\bf Differentiation:} commitment of the activated T-cell to an effector phenotype, $N^* \to E$,
    \item {\bf Expansion:} division of effector cells, $E \to 2E$,
    \item {\bf Contraction:} programmed loss of effector cells, $E \to \emptyset$.
\end{enumerate}
These response processes map onto partially distinct genetic programs: the MYC pathway drives activation and proliferation~\cite{Heinzel2017-qp}; the Wnt-Tcf7 pathway regulates effector versus memory fate decisions~\cite{Gattinoni2011-ft, Abadie2024-rd}; contraction can be attributed to effector cell apoptosis controlled by the Bcl-2-family of proteins, or the differentiation of effectors into effector-memory cells~\cite{Abadie2024-rd}. The latter is regulated by the balance of T-bet/Blimp-1 (effector) versus Bcl-6/Eomes (memory) programs, as well as metabolic pathways such as mTORC1~\cite{Cui2010-cz}. To reduce model degeneracy, we do not distinguish among these distinct biological sources of contraction in our analysis.

For each process $i$, we represent its genetic regulation by a time-dependent transition rate $r_i(t)$, modulated by the instantaneous signals $\vec\sigma(t)$ (eq.~\ref{eq:Ag-Cytokine-Signal}) received by the T-cell (Fig.~\ref{fig1}C), 
\begin{align}
\label{eq:rate_regulation}
    r_i(t) =& r^\text{max}\ g_i\left(\sigma_\Ag,\sigma_\infc, \sigma_\res;\vec{\psi}, \ell_{i}\right),
\end{align}
where the maximal rate $r^\text{max}$ is capped by the cell-cycle {speed}~\cite{Jenkins2008-if,Kretschmer2020-ad, Van_Stipdonk2001-nm}, and $g_i$ is a monotonic {\em response function} of signal input, chosen from the modified Monod-Wyman-Changeux functions (Hill-type family)~\cite{De_Ronde2012-py, Walczak2010-gs}:
\begin{align}
\begin{split}
&g_i(\sigma_\Ag,\sigma_\infc, \sigma_\res;\vec{\psi}_i, \ell_i) = \frac{1}{1 + \exp\left[-L_i(\vec\sigma;\vec{\psi}_i, \ell_i)\right]}, \\
&\text{with,} \\
&L_i(\vec\sigma;\vec{\psi}_i, \ell_i)= \ell_{i} + \sum_{\alpha \in\{\Ag,\infc,\res\}}\psi^\alpha_i \log\left[1 + {\sigma_\alpha}\right].
\end{split}
\label{eq:monod}
\end{align}
The parameter $\ell_i$ sets the baseline propensity of transition $i$: in the absence of external signals, the regulatory factor evaluates to $g_{0,i}= (1+\exp(-\ell_{i}))^{-1}$ (Fig.~\ref{fig1}C). The weights $\{\psi_i^{Ag}, \psi_i^\infc, \psi_i^\res\}$ encode the sensitivity and polarity (positive / negative = up /down-regulation) of response to the antigen ($\sigma_\Ag$) and harm signals ($\sigma_\infc$ and $\sigma_\res$). This formulation abstracts away biochemical details while implementing the fold-change response to stimuli observed in many biological circuits~\cite{Hart2014-ms, Oyler-Yaniv2017-yr, Achar2022-fl, Voisinne2015-ed, Adler2014-lz}.

To reflect shared regulatory programs while controlling complexity, we impose two simplifications. First, activation and early expansion are both MYC-driven~\cite{Heinzel2018-wr}; we therefore require the modulatory factors, $g_{N\to N^*}$ and $g_{E\to 2E}$, to share parameter values, even though their realized rates will differ because signals vary in time, as cells transition between naive and effector states. Second, logically, since extracellular cues should converge on coherent effector programs, we enforce a common set of signal sensitivities across modules. With the sign convention that positive sensitivities promote effector production, for each signal $\sigma$, we set $\psi_\sigma \equiv \psi^\sigma_{N\to N^*} = \psi^\sigma_{E\to 2E} = \psi^\sigma_{N^*\to E} = -\psi^\sigma_{E\to\emptyset}$.

\begin{figure*}[t!]
\centering
\includegraphics[width=\linewidth]{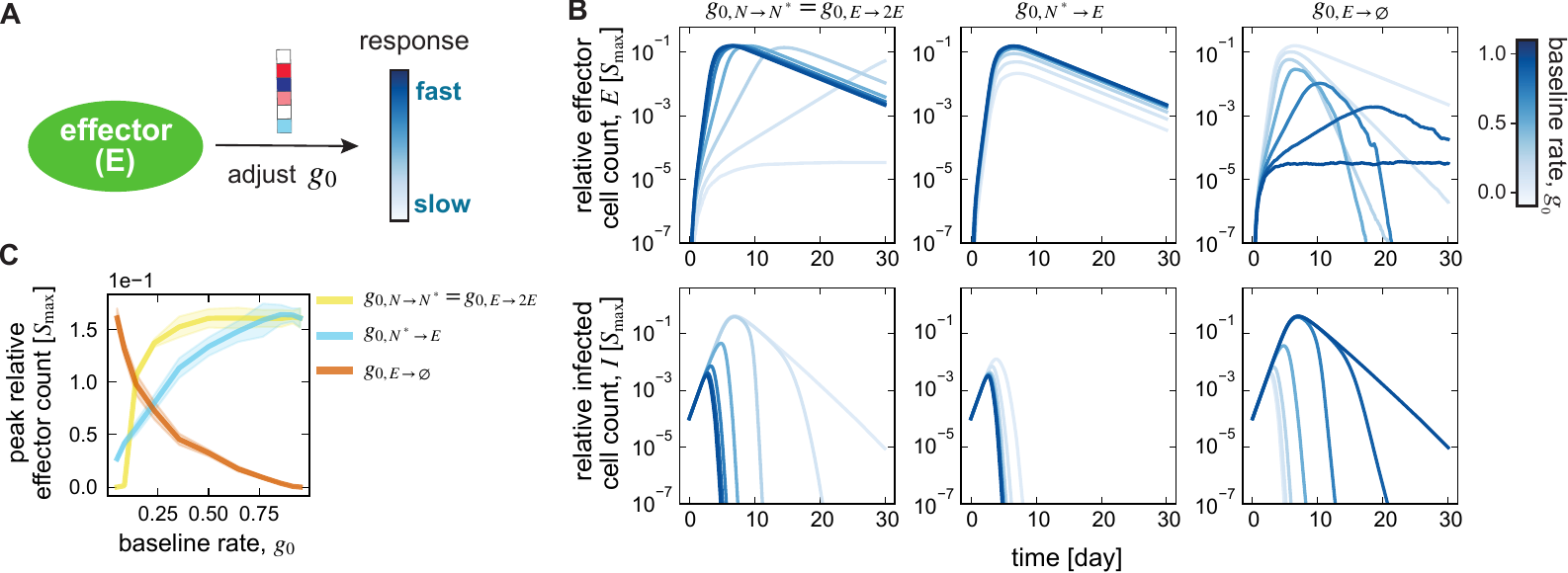}
\caption{{\bf Design parameters shape T-cell response and infection clearance.} {\bf (A)} Varying the design parameters influences effector trajectories. {\bf(B)} Trajectories are shown for the population size of effectors (top) and infected cells (bottom) for designs that vary a single baseline transition rate $g_{0,i}$ (color) indicated atop each panel. Both quantities are measured in units of the maximum number of susceptible cells, $S_\text{max}$, which is a convenient reference scale for different cell sub-populations. All other design parameters are fixed at: fast activation, $g_{0,N \to N^*} = 0.95$; fast differentiation, $g_{0,N^* \to E} = 0.95$; slow contraction, $g_{0,E\to \emptyset} = 0.05$; and zero signal feedback, $\psi_{\Ag} =\psi_\infc =\psi_\res = 0$. Trajectories reflecting the impact of varying signal sensitivities $\psi$'s are shown in SI Appendix, Fig.~\DesignResponseClearance. {\bf(C)} Peak effector count, measured in units of maximum number of susceptible cells $S_\text{max}$, is shown by varying each of the three baseline transition rates (colors), while keeping the rest of the parameters fixed (similar to (B)). Solid lines show the mean of effector counts across $n=5$ simulation replicates and shaded envelopes indicate one standard deviation (Materials and Methods). Parameters are the same as Fig.~\ref{fig1}, except for $ K_I = 10^{-2}S_\text{max}$.}
\label{fig2}
\end{figure*}

The parameters of these constrained response functions define a {\em response design} $\vec \theta$ as,
\begin{equation}
\label{eq:theta}
\vec \theta = (\underbrace{\psi_{\Ag},\psi_{\infc},\psi_{\res}}_\text{signal feedback}, \underbrace{\ell_{N\to N^*} = \ell_{E\to 2E}, \ell_{N^*\to E},\ell_{E\to \emptyset}}_\text{module-specific baselines}).
\end{equation}
We systematically sample $\vec \theta$'s to assess performance of designs against different infections (Fig.~\ref{fig1}D; SI Appendix). While $\vec{\theta}$ is broadly sampled, we hold the remaining biological rates and design parameters underlying the decision-making program in Fig.~\ref{fig1}A within empirically supported ranges reported in SI Appendix, Table~\ParameterTable.

\paragraph{Cell economics of T-cell response.} Tracking all clinical sequelae of an infection is rarely feasible, but tissue integrity degrades as cells are lost. Cumulative cell death correlates with macroscopic readouts such as weight change, pathogen burden, and temperature dynamics~\cite{Gupta2021-vo,Lebel2025-ah}. We therefore adopt a cell-economics objective in which harm is measured by the total number of cells lost over the course of infection. We evaluate the performance of a given design $\vec \theta$ by simulating the response dynamics (Fig.~\ref{fig1}D) and computing the {\em total cell death} (or inflicted harm) as
\begin{align}
\label{eq: clear_tox}
    \begin{split}
  H(\vec\theta) =
  \underbrace{\int _0^{T}\left[h_\infc\left(t, \vec\theta\right) + h_{E,I}\left(t, \vec\theta\right)\right]{\rm d}t}_{H_{\rm inf.}(\vec\theta)} + \underbrace{\int _0^{T} h_{E,S}\left(t, \vec\theta\right){\rm d}t }_{H_{\rm toxic.}(\vec\theta)},
    \end{split}
\end{align}
where $H_\text{inf.}(\vec\theta)$ aggregates infection-related harm (cells killed by the pathogen, and infected cells eliminated by effectors), $H_\text{toxic.}(\vec\theta)$ captures immunopathology (healthy bystander cells killed by effectors), over the fixed time window $T$. In our main simulations, we set $T = 30$ days to reflect the typical timescale of acute infections and the associated immune response (SI Appendix, Table~\ParameterTable), but we also assess the sensitivity of our results to variation in $T$ (Materials and Methods). This single, interpretable metric of harm (eq.~\ref{eq: clear_tox}) balances clearance benefits against collateral damage, enabling principled comparison across designs and infections (Fig.~\ref{fig1}D).

\begin{figure*}[t!]
\centering
\includegraphics[width=\linewidth]{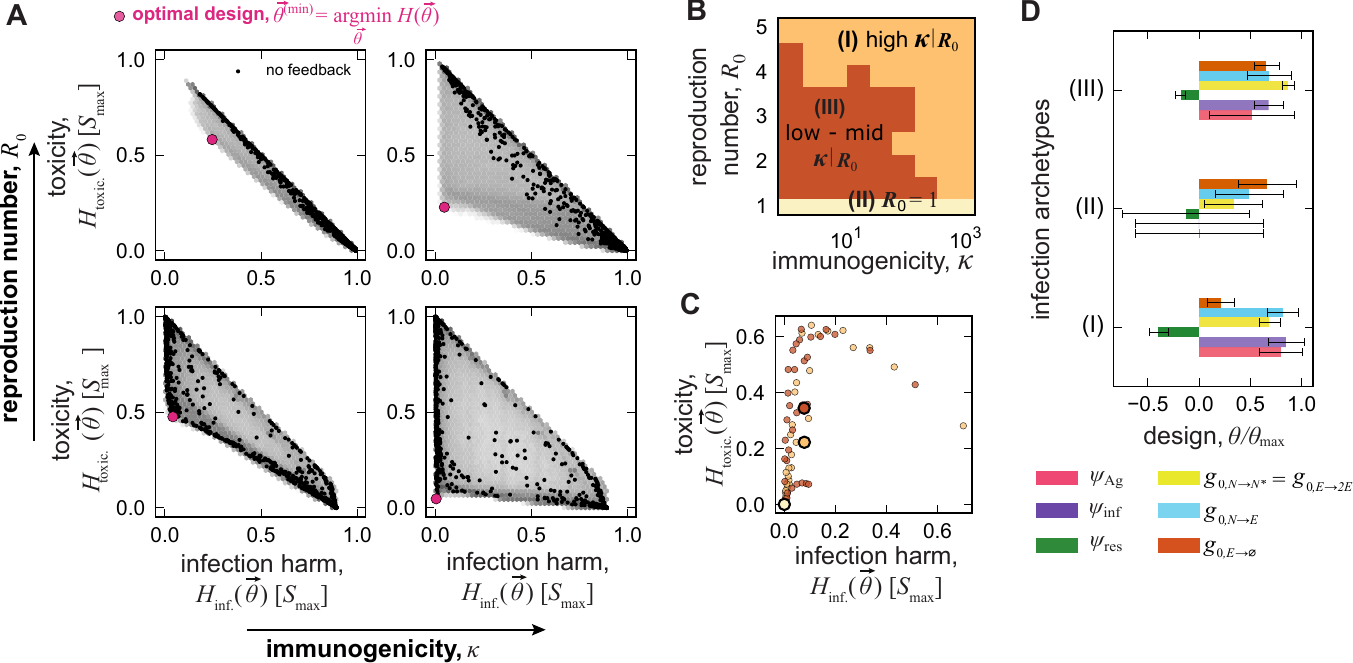}
\caption{{\bf Toxicity-infection harm trade-off across T-cell regulatory designs.} 
\textbf{(A)} Density plots of toxicity $H_\text{toxic.}$ versus infection harm $H_\text{inf.}$ (eq.~\ref{eq: clear_tox})---measured in units of the maximum number of susceptible cells $S_\text{max}$---for $4.8\times 10^6$ T-cell regulatory designs, $\vec\theta$, uniformly grid-sampled across four infection scenarios {(panels; columns: $\kappa= 5.6$ (left), $\kappa= 178$ (right); rows: $R_0 = 2.5$ (bottom), $R_0 = 5.0$ (top))}. Designs are sampled uniformly across signal sensitivities $\psi\,\text{'s} \in[-3,3]$ and baseline rates $\ell\,\text{'s} \in[-3,3]$ (eq.~\ref{eq:theta}). Infection scenarios are sampled by changing $K_I$ and $b_I$, while holding $S_{\max}$, $d_I$, and $K_S$ fixed. Black points denote no-feedback designs ($\vec\psi=0$), and the large magenta circle in each panel marks the design that minimizes total harm, $\vec\theta^{(\mathrm{min})}$.
{\bf (B)} {For many different infection scenarios $(\kappa,R_0)$, we identify an ensemble of near-optimal designs $\Theta(\kappa,R_0)$, whose total harm is within $10^{-2}\times S_\text{max}$ of the minimum sampled total harm (as in the magenta point in A). By computing the ensemble-averaged design parameters $\langle\vec\theta\rangle_{\Theta(\kappa,R_0)}$ for each infection scenario, and using these averages to cluster infections, we identify three pathogen ``archetypes" (colors): I--III, as indicated (SI Appendix).
{\bf (C)} Toxicity vs. infection harm associated with the optimal harm-minimizing designs for each pathogen archetype in (B) is shown.
Large points indicate archetype means; small semi-transparent points show individual immune challenges sampled in (B). 
Reducing infection harm (horizontal axis) comes at the expense of increased toxicity (vertical axis). {\bf (D)} Bars indicate the ensemble-mean of design parameters, $\langle\vec\theta\rangle_{\Theta^{(i)}}$, computed by identifying the ensemble of designs ${\Theta^{(i)}}$ which minimize cluster-averaged total harm for each archetype  within $10^{-2}\times S_\text{max}$ of the minimum sampled cluster-averaged total harm (similar to (B)). Error bars denote the standard deviation within this ensemble.} All design variables are normalized by their respective bounds ($\psi^{\max}=3$, $\ell^{\max}=3$).
Unless otherwise noted, results represent the average of $n = 5$ simulation replicates with  parameters: $S_\text{max} = 10^7, N_\text{lin.} = 10^{-5}S_\text{max}, K_S = S_\text{max}, K_H = 10^{-3}S_\text{max}, d_I = 0.5~\text{day}^{-1}, I_0 = 10^{-4}S_\text{max}.$ 
}
\label{fig3}
\end{figure*}

\section{Results}
\subsection*{Cell-fate transitions determine effector response and pathogen clearance}
How do rates of cell-fate transitions, determined by the regulatory design $\vec\theta$ (eq.~\ref{eq:theta}), impact the immune response and pathogen clearance? To isolate the effect of baseline propensities ($\vec\ell$) independently of signal feedback ($\vec\psi$), we first simulated ``hard-wired" designs in which all signal sensitivities were set to zero ($\vec\psi_i=0$). Each transition then happens at a fixed fraction of its maximal rate $g_{0,i} = [1+\exp(-\ell_i)]^{-1}$, independently of antigen or harm signals; here, $i\in\{N\to N^*/E\to 2E, N^* \to E, E\to \emptyset\}$ indicates the transition module.

Varying these baseline rates produced strikingly different effector trajectories (Fig.~\ref{fig2}). Promoting activation/proliferation ($g_{0,N\to N^*}, g_{0,E\to 2E}\, \uparrow$), or lowering the contraction baseline rate ($g_{0,E\to \emptyset}\,\downarrow$) generated a rapid, high-amplitude effector burst that cleared the pathogen quickly, after which effector numbers decay (Fig.~\ref{fig2}B, C). In contrast, slowing activation and division ($g_{0,N\to N^*} , g_{0,E\to 2E}\,\downarrow$), or accelerating contraction ($g_{0,E\to \emptyset}\,\uparrow$), yielded sluggish effector accumulation, delayed clearance, and in extreme cases pathogen persistence, accompanied by a prolonged plateau of few effectors. Interestingly, varying the baseline rate of differentiation $g_{0,N^*\to E}$ had a weaker impact on the overall effector profile and pathogen clearance (Fig.~\ref{fig2}B, C), suggesting {its relevance may lie elsewhere} in the response program.

Because effector and pathogen abundances determine the levels of antigen and cumulative harm-associated (cytokine) signals, these baseline choices not only shape the immediate effector response but also alter the available antigen and cytokine signals, which in turn impact the T-cell response through feedback regulation (SI Appendix, Fig.~\DesignResponseClearance).

\subsection*{Effector designs trade infection harm for toxicity} Previous research has identified a trade-off between maximizing infection clearance and minimizing response toxicity (immunopathology) in the immune response \cite{Johnson2011-ja,Cicchese2018-po, Medzhitov2021-to}. Consistently, by systematically sampling the space of designs (SI Appendix), we found the same trade-off: plotting infection harm $H_\text{inf.}(\theta)$ (i.e., infected cells killed by pathogen or by effector response) versus the toxicity $H_\text{toxic.} (\theta)$ (i.e., bystander cells killed by effector response) revealed that decreasing $H_\text{inf.} $ comes at the cost of increasing $H_\text{toxic.}$ (Fig.~\ref{fig3}A). The extent of this trade-off depends on pathogen characteristics. This trade-off is stronger for infections with a higher intra-host reproductive number $R_0$ or lower immunogenicity $\kappa$. Interestingly, for scenarios with low $R_0$, many of the hard-wired response programs without signal feedback ($\vec\psi_i  = 0$; black dots) perform near-optimally, laying close to the {\em Pareto front}. Here, the Pareto fronts at the boundaries of these graphs indicate the set of  strategies for which neither harm nor toxicity can be reduced without increasing the other, representing the best achievable balance between these competing objectives (see SI Appendix).
 For aggressive high-$R_0$ threats---be it acute infections (high $\kappa$), or highly metastatic cancers (mid to low $\kappa$)---these same hard-wired designs fall far from the Pareto front and well inside the sub-optimal region. Proportionate control in these cases requires designs with appropriate sensitivities to antigen and cytokine/harm signals ($\vec\psi\neq 0$) to minimize infection harm while limiting immunopathology.

\begin{figure*}[t!]
\centering
\includegraphics[width=0.9\linewidth]{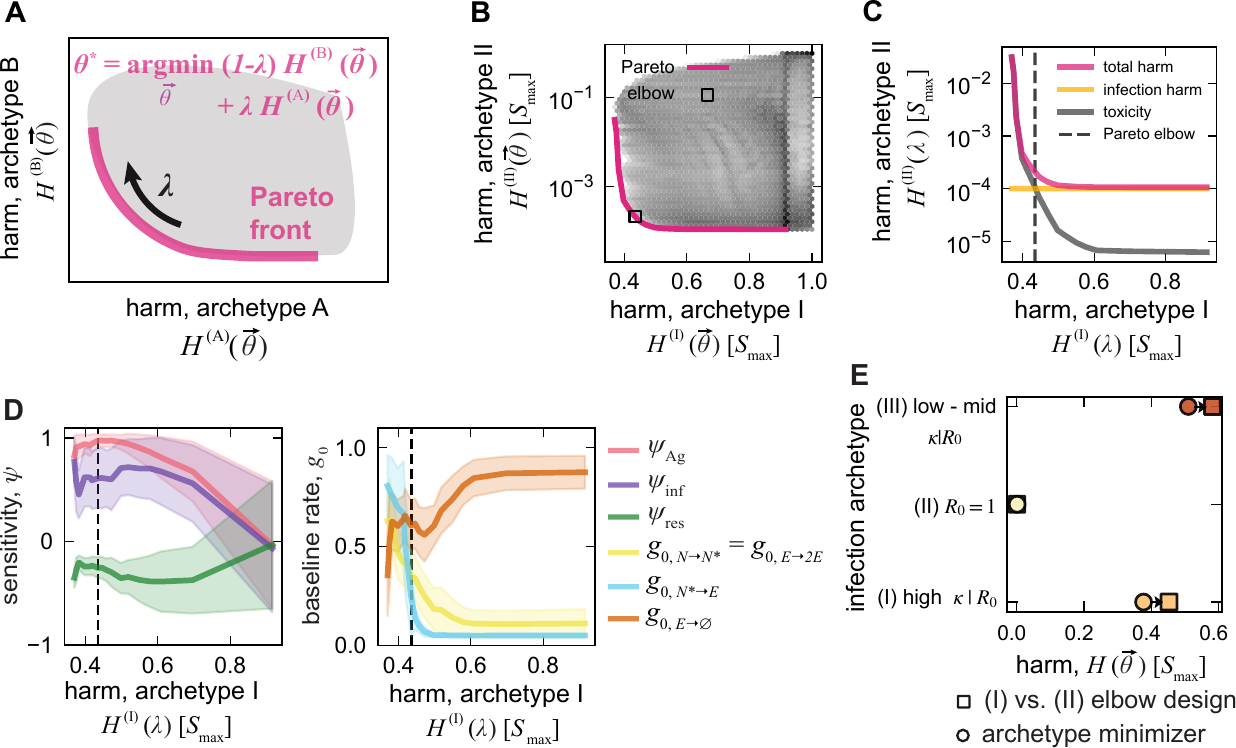}
\caption{{\bf Pareto-optimal designs balancing harm from acute infections and autoimmunity.}
{\bf(A)} Bi-objective optimization to minimize the weighted sum of harms associated with T-cell regulatory designs $\vec{\theta}$ across two pathogen archetypes ($A,B$) yields a Pareto front (magenta curve); eq.~\ref{eq: pareto optimization}. Points on the curve are Pareto-optimal designs, parameterized by the weight $\lambda \in [0,1]$, quantifying the emphasis on archetype A; moving left to right corresponds to decreasing $\lambda$ (less emphasis on archetype {A}). {\bf (B)} The density plot shows harms (in units of the maximum number of susceptible cells $S_\text{max})$  incurred by $4.8\times 10^6$ uniformly sampled designs in their response to archetype I (immunogenic infection; high $\kappa$) versus archetype II (non-replicating, self-like/ auto-immune inducing or allergenic antigens; $R_0=1$), as defined in Fig.~\ref{fig3}B. Designs are sampled on a uniform Cartesian grid of sensitivities $\psi_\sigma\in[-3,3]$ and baseline rates $\ell_i\in[-3,3]$ (as in Fig.~\ref{fig3}A). For many values of $\lambda$, we minimize the objective in (A), and trace out the Pareto front (magenta). For each point $\lambda$ on the front, we identify the ensemble of designs closest to the Pareto front ($\Theta(\lambda)$), where a design belongs to the near-Pareto ensemble at $\lambda$ if its incurred harm against archetypes I and II simultaneously lie within 5\% of the $\lambda$-Pareto-optimal harm of the respective archetypes. The black square indicates the elbow of the Pareto curve $\lambda^*$: the point where further reductions in archetype-{II} harm incur larger increases in archetype-{I} harm (SI Appendix). {\bf (C)} Decomposition of total harm along the Pareto front in (B) into infection-related harm and toxicity (eq.~\ref{eq: clear_tox}). The black dashed line indicates the position of the elbow from (B). {\bf(D)} The magnitudes of the signal sensitivities $\psi$ (left) and the baseline rates $g_0(\ell)$ (right), each scaled to variable bounds $(\psi^{\max}=3,\ \ell^{\max}=3)$, are shown as a function of harm incurred under archetype {I} along the Pareto curve (parameterized by $\lambda$). Solid and shaded envelopes show the mean and standard deviation of the design parameters, respectively, across the near-Pareto ensemble of designs $\Theta(\lambda),~\lambda \in [0,1]$ traced out by the Pareto front. {\bf (E)} For each archetype in Fig.~\ref{fig3}B, total harm is shown for the archetype-specific minimizer $\vec{\theta}^{(\mathrm{min})}$ (circle) and for the average over {the} near-elbow ensemble in (B). See SI Appendix, Fig.~\ParetoArchOne~for Pareto fronts formed by other pairs of archetypes. All simulation parameters other than the varied $K_I$, $b_I$, and $\vec{\theta}$ match Fig.~\ref{fig1}.
}
\label{fig4}
\end{figure*}

Next, we explored regulatory design space by evaluating responses of different designs against diverse infection scenarios, sampled on a grid of immunogenicities ($\kappa$) and reproduction numbers $(R_0)$ (Fig.~\ref{fig3}B). {For each infection scenario $(\kappa,R_0)$, we identified the ensemble of near-optimal designs, $\Theta(\kappa,R_0)$. A design $\vec\theta$ belongs to this ensemble if its incurred total harm (eq.~\ref{eq: clear_tox}) is within $10^{-2}\times S_\text{max}$ of the minimum realized total harm from all sampled designs for that infection; note that the maximum number of susceptible cells $S_\text{max}$ provides a convenient reference scale against which to quantify harm (cell death). Finally, we clustered infections based on their ensemble-averaged design parameters $\langle\vec\theta\rangle_{\Theta(\kappa,R_0)}$} (SI Appendix, Fig.~\ClusterInput). Three coherent design classes emerged, loosely corresponding to intuitive pathogen ``archetypes" (Fig.~\ref{fig3}B): {\bf I---} highly immunogenic bacteria and viruses (high $\kappa$ or high $R_0$),  associated with type-I immunity~\cite{Iwasaki2015-qi}; {\bf II---} non-replicating, self-like/auto-immune inducing or allergenic antigens ($R_0=1$), in part associated with type-II immunity~\cite{Iwasaki2015-qi}; {\bf III---} mid-range threats (low-medium $\kappa$ and $R_0$) including difficult to detect pathogens and slowly growing tumors. The balance between minimizing toxicity and clearing infection (i.e., reducing harm) differs across these archetypes, resulting in different design strategies: signal-sensitive designs optimized for archetypes I and III prioritize pathogen clearance at the cost of increased immunopathology, with the severity of the trade-off varying by scenario, whereas the largely pre-programmed designs optimized for archetype II suppress toxicity to avoid autoimmunity (Fig.~\ref{fig3}C).

Design parameters of effector programs minimizing harm vary across archetypes (Fig.~\ref{fig3}D). For archetype {II} (non-replicating antigens) the optimal design is pre-programmed suppression of the immune response: signal sensitivities $\vec\psi$ are close to 0, and the baseline rates are tuned to slow activation, proliferation, and differentiation, and enforce rapid effector deletion. This damped response completely avoids toxicity when the threat does not grow so infection harm is determined by the initial exposure and cannot be decreased. Across the other two archetypes a common pattern emerges: the response-derived cytokine signal $\sigma_\res$ acts as an anti-inflammatory brake ($\psi_\res<0$), with its magnitude tuned to context. Moreover, antigen sensitivity $\psi_{\rm Ag}$ grows with immunogenicity, with the designs optimized for archetype {I} (strong immunogenicity) showing the largest sensitivity to antigens, while those for archetype {II} (non-immunogenic) show no antigen sensitivity, and are largely hard-wired.
Overall, across all archetypes, near-optimal designs distinctively tune both the baseline rates $g_0$'s and the infection-signal sensitivity $\psi_\infc$ to balance clearance against toxicity and minimize harm (SI Appendix, Figs.~\BaselineOnHarm,~\PsiOnHarm).

\subsection*{Endogenous T-cell programs are shaped by minimizing harm from autoimmunity and acute infections} 
Although the relative evolutionary pressures exerted by different threats are unknown, we can still pose a principled multi-objective problem for the effector program, i.e., find a design {$\vec \theta_\text{min}$} (eq.~\ref{eq:theta}) that minimizes the weighted sum of harms across a desired set of pathogen archetypes:
\begin{equation}
  \vec { \theta}_\text{min}= \underset{\vec\theta}{\arg\min} \sum_{i: \text{archetypes}} \lambda_i H^{(i)} (\vec \theta),
 \label{eq: pareto optimization}
\end{equation}
where $\lambda _i\in [0, 1]$ with the constraint $\sum_i \lambda_i=1$. The weight $\lambda_i$ specifies how strongly we prioritize minimizing harm associated with the  $i^{th}$ pathogen archetype {(i.e., an objective)}, when identifying the optimal design $\vec { \theta}_\text{min}$. {For $m$ archetypes, the solution to this family of problems maps an $m-1$-dimensional simplex of $\vec\lambda = ( \lambda_1, \dots,  \lambda_m)$ values to a set of designs that achieve the best possible trade-off between the different archetypes (Fig.~\ref{fig4}A; SI Appendix). When optimizing performance against one archetype necessarily worsens performance against another (i.e., {a} tradeoff), these optimal designs form a \textit{Pareto front}: a set of {solutions} for which no objective can be improved without degrading at least one other (Fig.~\ref{fig4}A). In contrast, if there are no such conflicts among archetypes, the Pareto front collapses to a single design that is simultaneously optimal for all objectives.  This framework naturally accommodates uncertainty in the  prevalence of pathogenic challenges  and in the effective ``preferences" imposed by evolution or environment, in finding the optimal solution.} Similar approaches have been successfully applied in other biological systems~\cite{Shoval2012-ln, Kocillari2018-pf}.

Even though optimization can be done simultaneously over all {three} archetypes, we proceed by examining pairwise trade-offs that generate a one dimensional Pareto front, parameterized by {a single scalar} $\lambda$ (Fig.~\ref{fig4}A). A biologically salient case contrasts {archetype I (immunogenic pathogens) with archetype II (non-replicating antigens).} Non-replicating antigens plausibly arise from un-programmed cell deaths or external trauma inducing inflammation and may be implicated in autoimmune flares. This bi-objective harm minimization aims to find designs balancing effective clearance of immunogenic viral or bacterial infections and suppression of autoimmune responses; see SI Appendix, Fig.~\ParetoArchOne~for Pareto fronts formed by optimizing other pairs of clusters.

Fig.~\ref{fig4}B shows harm from archetype {II} plotted against that of archetype {I} for a broad range of sampled designs and highlights the Pareto front (magenta).  Designs close to the Pareto front achieve near-best total protection for a given compromise weight $\lambda$; moving to the right along the Pareto front shifts the emphasis from clearance of immunogenic pathogens ($\lambda \uparrow$) to tolerance ($\lambda \downarrow$). Decomposing total harm into infection-related harm and toxicity (eq.~\ref{eq: clear_tox}) shows that along this Pareto front it is relatively easy to reduce toxicity to non-replicating antigens without substantially worsening outcomes against immunogenic infections; this is demonstrated in Fig.~\ref{fig4}C as substantial decreases in toxicity are achieved with only minor changes in infection harm.

\begin{figure}[t!]
\centering
\includegraphics[width=1.0\linewidth]{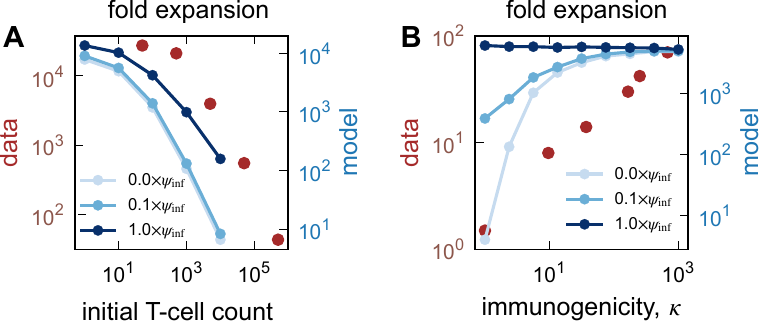}
\caption{{\bf Elbow-optimal designs recapitulating macro-dynamics of T-cell expansion.} {\bf (A,B)} In maroon (left vertical axis), fold expansion (i.e, peak clone size divided by the initial precursor number) of antigen-specific CD8$^+$ T-cells in mice plotted against {(A)} the initial precursor number (data from~\cite{Badovinac2007-ii}), and {(B)} the relative pMHC concentration required for half-maximal activation EC$_{50}$ (data from \cite{Zehn2009-up}); in our model, EC$_{50}$ is proxied by $K_I$. In blue (right vertical axis), simulated mean fold expansion for designs belonging to the near-elbow ensemble from Fig.~\ref{fig4}B ($\Theta(\lambda^*)$) shown as a function of the initial precursor T-cell number (A), and the immunogenicity of the antigen $\kappa$ (B). Their sensitivity to the infection-harm signal ($\psi_{\text{inf}}$) is varied (shades of blue). For each design in the near-elbow ensemble identified in Fig.~\ref{fig4}, $\vec\theta \in \Theta(\lambda^*)$, we change $\psi_\infc \to c\times \psi_\infc$ for constants $c \in \{0, 0.1,1.0\}$, and simulate responses against the broad sample of infections (Fig.~\ref{fig3}B). Trend lines are obtained by averaging over infections and designs in the near-elbow ensemble (SI Appendix, Figs.~\BaselineOnMacroDyn,~\PsieOnMacroDyn). Elbow designs with reduced sensitivity to infection-derived harm qualitatively match the experimental macro-dynamics of T-cell expansion. Unless otherwise indicated, simulation parameters: $S_\text{max} = 10^7, K_S = S_\text{max}, K_H = 10^{-3}S_\text{max}, d_I = 0.5~\text{day}^{-1}, I_0 = 10^{-4}S_\text{max}$.}
\label{fig5}
\end{figure}

\begin{table}[b!]
\centering
\begin{tabular}{lrrr}
{\bf Design Parameter} & {\bf Effect} & {\bf Source}\\
\midrule
{\bf baseline rates:}\\
\,\,1. activation \& prolif., $g_{N\to N^*},g_{E\to 2E}$ & low & \cite{Van_Stipdonk2003-ui, Heinzel2017-qp}\\
\,\,2. differentiation, $g_{N^*\to E}$ & low & \cite{Abadie2024-rd}\\
\,\,3. contraction, $g_{E\to\emptyset}$ & moderate & \cite{De_Boer2001-wl, Heinzel2017-qp}\\\\
{\bf sensitivity to signal:}\\
\,\,4. antigen, $\psi_{\Ag}$ & promote & \cite{Van_Stipdonk2003-ui, Zehn2009-up, Marchingo2014-dh}\\
\,\,5. infection, $\psi_\infc$ & promote & \cite{Blattman2003-sc,Joshi2007-fz, Medzhitov2021-to}\\
\,\,6. response, $\psi_\res$ & suppress & \cite{Medzhitov2021-to, Cicchese2018-po}\\
\bottomrule
\end{tabular}
\centering\caption{Endogenous design of CD8$^+$ T-cell response.
}
\label{table1}
\end{table}

Next, we evaluated immune designs near the Pareto front (Fig.~\ref{fig4}D; SI Appendix). For each point $\lambda$, we identified an ensemble of designs near the Pareto front, $\Theta(\lambda)$. For a design to belong to the near-Pareto ensemble, its incurred harm against archetypes {I and II} must simultaneously lie within 5\% of the Pareto optimal harm of those respective archetypes at the same compromise value $\lambda$.  The \emph{elbow} of the Pareto front represents the point on the curve, $\lambda^*$, at which further decreases in harm from archetype {II} will incur disproportionally large increases in harm from archetype {I} (Fig.~\ref{fig4}B; SI Appendix). The ensemble of designs near the elbow $\Theta(\lambda^*)$ encodes a compelling logic: sensitivity from antigen and infection cues promotes effector production ($\psi_{\Ag}, \psi_{I} >0$), whereas response-derived cues act like a brake  with a predominantly anti-inflammatory effect on the immune response ($\psi_\res <0$);  Fig.~\ref{fig4}D. Sensitivity to these signals diminishes as $\lambda$ decreases (i.e., as avoidance of autoimmunity is prioritized); see Fig.~\ref{fig4}D. Baseline rates ($g_0$'s) of the designs at the elbow favor moderate to slow activation and proliferation of effectors, slow differentiation to effector state, and moderate contraction (Fig.~\ref{fig4}D). As $\lambda$ increases ({leftward; clearance weighes more}), all baseline rates shift toward promoting more effector cells. As $\lambda$ decreases below the elbow ({rightward; toxicity weighs more}), the already slow differentiation baseline rate falls toward zero, while activation/proliferation and contraction are tuned to suppress effector production, reflecting designs that minimize collateral damage even at the expense of incomplete clearance. Notably, despite being optimized only for archetypes I and {II}, designs in the near-elbow ensemble $\Theta(\lambda^*)$ perform competitively for {archetype III}, incurring harm close to the archetype-specific optimal designs (Fig.~\ref{fig4}E; SI Appendix, Fig.~\ParetoArchOne).

From these analyses alone, it is not possible to ascertain the compromises encoded in the endogenous T-cell response program. However, experiments on T-cell fate decisions in endogenous immunological programs (Table~\ref{table1}) are qualitatively consistent with designs that jointly minimize harm from immunogenic pathogens and auto-immune responses to transient injuries (elbow in Fig. \ref{fig4}B).

\begin{figure*}[t!]
\centering
\includegraphics[width=\textwidth]{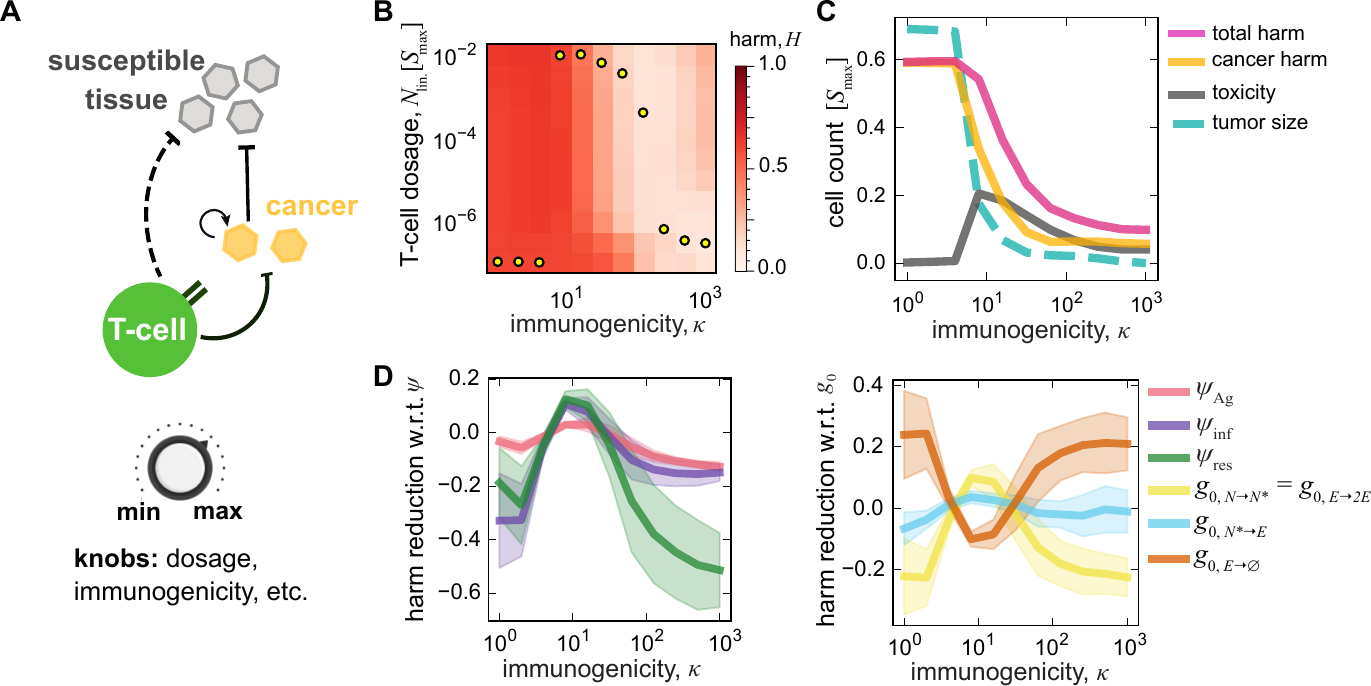}
\caption{{\bf Engineering immune response for cancer immunotherapy.} {\bf(A)} The schematic shows the potential clearance of replicating cancer tumor cells by T-cells (e.g., infused during immunotherapy), and the potential collateral damage of therapy {to} susceptible tissue cells. The accessible levers in an immunotherapy include the dosage and antigenicity of the infused cells. Antigenicity is a property of the interaction between a specific TCR-antigen pair, and it can be increased by either changing the antigen or the TCR. {Given a specific cognate T-cell, immunogenicity is the ratio of the antigenicity of a cancer cell to that of a healthy cell.} {\bf(B)} {Averaged over the tuned near-elbow ensemble $\tilde{\Theta}(\lambda^*)$}, the heatmap displays harm (in units of the maximum number of susceptible cells $S_\text{max})$ as a function of cancer antigenicity and dosage of administered T-cells. Yellow points correspond to the optimal dosage of T-cells for a given immunogenicity/antigenicity. {\bf(C)} Total harm, harm from cancer, immune toxicity, and the final tumor size following the immune response are shown as functions of cancer immunogenicity ($\kappa$) at the optimal T-cell dosage (yellow circles in B). {\bf(D)} Clearance response---defined as the derivative of negative harm (i.e., protection) with respect to each design parameter $\theta_i$, $-\partial H(\vec\theta)/\partial {\theta_i}$ for $\vec\theta \in \tilde{\Theta}(\lambda^*)$---averaged over the tuned near-elbow designs $\tilde{\Theta}(\lambda^*)$, is shown for different levels of tumor immunogenicity $\kappa$; Left: $\vec\psi$ parameters; right: $\vec g_0$ parameters (colors). Shaded envelopes denote the standard deviation of outcomes over $\theta \in \tilde{\Theta}(\lambda^*)$. For this analysis, we assume that T-cells are administered at high dosage ($N_\text{lin.} = 10^{-2} S_\text{max}$). Simulation parameters: $S_\text{max} = 10^7, K_S = S_\text{max}, K_H = 10^{-3}S_\text{max}, d_I = 0.0~\text{day}^{-1}, I_0 = 10^{-1}S_\text{max}, b_I = 0.1~\text{day}^{-1}$. 
\label{fig6}}
\end{figure*}

The choice of design impacts the dynamics of the immune response, which we can compare to experimental data. Prior experiments in mice~\cite{Zehn2009-up, Badovinac2007-ii} indicate that peak fold-expansion of T-cell populations in response to a pathogen scales inversely with the initial number of cognate T-cells, and with the pMHC concentration for half-maximal response (proxied by $K_I$ in our model); see Fig.~\ref{fig5}A,~B. This pattern was recapitulated by explicitly modeling competition for antigens in polyclonal T-cell responses \cite{Mayer2019-pj}. Without any parameter fitting, near-elbow designs $\vec\Theta(\lambda^*)$ from Fig.~\ref{fig4} reproduce this inverse dependence on initial clone size but underestimate the steepness, and overestimate the magnitude of the trend with respect to immunogenicity ($\kappa$) (Fig.~\ref{fig5}A,~B).  

Reducing the sensitivity to infection harm $\psi_\infc$ of these designs by an order of magnitude restores the observed dynamics range (Fig. \ref{fig5}B) while having only a minor effect on total harm, and hence, our optimization objective (Fig.~\PsieOnMacroDyn). Interestingly, adjusting the other sensitivities ($\psi_\Ag, \psi_\res$), or the baseline rates $g_0$'s does not reproduce the dynamic range observed in the experiments (Figs.~\BaselineOnMacroDyn,~\PsieOnMacroDyn). Improved quantitative agreement may be achieved by adjusting a subset of the model's non-design parameters (SI Appendix, Table~\ParameterTable), an analysis that would benefit from a larger dataset to enable robust fitting and validation; see SI  Appendix and Figs.~\paramsensitivityFigureone--\noisesensitivityFigureone~for the sensitivity of  model performance to variation in non-design parameters and to simulation noise.

To align responses with the observed macro-dynamic trends, subsequent analyses consider near-elbow designs $\vec\theta \in \Theta(\lambda^*)$ with their infection-signal sensitivity reduced to one-tenth of their original value: for $\psi_\infc \mapsto 0.1\,\psi_\infc$. This ensemble of \emph{tuned} designs is indicated by $\tilde{\Theta}(\lambda^*)$.

\subsection*{Designing an effector response to a growing tumor}

CD8$^+$ T-cells are central to cancer surveillance and elimination of mutated pre-cancerous cells. We therefore asked how the tuned elbow-optimal designs $\vec\theta \in \tilde{\Theta}(\lambda^*)$ identified for infectious threats (Fig. \ref{fig4}B, Fig. \ref{fig5}) respond to cancer, which we modeled as a growing, low-immunogenic tumor that kills healthy cells as it expands (Fig.~\ref{fig6}A; SI Appendix). We quantified harm as the total number of healthy cells lost, combining cell death caused directly by tumor expansion (replacing healthy cells) and collateral killing mediated by the immune response (Fig.~\ref{fig6}B,C).

Similar to the case of infection control, a clearance-toxicity trade-off arises, but it is sharpened in cancer because many tumor antigens are only marginally distinct from self. Even high-affinity recognition risks cross-reactivity and bystander damage. Tuned near-elbow designs incur high total harm for low-immunogenic tumors (Fig.~\ref{fig6}B,~C). As tumor immunogenicity increases, harm decreases due to reduced cross-reactive toxicity and enhanced {tumor} clearance (Fig.~\ref{fig6}C).

Cancer immunotherapy leverages and improves upon the body's own immune system to target tumors~\cite{Baulu2023-zd}. Patient-specific tumor-infiltrating lymphocytes (TILs) and TCR-engineered T-cells (TCR-T's) purified and expanded ex-vivo, and infused in patients to recognize tumor peptide-MHC's have shown clinical benefits. Chimeric antigen receptor (CAR) T-cells, on the other hand, target surface antigens without MHC restriction and can be deployed more broadly when targets are shared across patients. {Contemporary CARs incorporate co-stimulatory domains (e.g., CD28 or 4-1BB) that enhance activation, proliferation, survival, and effector function~\cite{Zugasti2025-dj}.} In practice, three levers are pulled to engineer a T-cell therapy: (i) the antigen-recognition module (native TCR, TCR-T, or CAR) to specifically target the tumor; (ii) the intracellular signaling architecture (co-stimulation and engineered circuits) to sustains a robust and durable response; and (iii) the infused T-cell dose/schedule, tuned to maximize tumor control while limiting toxicity.

Our model predicts that the optimal dose should be tuned to tumor immunogenicity (Fig. \ref{fig6}B), consistent with prior findings on antigen-dependent T-cell dosing~\cite{Rotte2022-kq}. Examining the tuned near-elbow designs, we find: at low immunogenicity, clearance is inefficient and toxicity dominates, so smaller infusion doses minimize harm; at high immunogenicity, recognition and in-situ expansion are efficient, so even modest doses suffice; at intermediate immunogenicity, larger initial doses minimize total harm by accelerating tumor elimination while keeping toxicity tolerable (Fig.~\ref{fig6}C).

TCR engineering raises an additional nuance: increasing TCR-tumor affinity often increases affinity to self ~\cite{Hoffmann2020-yg}, so that increasing tumor immunogenicity is difficult. To capture this, we evaluated treatment scenarios where the affinities/reactivities of tumor and self to the engineered TCR/CAR are proportional, leaving tumor immunogenicity unchanged ($\kappa = K_I^{-1}K_S = \text{constant}$; SI Appendix, Fig.~\CancerAntigenicity). We quantify affinity or reactivity of a cell by the ratio of the maximum healthy cell counts $S_\text{max}$ to the number of target cells required for half-maximal T-cell response (EC$_{50}$), i.e., $K_{I}^{-1}S_\text{max}$ for tumor cells, and $K_{S}^{-1}S_\text{max}$ for self targets. Provided that tumor immunogenicity $\kappa$ is not too small and self-reactivity remains moderate, the dose-response as a function of tumor antigenicity (Fig.~\CancerAntigenicity A) mirrors its dependence on immunogenicity in Fig.~\ref{fig6}B. Nevertheless, these cross-reactive regimes ({Fig.~\CancerAntigenicity}B) yield higher total harm than scenarios where self-interaction is fixed so that increasing tumor antigenicity increases immunogenicity (Fig.~\ref{fig6}C). {This highlights the cost of affinity-driven cross-reactivity in TCR engineering for cancer therapy~\cite{Liu2025-qk, Kondo2025-nl, Visani2025-wx}.}

Finally, we asked which aspects of the endogenous CD8$^+$ T-cell program could be modulated to improve tumor control. Using high-dose infusions for a broad range of immunogenicities, we computed the sensitivity of negative harm (i.e., protection) to small changes in design parameter values, i.e., $-\nabla H(\vec\theta)$ for $\vec\theta \in \tilde{\Theta}(\lambda^*)$ (Fig.~\ref{fig6}D). For very high-affinity T-cells (those able to respond at $\leq 1/100$ of the number of healthy cells), gains from further engineering are limited. For intermediate affinities---arguably more relevant clinically---several levers are impactful: increasing activation/proliferation baseline rates ($g_{0,N\to N^*}, g_{0,E\to 2E} \uparrow$)~\cite{Escobar2023-ot} and decreasing effector death ($g_{0,E\to\emptyset} \downarrow$) provide strong improvements but may entail perturbing oncogene-linked pathways (e.g., MYC, BIM) and warrant caution~\cite{Muthalagu2014-ni}. Boosting antigen and infection sensitivities ($\psi_{\Ag}, \psi_\infc\, \uparrow$), and tempering the anti-inflammatory brake ($|\psi_\res| \downarrow$) also enhance clearance, albeit with a risk of runaway inflammation~\cite{Tisoncik2012-kx}. Increasing the baseline rate of differentiation ($g_{0,N^*\to E} \uparrow$) yields smaller gains. This suggests that reducing effector differentiation---e.g., via Tcf7 modulation~\cite{Abadie2024-rd, Escobar2023-ot}---may provide a tractable knob to promote memory production and reduce relapse risk. Reduced effector differentiation may then be compensated for by tuning effector expansion and longevity. 

Interestingly, recent CRISPR screens show that among other candidates, knocking out FAS and PTPN2 genes in CAR T-cells reduces apoptosis ($g_{0,E\to \emptyset} \downarrow$) and redirects differentiation toward effector programs ($g_{0,E\to 2E} \uparrow$), thereby enhancing CAR T-cell efficacy~\cite{Yoshikawa2022-hd,Yi2025-bb,Knudsen2025-xb, Datlinger2025-oi}. Our modeling framework can help prioritize targets for such screens to rationally identify candidate genes that improve T-cell immunotherapies. It should be noted that the achievable benefit of any perturbation for T-cell engineering will depend on patient- and tumor-specific constraints. {These include the extent to which affinity-driven TCR cross-reactivity can be mitigated (Fig.~\CancerAntigenicity) and may require patient-specific measurements of T-cell interactions and signaling to calibrate designs for robust, durable therapy.}

\section{Discussion}
In this work, we modeled the CD8$^+$ T-cell response as a feedback-controlled program that integrates three cues (i.e., instantaneous antigen, time-integrated infection-derived harm, and time-integrated response-derived harm) to minimize cumulative cell death during an immune challenge. Modulation of response proportional to the antigen signal and the time-integrated harm signals mirrors proportional-integral-derivative (PID) control of engineered systems, a strategy known for its stable and robust steering of dynamical processes toward desired outcomes \cite{Ang2005-qv}.

{Exploring an ensemble of biologically motivated controllers (designs) and optimizing for pathogen clearance while minimizing immunopathology exposed a Pareto trade-off: a reduction in infection harm generally leads to an increased collateral damage to healthy tissues. This pattern persisted across diverse {immune} challenges defined by their immunogenicity and within-host basic reproduction number.}

Three design principles emerged from our analyses: (i) Pre-programmed (baseline-tuned) designs help restrain autoimmunity and slowly proliferating threats but are inefficient against fast-proliferating infections or metastatic tumors; effective control there requires signal integration and feedback. (ii) Among signal sources, antigen and response-harm dominate; sensitivity to infection-harm contributes comparatively little once CD8$^+$ effector T-cell expansion is underway. (iii) Feedback from response-harm acts as an anti-inflammatory brake, enabling strong antigen-driven responses while limiting collateral damage when the response overshoots---a characteristic consistent with endogenous effector programs~\cite{Medzhitov2021-to, Cicchese2018-po}. Table~\ref{table1} further summarizes the characteristics of  our inferred optimal designs and relates them to supporting evidence from endogenous effector programs.

We posited that effector programs should attempt to solve a multi-objective problem to minimize harm (cell death) inflicted by diverse pathogenic challenges. Interestingly, we found that a Pareto-optimal design that suppresses responses to frequent, self-like injuries while clearing recognizable, replicating (e.g., acute) infections mirrors the endogenous T-cell programs in vertebrates: strongly antigen-driven with an anti-inflammatory brake, and conservative cell-fate transition speeds that prevent runaway effector responses. We note that, beyond minimizing harm due to cell death, additional factors may also shape the objectives optimized by effector T-cell programs, including the metabolic costs of effector activity and cytokine production, the cumulative burden of inflammation, and the long-term value of generating memory cells. Although the relative importance of these contributions remains unclear, they could be incorporated and analyzed within the same Pareto-optimization framework we introduced in this work.

Memory formation is a key feature of CD8$^+$ T-cell immunity, but here we focus on acute infection control and therefore do not model memory expansion or recall explicitly. In endogenous responses, immunological memory arises in part from the differentiation of a small fraction of effector cells ($\sim 5-10\%$ of the peak response~\cite{Kaech2012-ye, Abadie2024-rd}) into effector-memory cells, effectively removing them from the active response. In our model, this loss of effector cells is captured by the contraction step, which can represent either effector apoptosis or differentiation into memory. By aggregating these processes into a single contraction term, we reduce degeneracy in the design parameters while retaining interpretability. Nonetheless, memory ``quality" depends on differentiation timing and clone avidity~\cite{Bresser2022-iw, Youngblood2017-my, Kretschmer2020-ad, Grassmann2020-ek}, which would be relevant for the strength and durability of protection upon secondary immune challenges. Extending our framework to an explicit allocation trade-off, where diverting cells to memory reduces primary effector output but seeds future protection, will be an important next step toward linking control designs to memory formation and the evolutionary pressures shaping effector programs.

Using a minimal tumor model, we showed how the design objective informs rational immunotherapy protocols. First, it suggests that infused T-cell dosage should depend on tumor immunogenicity---balancing clearance against collateral damage, as previously emphasized in the literature~\cite{Baulu2023-zd,Kondo2025-nl,Rotte2022-kq}. Second, treating the design parameters as engineering targets yields testable strategies to improve efficacy and safety, which is highly desirable~\cite{Xia2024-fr, Li2022-rf, Allen2022-lo}. {For moderately immunogenic tumors, promoting effector output by raising baseline rates of activation/proliferation or increasing antigen and infection sensitivities, while tempering the anti-inflammatory brake, may produce robust responses.}

To operationalize these ideas, abstract signal sensitivities and baseline rates must be mapped to gene programs as targets of cell engineering (Table \ref{table1}). Combining principled control-theoretic design with patient-specific measurements of antigenicity and damage signaling may enable tailored effector programs that respect the fundamental trade-offs of response from this study, while bending them toward better outcomes \cite{Lassig2023-ky}.

Key questions follow from our study. Our coarse-grained model elides many details of tissue biology and micro-environment: how do tissue contexts impact the state of resident cells and shift clearance-toxicity trade-offs \cite{van-Dorp2025-my, Jerison2025-bl}? Endogenous immune responses are polyclonal \cite{Straub2023-oa}: how do the proposed designs shape information flow and selection of dominant clones in polyclonal responses? Do the optimization principles posited here predispose the immune response to certain vulnerabilities, and what mechanisms might have evolved to mask them? Would explicitly modeling interactions between CD8$^+$ T-cells and other immune compartments, such as regulatory T-cells and innate immune system, enhance the robustness of cytotoxic responses across different pathogenic challenges? Addressing these issues through the lens of cell economics and immunological trade-offs may clarify the roles of chance, contingency, and necessity in the evolution of effector programs.

\section{Materials and Methods}
\noindent Data and code for all the analyses can be found in the Github
repository:~\url{https://github.com/StatPhysBio/designimmune}. Previously published data from refs.~\cite{Badovinac2007-ii,Zehn2009-up} were used for this work.\\

\noindent{\bf Intra-host infection dynamics.} We model a population of healthy, susceptible cells $S(t)$, initially consisting of $S_{\text{max}}$ identical cells. Pathogen exposure seeds an infected subpopulation $I(t)$, which grows via encounters with susceptible cells at rate $b_I$, and declines by infected-cell death at rate $d_I$, triggering an immune response (Figs.~\ref{fig1}A,B,~\DiffSchematic).
The dynamics of the susceptible and infected cells follow,
\begin{align}
\label{infection-dynamics}
    \begin{split}
        \dot{S} &= - b_I I S -d_E\frac{E}{K_{S} + I\frac{K_S + E}{K_I + E} + E + S}S 
        \\
        \dot{I} &= b_II S - d_I I -d_E\frac{E}{K_{I} + S\frac{K_I + E}{K_S + E} + E + I}I,
    \end{split}
\end{align}
where $d_E$ is the maximal clearance rate of infected/susceptible cells by effector T-cells $E$, {and $K_{I}$ and $K_{S}$ are \emph{recognition thresholds}, quantifying how readily effector cells can recognize infected and susceptible cells, respectively, and consequently kill them; a larger $K_{I}$ or $K_{S}$ implies more effector cells are needed for recognition}~\cite{Chao2004-jq, Ganusov2011-bl, Halle2016-kw}. The functional form of the last terms in these equations is motivated by {considering competitive binding of T-cells to infected and healthy cells} (SI Appendix). Note that these equations describe dynamics during the immune challenge and exclude the slow  birth-death processes that homeostatically maintain the susceptible cells in the absence of infection.\\

\noindent {\bf Stochastic T-cell response model.}  We model the CD8$^+$ T-cell response to an immune challenge as a time-inhomogeneous birth-death process coupled to deterministic dynamics of susceptible and infected cells. Beginning with an initial population of monoclonal na\"ive T-cells, we track individual T-cell lineages through recruitment of na\"ive cells to interact with antigen-presenting cells (APCs; $N\to \textbf{APC}\cdot N$), activation ($\textbf{APC}\cdot N \to N^*$), differentiation to effector ($N^*\to E$), expansion ($E\to 2E$), and contraction ($E\to\emptyset$). Na\"ive T-cells are recruited to APCs at a rate modulated by {infection} harm signals, and upon sustained engagement, become activated and undergo 2-3 burst-like divisions largely independent of continued antigenic input. During this early proliferative window, each activated cell commits to an effector or memory fate according to the first arrival of a time-inhomogeneous Poisson process with a signal-dependent rate; the number differentiating into effectors is drawn as a binomial random variable, with remaining cells adopting a memory phenotype. Effector cells subsequently divide and die at rates governed by time-varying antigen and harm signals. Proliferation is capped by a maximum division number~\cite{Buchholz2013-ab,Marchingo2016-ao,Badovinac2007-ii, Zhang2011-dv}, which we set to 15 rounds, beyond which cells undergo terminal arrest (see SI Appendix for details).

In our model, inter-lineage variability arises solely from stochastic differences in transition timing, capturing with prior observations~\cite{Heinzel2017-qp, Cho2017-ul, Abadie2024-rd, Plambeck2022-dt, Xin2016-ut, Youngblood2017-my, Buchholz2013-ab, Kretschmer2020-ad}. To quantify this stochasticity and demonstrate robustness, results are averaged over five independent full-simulation replicates, with variability reported. The coefficient of variation averaged over immune challenges is small (C.V. = 0.05; Fig.~\noisesensitivityFigureone), indicating that macroscopic quantities such as total harm and effector peak are robust to stochasticity in T-cell response dynamics. We also characterize sensitivity to simulation parameters (Figs.~\paramsensitivityFigureone,~\paramsensitivityFiguretwo), which vary across hosts and infection scenarios.\\

\noindent {\bf Infection and response signals.} 
T-cell activation relies on the integration of antigen and harm signals. Cell deaths can induce harm in the host and produce pro- or anti-inflammatory signals, depending on the underlying cause. We simplify this by distinguishing infection-induced and response-induced harms, which in a cumulative form, generate signals $\sigma_{\text{inf}}(t)$ and $\sigma_{\text{res}}(t)$, respectively. The instantaneous level of presented antigens, proxied by the infected-cell burden $I(t)$, defines a separate antigen signal $\sigma_{\text{Ag}}(t)$. Together, {scaled} by their recognition thresholds, $K_I$ for antigens and $K_H$ for non-antigens (harm), the three signals available to T-cells follow,
\begin{align}
    \begin{split}
     &\text{antigen signal:}~~ \sigma_{\Ag}(t) ={K_I}^{-1} {I}(t)\\
    &\text{harm signals:}~~\sigma_{\infc/\res}(t) = K_H^{-1}\int_0^t e^{-(t-s) d_H}\,h_{\infc/\res}(s)\,{\rm d}s.
    \label{eq:Ag-Cytokine-Signal}
    \end{split}
\end{align}
where   $h_{\infc} (s)$ is the number of cells killed by the pathogen (infection),  and $h_\res (s)$ denotes the number of cells killed by the effector T-cell response at time $s$ (eq.~\ref{eq:harm}). The time-integrated harm-induced signals are exponentially discounted at rate $d_H$, reflecting the decay of cytokines in the absence of ongoing stimulation (SI Appendix).

\section*{Acknowledgement}
This work has been supported by the CAREER award from the National Science Foundation grant 2045054 (A.N., O.A.U., Z.M.), the National Institutes of Health MIRA award R35~GM142795 (A.N.), and the Intramural Research Program of the National Cancer Institute, Project \# ZIA BC 012007 (G.A.-B.). This work is also supported, in part, through the Departments of Physics and Applied Mathematics and the College of Arts and Sciences at the University of Washington. The numerical analyses in this work were completed on Hyak, the UW's high performance computing cluster, which is funded by the UW student technology fee. This work benefited from discussions during the 2023 summer workshop ``Statistical Physics and Adaptive Immunity" at the Aspen Center for Physics, which is supported by National Science Foundation grant PHY-2210452, and the 2024 program ``Interactions and Co-evolution between Viruses and Immune Systems" at the Kavli Institute for theoretical physics (KITP), which is supported by National Science Foundation grant PHY-2309135, and the Gordon and Betty Moore Foundation Grant No. 2919.02.

\bibliographystyle{apsrev4-2}

\begin{thebibliography}{0}%
\makeatletter
\providecommand \@ifxundefined [1]{%
 \@ifx{#1\undefined}
}%
\providecommand \@ifnum [1]{%
 \ifnum #1\expandafter \@firstoftwo
 \else \expandafter \@secondoftwo
 \fi
}%
\providecommand \@ifx [1]{%
 \ifx #1\expandafter \@firstoftwo
 \else \expandafter \@secondoftwo
 \fi
}%
\providecommand \natexlab [1]{#1}%
\providecommand \enquote  [1]{``#1''}%
\providecommand \bibnamefont  [1]{#1}%
\providecommand \bibfnamefont [1]{#1}%
\providecommand \citenamefont [1]{#1}%
\providecommand \href@noop [0]{\@secondoftwo}%
\providecommand \href [0]{\begingroup \@sanitize@url \@href}%
\providecommand \@href[1]{\@@startlink{#1}\@@href}%
\providecommand \@@href[1]{\endgroup#1\@@endlink}%
\providecommand \@sanitize@url [0]{\catcode `\\12\catcode `\$12\catcode
  `\&12\catcode `\#12\catcode `\^12\catcode `\_12\catcode `\%12\relax}%
\providecommand \@@startlink[1]{}%
\providecommand \@@endlink[0]{}%
\providecommand \url  [0]{\begingroup\@sanitize@url \@url }%
\providecommand \@url [1]{\endgroup\@href {#1}{\urlprefix }}%
\providecommand \urlprefix  [0]{URL }%
\providecommand \Eprint [0]{\href }%
\providecommand \doibase [0]{https://doi.org/}%
\providecommand \selectlanguage [0]{\@gobble}%
\providecommand \bibinfo  [0]{\@secondoftwo}%
\providecommand \bibfield  [0]{\@secondoftwo}%
\providecommand \translation [1]{[#1]}%
\providecommand \BibitemOpen [0]{}%
\providecommand \bibitemStop [0]{}%
\providecommand \bibitemNoStop [0]{.\EOS\space}%
\providecommand \EOS [0]{\spacefactor3000\relax}%
\providecommand \BibitemShut  [1]{\csname bibitem#1\endcsname}%
\let\auto@bib@innerbib\@empty
\end{thebibliography}%


%


\begin{thebibliography}{94}%
\makeatletter
\providecommand \@ifxundefined [1]{%
 \@ifx{#1\undefined}
}%
\providecommand \@ifnum [1]{%
 \ifnum #1\expandafter \@firstoftwo
 \else \expandafter \@secondoftwo
 \fi
}%
\providecommand \@ifx [1]{%
 \ifx #1\expandafter \@firstoftwo
 \else \expandafter \@secondoftwo
 \fi
}%
\providecommand \natexlab [1]{#1}%
\providecommand \enquote  [1]{``#1''}%
\providecommand \bibnamefont  [1]{#1}%
\providecommand \bibfnamefont [1]{#1}%
\providecommand \citenamefont [1]{#1}%
\providecommand \href@noop [0]{\@secondoftwo}%
\providecommand \href [0]{\begingroup \@sanitize@url \@href}%
\providecommand \@href[1]{\@@startlink{#1}\@@href}%
\providecommand \@@href[1]{\endgroup#1\@@endlink}%
\providecommand \@sanitize@url [0]{\catcode `\\12\catcode `\$12\catcode
  `\&12\catcode `\#12\catcode `\^12\catcode `\_12\catcode `\%12\relax}%
\providecommand \@@startlink[1]{}%
\providecommand \@@endlink[0]{}%
\providecommand \url  [0]{\begingroup\@sanitize@url \@url }%
\providecommand \@url [1]{\endgroup\@href {#1}{\urlprefix }}%
\providecommand \urlprefix  [0]{URL }%
\providecommand \Eprint [0]{\href }%
\providecommand \doibase [0]{https://doi.org/}%
\providecommand \selectlanguage [0]{\@gobble}%
\providecommand \bibinfo  [0]{\@secondoftwo}%
\providecommand \bibfield  [0]{\@secondoftwo}%
\providecommand \translation [1]{[#1]}%
\providecommand \BibitemOpen [0]{}%
\providecommand \bibitemStop [0]{}%
\providecommand \bibitemNoStop [0]{.\EOS\space}%
\providecommand \EOS [0]{\spacefactor3000\relax}%
\providecommand \BibitemShut  [1]{\csname bibitem#1\endcsname}%
\let\auto@bib@innerbib\@empty
\bibitem [{\citenamefont {Qi}\ \emph {et~al.}(2014)\citenamefont {Qi},
  \citenamefont {Liu}, \citenamefont {Cheng}, \citenamefont {Glanville},
  \citenamefont {Zhang}, \citenamefont {Lee}, \citenamefont {Olshen},
  \citenamefont {Weyand}, \citenamefont {Boyd},\ and\ \citenamefont
  {Goronzy}}]{Qi2014-pz}%
  \BibitemOpen
  \bibfield  {author} {\bibinfo {author} {\bibfnamefont {Q.}~\bibnamefont
  {Qi}}, \bibinfo {author} {\bibfnamefont {Y.}~\bibnamefont {Liu}}, \bibinfo
  {author} {\bibfnamefont {Y.}~\bibnamefont {Cheng}}, \bibinfo {author}
  {\bibfnamefont {J.}~\bibnamefont {Glanville}}, \bibinfo {author}
  {\bibfnamefont {D.}~\bibnamefont {Zhang}}, \bibinfo {author} {\bibfnamefont
  {J.-Y.}\ \bibnamefont {Lee}}, \bibinfo {author} {\bibfnamefont {R.~A.}\
  \bibnamefont {Olshen}}, \bibinfo {author} {\bibfnamefont {C.~M.}\
  \bibnamefont {Weyand}}, \bibinfo {author} {\bibfnamefont {S.~D.}\
  \bibnamefont {Boyd}},\ and\ \bibinfo {author} {\bibfnamefont {J.~J.}\
  \bibnamefont {Goronzy}},\ }\href@noop {} {\bibfield  {journal} {\bibinfo
  {journal} {Proc. Natl. Acad. Sci. U. S. A.}\ }\textbf {\bibinfo {volume}
  {111}},\ \bibinfo {pages} {13139} (\bibinfo {year} {2014})}\BibitemShut
  {NoStop}%
\bibitem [{\citenamefont {Lythe}\ \emph {et~al.}(2016)\citenamefont {Lythe},
  \citenamefont {Callard}, \citenamefont {Hoare},\ and\ \citenamefont
  {Molina-París}}]{Lythe2016-xv}%
  \BibitemOpen
  \bibfield  {author} {\bibinfo {author} {\bibfnamefont {G.}~\bibnamefont
  {Lythe}}, \bibinfo {author} {\bibfnamefont {R.~E.}\ \bibnamefont {Callard}},
  \bibinfo {author} {\bibfnamefont {R.~L.}\ \bibnamefont {Hoare}},\ and\
  \bibinfo {author} {\bibfnamefont {C.}~\bibnamefont {Molina-París}},\
  }\href@noop {} {\bibfield  {journal} {\bibinfo  {journal} {J. Theor. Biol.}\
  }\textbf {\bibinfo {volume} {389}},\ \bibinfo {pages} {214} (\bibinfo {year}
  {2016})}\BibitemShut {NoStop}%
\bibitem [{\citenamefont {Mora}\ and\ \citenamefont
  {Walczak}(2019)}]{Mora2019-zn}%
  \BibitemOpen
  \bibfield  {author} {\bibinfo {author} {\bibfnamefont {T.}~\bibnamefont
  {Mora}}\ and\ \bibinfo {author} {\bibfnamefont {A.~M.}\ \bibnamefont
  {Walczak}},\ }\href@noop {} {\bibfield  {journal} {\bibinfo  {journal} {Curr.
  Opin. Syst. Biol.}\ }\textbf {\bibinfo {volume} {18}},\ \bibinfo {pages}
  {104} (\bibinfo {year} {2019})}\BibitemShut {NoStop}%
\bibitem [{\citenamefont {van Stipdonk}\ \emph {et~al.}(2003)\citenamefont {van
  Stipdonk}, \citenamefont {Hardenberg}, \citenamefont {Bijker}, \citenamefont
  {Lemmens}, \citenamefont {Droin}, \citenamefont {Green},\ and\ \citenamefont
  {Schoenberger}}]{Van_Stipdonk2003-ui}%
  \BibitemOpen
  \bibfield  {author} {\bibinfo {author} {\bibfnamefont {M.~J.~B.}\
  \bibnamefont {van Stipdonk}}, \bibinfo {author} {\bibfnamefont
  {G.}~\bibnamefont {Hardenberg}}, \bibinfo {author} {\bibfnamefont {M.~S.}\
  \bibnamefont {Bijker}}, \bibinfo {author} {\bibfnamefont {E.~E.}\
  \bibnamefont {Lemmens}}, \bibinfo {author} {\bibfnamefont {N.~M.}\
  \bibnamefont {Droin}}, \bibinfo {author} {\bibfnamefont {D.~R.}\ \bibnamefont
  {Green}},\ and\ \bibinfo {author} {\bibfnamefont {S.~P.}\ \bibnamefont
  {Schoenberger}},\ }\href@noop {} {\bibfield  {journal} {\bibinfo  {journal}
  {Nat. Immunol.}\ }\textbf {\bibinfo {volume} {4}},\ \bibinfo {pages} {361}
  (\bibinfo {year} {2003})}\BibitemShut {NoStop}%
\bibitem [{\citenamefont {Joshi}\ \emph {et~al.}(2007)\citenamefont {Joshi},
  \citenamefont {Cui}, \citenamefont {Chandele}, \citenamefont {Lee},
  \citenamefont {Urso}, \citenamefont {Hagman}, \citenamefont {Gapin},\ and\
  \citenamefont {Kaech}}]{Joshi2007-fz}%
  \BibitemOpen
  \bibfield  {author} {\bibinfo {author} {\bibfnamefont {N.~S.}\ \bibnamefont
  {Joshi}}, \bibinfo {author} {\bibfnamefont {W.}~\bibnamefont {Cui}}, \bibinfo
  {author} {\bibfnamefont {A.}~\bibnamefont {Chandele}}, \bibinfo {author}
  {\bibfnamefont {H.~K.}\ \bibnamefont {Lee}}, \bibinfo {author} {\bibfnamefont
  {D.~R.}\ \bibnamefont {Urso}}, \bibinfo {author} {\bibfnamefont
  {J.}~\bibnamefont {Hagman}}, \bibinfo {author} {\bibfnamefont
  {L.}~\bibnamefont {Gapin}},\ and\ \bibinfo {author} {\bibfnamefont {S.~M.}\
  \bibnamefont {Kaech}},\ }\href@noop {} {\bibfield  {journal} {\bibinfo
  {journal} {Immunity}\ }\textbf {\bibinfo {volume} {27}},\ \bibinfo {pages}
  {281} (\bibinfo {year} {2007})}\BibitemShut {NoStop}%
\bibitem [{\citenamefont {Marchingo}\ \emph {et~al.}(2014)\citenamefont
  {Marchingo}, \citenamefont {Kan}, \citenamefont {Sutherland}, \citenamefont
  {Duffy}, \citenamefont {Wellard}, \citenamefont {Belz}, \citenamefont {Lew},
  \citenamefont {Dowling}, \citenamefont {Heinzel},\ and\ \citenamefont
  {Hodgkin}}]{Marchingo2014-dh}%
  \BibitemOpen
  \bibfield  {author} {\bibinfo {author} {\bibfnamefont {J.~M.}\ \bibnamefont
  {Marchingo}}, \bibinfo {author} {\bibfnamefont {A.}~\bibnamefont {Kan}},
  \bibinfo {author} {\bibfnamefont {R.~M.}\ \bibnamefont {Sutherland}},
  \bibinfo {author} {\bibfnamefont {K.~R.}\ \bibnamefont {Duffy}}, \bibinfo
  {author} {\bibfnamefont {C.~J.}\ \bibnamefont {Wellard}}, \bibinfo {author}
  {\bibfnamefont {G.~T.}\ \bibnamefont {Belz}}, \bibinfo {author}
  {\bibfnamefont {A.~M.}\ \bibnamefont {Lew}}, \bibinfo {author} {\bibfnamefont
  {M.~R.}\ \bibnamefont {Dowling}}, \bibinfo {author} {\bibfnamefont
  {S.}~\bibnamefont {Heinzel}},\ and\ \bibinfo {author} {\bibfnamefont {P.~D.}\
  \bibnamefont {Hodgkin}},\ }\href@noop {} {\bibfield  {journal} {\bibinfo
  {journal} {Science}\ }\textbf {\bibinfo {volume} {346}},\ \bibinfo {pages}
  {1123} (\bibinfo {year} {2014})}\BibitemShut {NoStop}%
\bibitem [{\citenamefont {Marchingo}\ \emph {et~al.}(2016)\citenamefont
  {Marchingo}, \citenamefont {Prevedello}, \citenamefont {Kan}, \citenamefont
  {Heinzel}, \citenamefont {Hodgkin},\ and\ \citenamefont
  {Duffy}}]{Marchingo2016-ao}%
  \BibitemOpen
  \bibfield  {author} {\bibinfo {author} {\bibfnamefont {J.~M.}\ \bibnamefont
  {Marchingo}}, \bibinfo {author} {\bibfnamefont {G.}~\bibnamefont
  {Prevedello}}, \bibinfo {author} {\bibfnamefont {A.}~\bibnamefont {Kan}},
  \bibinfo {author} {\bibfnamefont {S.}~\bibnamefont {Heinzel}}, \bibinfo
  {author} {\bibfnamefont {P.~D.}\ \bibnamefont {Hodgkin}},\ and\ \bibinfo
  {author} {\bibfnamefont {K.~R.}\ \bibnamefont {Duffy}},\ }\href@noop {}
  {\bibfield  {journal} {\bibinfo  {journal} {Nat. Commun.}\ }\textbf {\bibinfo
  {volume} {7}},\ \bibinfo {pages} {13540} (\bibinfo {year}
  {2016})}\BibitemShut {NoStop}%
\bibitem [{\citenamefont {Eizenberg-Magar}\ \emph {et~al.}(2017)\citenamefont
  {Eizenberg-Magar}, \citenamefont {Rimer}, \citenamefont {Zaretsky},
  \citenamefont {Lara-Astiaso}, \citenamefont {Reich-Zeliger},\ and\
  \citenamefont {Friedman}}]{Eizenberg-Magar2017-cp}%
  \BibitemOpen
  \bibfield  {author} {\bibinfo {author} {\bibfnamefont {I.}~\bibnamefont
  {Eizenberg-Magar}}, \bibinfo {author} {\bibfnamefont {J.}~\bibnamefont
  {Rimer}}, \bibinfo {author} {\bibfnamefont {I.}~\bibnamefont {Zaretsky}},
  \bibinfo {author} {\bibfnamefont {D.}~\bibnamefont {Lara-Astiaso}}, \bibinfo
  {author} {\bibfnamefont {S.}~\bibnamefont {Reich-Zeliger}},\ and\ \bibinfo
  {author} {\bibfnamefont {N.}~\bibnamefont {Friedman}},\ }\href@noop {}
  {\bibfield  {journal} {\bibinfo  {journal} {Proc. Natl. Acad. Sci. U. S. A.}\
  }\textbf {\bibinfo {volume} {114}},\ \bibinfo {pages} {E6447} (\bibinfo
  {year} {2017})}\BibitemShut {NoStop}%
\bibitem [{\citenamefont {Badovinac}\ \emph {et~al.}(2007)\citenamefont
  {Badovinac}, \citenamefont {Haring},\ and\ \citenamefont
  {Harty}}]{Badovinac2007-ii}%
  \BibitemOpen
  \bibfield  {author} {\bibinfo {author} {\bibfnamefont {V.~P.}\ \bibnamefont
  {Badovinac}}, \bibinfo {author} {\bibfnamefont {J.~S.}\ \bibnamefont
  {Haring}},\ and\ \bibinfo {author} {\bibfnamefont {J.~T.}\ \bibnamefont
  {Harty}},\ }\href@noop {} {\bibfield  {journal} {\bibinfo  {journal}
  {Immunity}\ }\textbf {\bibinfo {volume} {26}},\ \bibinfo {pages} {827}
  (\bibinfo {year} {2007})}\BibitemShut {NoStop}%
\bibitem [{\citenamefont {Zhang}\ and\ \citenamefont
  {Bevan}(2011)}]{Zhang2011-dv}%
  \BibitemOpen
  \bibfield  {author} {\bibinfo {author} {\bibfnamefont {N.}~\bibnamefont
  {Zhang}}\ and\ \bibinfo {author} {\bibfnamefont {M.~J.}\ \bibnamefont
  {Bevan}},\ }\href@noop {} {\bibfield  {journal} {\bibinfo  {journal}
  {Immunity}\ }\textbf {\bibinfo {volume} {35}},\ \bibinfo {pages} {161}
  (\bibinfo {year} {2011})}\BibitemShut {NoStop}%
\bibitem [{\citenamefont {Buchholz}\ \emph {et~al.}(2013)\citenamefont
  {Buchholz}, \citenamefont {Flossdorf}, \citenamefont {Hensel}, \citenamefont
  {Kretschmer}, \citenamefont {Weissbrich}, \citenamefont {Gräf},
  \citenamefont {Verschoor}, \citenamefont {Schiemann}, \citenamefont
  {Höfer},\ and\ \citenamefont {Busch}}]{Buchholz2013-ab}%
  \BibitemOpen
  \bibfield  {author} {\bibinfo {author} {\bibfnamefont {V.~R.}\ \bibnamefont
  {Buchholz}}, \bibinfo {author} {\bibfnamefont {M.}~\bibnamefont {Flossdorf}},
  \bibinfo {author} {\bibfnamefont {I.}~\bibnamefont {Hensel}}, \bibinfo
  {author} {\bibfnamefont {L.}~\bibnamefont {Kretschmer}}, \bibinfo {author}
  {\bibfnamefont {B.}~\bibnamefont {Weissbrich}}, \bibinfo {author}
  {\bibfnamefont {P.}~\bibnamefont {Gräf}}, \bibinfo {author} {\bibfnamefont
  {A.}~\bibnamefont {Verschoor}}, \bibinfo {author} {\bibfnamefont
  {M.}~\bibnamefont {Schiemann}}, \bibinfo {author} {\bibfnamefont
  {T.}~\bibnamefont {Höfer}},\ and\ \bibinfo {author} {\bibfnamefont {D.~H.}\
  \bibnamefont {Busch}},\ }\href@noop {} {\bibfield  {journal} {\bibinfo
  {journal} {Science}\ }\textbf {\bibinfo {volume} {340}},\ \bibinfo {pages}
  {630} (\bibinfo {year} {2013})}\BibitemShut {NoStop}%
\bibitem [{\citenamefont {Sarkar}\ \emph {et~al.}(2008)\citenamefont {Sarkar},
  \citenamefont {Kalia}, \citenamefont {Haining}, \citenamefont {Konieczny},
  \citenamefont {Subramaniam},\ and\ \citenamefont {Ahmed}}]{Sarkar2008-ob}%
  \BibitemOpen
  \bibfield  {author} {\bibinfo {author} {\bibfnamefont {S.}~\bibnamefont
  {Sarkar}}, \bibinfo {author} {\bibfnamefont {V.}~\bibnamefont {Kalia}},
  \bibinfo {author} {\bibfnamefont {W.~N.}\ \bibnamefont {Haining}}, \bibinfo
  {author} {\bibfnamefont {B.~T.}\ \bibnamefont {Konieczny}}, \bibinfo {author}
  {\bibfnamefont {S.}~\bibnamefont {Subramaniam}},\ and\ \bibinfo {author}
  {\bibfnamefont {R.}~\bibnamefont {Ahmed}},\ }\href@noop {} {\bibfield
  {journal} {\bibinfo  {journal} {J. Exp. Med.}\ }\textbf {\bibinfo {volume}
  {205}},\ \bibinfo {pages} {625} (\bibinfo {year} {2008})}\BibitemShut
  {NoStop}%
\bibitem [{\citenamefont {Jenkins}\ \emph {et~al.}(2008)\citenamefont
  {Jenkins}, \citenamefont {Mintern}, \citenamefont {La~Gruta}, \citenamefont
  {Kedzierska}, \citenamefont {Doherty},\ and\ \citenamefont
  {Turner}}]{Jenkins2008-if}%
  \BibitemOpen
  \bibfield  {author} {\bibinfo {author} {\bibfnamefont {M.~R.}\ \bibnamefont
  {Jenkins}}, \bibinfo {author} {\bibfnamefont {J.}~\bibnamefont {Mintern}},
  \bibinfo {author} {\bibfnamefont {N.~L.}\ \bibnamefont {La~Gruta}}, \bibinfo
  {author} {\bibfnamefont {K.}~\bibnamefont {Kedzierska}}, \bibinfo {author}
  {\bibfnamefont {P.~C.}\ \bibnamefont {Doherty}},\ and\ \bibinfo {author}
  {\bibfnamefont {S.~J.}\ \bibnamefont {Turner}},\ }\href@noop {} {\bibfield
  {journal} {\bibinfo  {journal} {J. Immunol.}\ }\textbf {\bibinfo {volume}
  {181}},\ \bibinfo {pages} {3818} (\bibinfo {year} {2008})}\BibitemShut
  {NoStop}%
\bibitem [{\citenamefont {Seder}\ \emph {et~al.}(2008)\citenamefont {Seder},
  \citenamefont {Darrah},\ and\ \citenamefont {Roederer}}]{Seder2008-xh}%
  \BibitemOpen
  \bibfield  {author} {\bibinfo {author} {\bibfnamefont {R.~A.}\ \bibnamefont
  {Seder}}, \bibinfo {author} {\bibfnamefont {P.~A.}\ \bibnamefont {Darrah}},\
  and\ \bibinfo {author} {\bibfnamefont {M.}~\bibnamefont {Roederer}},\
  }\href@noop {} {\bibfield  {journal} {\bibinfo  {journal} {Nat. Rev.
  Immunol.}\ }\textbf {\bibinfo {volume} {8}},\ \bibinfo {pages} {247}
  (\bibinfo {year} {2008})}\BibitemShut {NoStop}%
\bibitem [{\citenamefont {Youngblood}\ \emph {et~al.}(2017)\citenamefont
  {Youngblood}, \citenamefont {Hale}, \citenamefont {Kissick}, \citenamefont
  {Ahn}, \citenamefont {Xu}, \citenamefont {Wieland}, \citenamefont {Araki},
  \citenamefont {West}, \citenamefont {Ghoneim}, \citenamefont {Fan},
  \citenamefont {Dogra}, \citenamefont {Davis}, \citenamefont {Konieczny},
  \citenamefont {Antia}, \citenamefont {Cheng},\ and\ \citenamefont
  {Ahmed}}]{Youngblood2017-my}%
  \BibitemOpen
  \bibfield  {author} {\bibinfo {author} {\bibfnamefont {B.}~\bibnamefont
  {Youngblood}}, \bibinfo {author} {\bibfnamefont {J.~S.}\ \bibnamefont
  {Hale}}, \bibinfo {author} {\bibfnamefont {H.~T.}\ \bibnamefont {Kissick}},
  \bibinfo {author} {\bibfnamefont {E.}~\bibnamefont {Ahn}}, \bibinfo {author}
  {\bibfnamefont {X.}~\bibnamefont {Xu}}, \bibinfo {author} {\bibfnamefont
  {A.}~\bibnamefont {Wieland}}, \bibinfo {author} {\bibfnamefont
  {K.}~\bibnamefont {Araki}}, \bibinfo {author} {\bibfnamefont {E.~E.}\
  \bibnamefont {West}}, \bibinfo {author} {\bibfnamefont {H.~E.}\ \bibnamefont
  {Ghoneim}}, \bibinfo {author} {\bibfnamefont {Y.}~\bibnamefont {Fan}},
  \bibinfo {author} {\bibfnamefont {P.}~\bibnamefont {Dogra}}, \bibinfo
  {author} {\bibfnamefont {C.~W.}\ \bibnamefont {Davis}}, \bibinfo {author}
  {\bibfnamefont {B.~T.}\ \bibnamefont {Konieczny}}, \bibinfo {author}
  {\bibfnamefont {R.}~\bibnamefont {Antia}}, \bibinfo {author} {\bibfnamefont
  {X.}~\bibnamefont {Cheng}},\ and\ \bibinfo {author} {\bibfnamefont
  {R.}~\bibnamefont {Ahmed}},\ }\href@noop {} {\bibfield  {journal} {\bibinfo
  {journal} {Nature}\ }\textbf {\bibinfo {volume} {552}},\ \bibinfo {pages}
  {404} (\bibinfo {year} {2017})}\BibitemShut {NoStop}%
\bibitem [{\citenamefont {Abadie}\ \emph {et~al.}(2024)\citenamefont {Abadie},
  \citenamefont {Clark}, \citenamefont {Valanparambil}, \citenamefont {Ukogu},
  \citenamefont {Yang}, \citenamefont {Daza}, \citenamefont {Ng}, \citenamefont
  {Fathima}, \citenamefont {Wang}, \citenamefont {Lee}, \citenamefont {Nasti},
  \citenamefont {Bhandoola}, \citenamefont {Nourmohammad}, \citenamefont
  {Ahmed}, \citenamefont {Shendure}, \citenamefont {Cao},\ and\ \citenamefont
  {Kueh}}]{Abadie2024-rd}%
  \BibitemOpen
  \bibfield  {author} {\bibinfo {author} {\bibfnamefont {K.}~\bibnamefont
  {Abadie}}, \bibinfo {author} {\bibfnamefont {E.~C.}\ \bibnamefont {Clark}},
  \bibinfo {author} {\bibfnamefont {R.~M.}\ \bibnamefont {Valanparambil}},
  \bibinfo {author} {\bibfnamefont {O.}~\bibnamefont {Ukogu}}, \bibinfo
  {author} {\bibfnamefont {W.}~\bibnamefont {Yang}}, \bibinfo {author}
  {\bibfnamefont {R.~M.}\ \bibnamefont {Daza}}, \bibinfo {author}
  {\bibfnamefont {K.~K.~H.}\ \bibnamefont {Ng}}, \bibinfo {author}
  {\bibfnamefont {J.}~\bibnamefont {Fathima}}, \bibinfo {author} {\bibfnamefont
  {A.~L.}\ \bibnamefont {Wang}}, \bibinfo {author} {\bibfnamefont
  {J.}~\bibnamefont {Lee}}, \bibinfo {author} {\bibfnamefont {T.~H.}\
  \bibnamefont {Nasti}}, \bibinfo {author} {\bibfnamefont {A.}~\bibnamefont
  {Bhandoola}}, \bibinfo {author} {\bibfnamefont {A.}~\bibnamefont
  {Nourmohammad}}, \bibinfo {author} {\bibfnamefont {R.}~\bibnamefont {Ahmed}},
  \bibinfo {author} {\bibfnamefont {J.}~\bibnamefont {Shendure}}, \bibinfo
  {author} {\bibfnamefont {J.}~\bibnamefont {Cao}},\ and\ \bibinfo {author}
  {\bibfnamefont {H.~Y.}\ \bibnamefont {Kueh}},\ }\href@noop {} {\bibfield
  {journal} {\bibinfo  {journal} {Immunity}\ }\textbf {\bibinfo {volume}
  {57}},\ \bibinfo {pages} {271} (\bibinfo {year} {2024})}\BibitemShut
  {NoStop}%
\bibitem [{\citenamefont {Heinzel}\ \emph {et~al.}(2017)\citenamefont
  {Heinzel}, \citenamefont {Binh~Giang}, \citenamefont {Kan}, \citenamefont
  {Marchingo}, \citenamefont {Lye}, \citenamefont {Corcoran},\ and\
  \citenamefont {Hodgkin}}]{Heinzel2017-qp}%
  \BibitemOpen
  \bibfield  {author} {\bibinfo {author} {\bibfnamefont {S.}~\bibnamefont
  {Heinzel}}, \bibinfo {author} {\bibfnamefont {T.}~\bibnamefont {Binh~Giang}},
  \bibinfo {author} {\bibfnamefont {A.}~\bibnamefont {Kan}}, \bibinfo {author}
  {\bibfnamefont {J.~M.}\ \bibnamefont {Marchingo}}, \bibinfo {author}
  {\bibfnamefont {B.~K.}\ \bibnamefont {Lye}}, \bibinfo {author} {\bibfnamefont
  {L.~M.}\ \bibnamefont {Corcoran}},\ and\ \bibinfo {author} {\bibfnamefont
  {P.~D.}\ \bibnamefont {Hodgkin}},\ }\href@noop {} {\bibfield  {journal}
  {\bibinfo  {journal} {Nat. Immunol.}\ }\textbf {\bibinfo {volume} {18}},\
  \bibinfo {pages} {96} (\bibinfo {year} {2017})}\BibitemShut {NoStop}%
\bibitem [{\citenamefont {Cho}\ \emph {et~al.}(2017)\citenamefont {Cho},
  \citenamefont {Flossdorf}, \citenamefont {Kretschmer}, \citenamefont
  {Höfer}, \citenamefont {Busch},\ and\ \citenamefont
  {Buchholz}}]{Cho2017-ul}%
  \BibitemOpen
  \bibfield  {author} {\bibinfo {author} {\bibfnamefont {Y.-L.}\ \bibnamefont
  {Cho}}, \bibinfo {author} {\bibfnamefont {M.}~\bibnamefont {Flossdorf}},
  \bibinfo {author} {\bibfnamefont {L.}~\bibnamefont {Kretschmer}}, \bibinfo
  {author} {\bibfnamefont {T.}~\bibnamefont {Höfer}}, \bibinfo {author}
  {\bibfnamefont {D.~H.}\ \bibnamefont {Busch}},\ and\ \bibinfo {author}
  {\bibfnamefont {V.~R.}\ \bibnamefont {Buchholz}},\ }\href@noop {} {\bibfield
  {journal} {\bibinfo  {journal} {Cell Rep.}\ }\textbf {\bibinfo {volume}
  {20}},\ \bibinfo {pages} {806} (\bibinfo {year} {2017})}\BibitemShut
  {NoStop}%
\bibitem [{\citenamefont {Plambeck}\ \emph {et~al.}(2022)\citenamefont
  {Plambeck}, \citenamefont {Kazeroonian}, \citenamefont {Loeffler},
  \citenamefont {Kretschmer}, \citenamefont {Salinno}, \citenamefont
  {Schroeder}, \citenamefont {Busch}, \citenamefont {Flossdorf},\ and\
  \citenamefont {Buchholz}}]{Plambeck2022-dt}%
  \BibitemOpen
  \bibfield  {author} {\bibinfo {author} {\bibfnamefont {M.}~\bibnamefont
  {Plambeck}}, \bibinfo {author} {\bibfnamefont {A.}~\bibnamefont
  {Kazeroonian}}, \bibinfo {author} {\bibfnamefont {D.}~\bibnamefont
  {Loeffler}}, \bibinfo {author} {\bibfnamefont {L.}~\bibnamefont
  {Kretschmer}}, \bibinfo {author} {\bibfnamefont {C.}~\bibnamefont {Salinno}},
  \bibinfo {author} {\bibfnamefont {T.}~\bibnamefont {Schroeder}}, \bibinfo
  {author} {\bibfnamefont {D.~H.}\ \bibnamefont {Busch}}, \bibinfo {author}
  {\bibfnamefont {M.}~\bibnamefont {Flossdorf}},\ and\ \bibinfo {author}
  {\bibfnamefont {V.~R.}\ \bibnamefont {Buchholz}},\ }\href@noop {} {\bibfield
  {journal} {\bibinfo  {journal} {Proceedings of the National Academy of
  Sciences}\ }\textbf {\bibinfo {volume} {119}},\ \bibinfo {pages}
  {e2116260119} (\bibinfo {year} {2022})}\BibitemShut {NoStop}%
\bibitem [{\citenamefont {Xin}\ \emph {et~al.}(2016)\citenamefont {Xin},
  \citenamefont {Masson}, \citenamefont {Liao}, \citenamefont {Preston},
  \citenamefont {Guan}, \citenamefont {Gloury}, \citenamefont {Olshansky},
  \citenamefont {Lin}, \citenamefont {Li}, \citenamefont {Speed}, \citenamefont
  {Smyth}, \citenamefont {Ernst}, \citenamefont {Leonard}, \citenamefont
  {Pellegrini}, \citenamefont {Kaech}, \citenamefont {Nutt}, \citenamefont
  {Shi}, \citenamefont {Belz},\ and\ \citenamefont {Kallies}}]{Xin2016-ut}%
  \BibitemOpen
  \bibfield  {author} {\bibinfo {author} {\bibfnamefont {A.}~\bibnamefont
  {Xin}}, \bibinfo {author} {\bibfnamefont {F.}~\bibnamefont {Masson}},
  \bibinfo {author} {\bibfnamefont {Y.}~\bibnamefont {Liao}}, \bibinfo {author}
  {\bibfnamefont {S.}~\bibnamefont {Preston}}, \bibinfo {author} {\bibfnamefont
  {T.}~\bibnamefont {Guan}}, \bibinfo {author} {\bibfnamefont {R.}~\bibnamefont
  {Gloury}}, \bibinfo {author} {\bibfnamefont {M.}~\bibnamefont {Olshansky}},
  \bibinfo {author} {\bibfnamefont {J.-X.}\ \bibnamefont {Lin}}, \bibinfo
  {author} {\bibfnamefont {P.}~\bibnamefont {Li}}, \bibinfo {author}
  {\bibfnamefont {T.~P.}\ \bibnamefont {Speed}}, \bibinfo {author}
  {\bibfnamefont {G.~K.}\ \bibnamefont {Smyth}}, \bibinfo {author}
  {\bibfnamefont {M.}~\bibnamefont {Ernst}}, \bibinfo {author} {\bibfnamefont
  {W.~J.}\ \bibnamefont {Leonard}}, \bibinfo {author} {\bibfnamefont
  {M.}~\bibnamefont {Pellegrini}}, \bibinfo {author} {\bibfnamefont {S.~M.}\
  \bibnamefont {Kaech}}, \bibinfo {author} {\bibfnamefont {S.~L.}\ \bibnamefont
  {Nutt}}, \bibinfo {author} {\bibfnamefont {W.}~\bibnamefont {Shi}}, \bibinfo
  {author} {\bibfnamefont {G.~T.}\ \bibnamefont {Belz}},\ and\ \bibinfo
  {author} {\bibfnamefont {A.}~\bibnamefont {Kallies}},\ }\href@noop {}
  {\bibfield  {journal} {\bibinfo  {journal} {Nat. Immunol.}\ }\textbf
  {\bibinfo {volume} {17}},\ \bibinfo {pages} {422} (\bibinfo {year}
  {2016})}\BibitemShut {NoStop}%
\bibitem [{\citenamefont {Kretschmer}\ \emph {et~al.}(2020)\citenamefont
  {Kretschmer}, \citenamefont {Flossdorf}, \citenamefont {Mir}, \citenamefont
  {Cho}, \citenamefont {Plambeck}, \citenamefont {Treise}, \citenamefont
  {Toska}, \citenamefont {Heinzel}, \citenamefont {Schiemann}, \citenamefont
  {Busch},\ and\ \citenamefont {Buchholz}}]{Kretschmer2020-ad}%
  \BibitemOpen
  \bibfield  {author} {\bibinfo {author} {\bibfnamefont {L.}~\bibnamefont
  {Kretschmer}}, \bibinfo {author} {\bibfnamefont {M.}~\bibnamefont
  {Flossdorf}}, \bibinfo {author} {\bibfnamefont {J.}~\bibnamefont {Mir}},
  \bibinfo {author} {\bibfnamefont {Y.-L.}\ \bibnamefont {Cho}}, \bibinfo
  {author} {\bibfnamefont {M.}~\bibnamefont {Plambeck}}, \bibinfo {author}
  {\bibfnamefont {I.}~\bibnamefont {Treise}}, \bibinfo {author} {\bibfnamefont
  {A.}~\bibnamefont {Toska}}, \bibinfo {author} {\bibfnamefont
  {S.}~\bibnamefont {Heinzel}}, \bibinfo {author} {\bibfnamefont
  {M.}~\bibnamefont {Schiemann}}, \bibinfo {author} {\bibfnamefont {D.~H.}\
  \bibnamefont {Busch}},\ and\ \bibinfo {author} {\bibfnamefont {V.~R.}\
  \bibnamefont {Buchholz}},\ }\href@noop {} {\bibfield  {journal} {\bibinfo
  {journal} {Nat. Commun.}\ }\textbf {\bibinfo {volume} {11}},\ \bibinfo
  {pages} {113} (\bibinfo {year} {2020})}\BibitemShut {NoStop}%
\bibitem [{\citenamefont {Downie}\ \emph {et~al.}(2021)\citenamefont {Downie},
  \citenamefont {Mayer}, \citenamefont {Metcalf},\ and\ \citenamefont
  {Graham}}]{Downie2021-xe}%
  \BibitemOpen
  \bibfield  {author} {\bibinfo {author} {\bibfnamefont {A.~E.}\ \bibnamefont
  {Downie}}, \bibinfo {author} {\bibfnamefont {A.}~\bibnamefont {Mayer}},
  \bibinfo {author} {\bibfnamefont {C.~J.~E.}\ \bibnamefont {Metcalf}},\ and\
  \bibinfo {author} {\bibfnamefont {A.~L.}\ \bibnamefont {Graham}},\
  }\href@noop {} {\bibfield  {journal} {\bibinfo  {journal} {PLoS Comput.
  Biol.}\ }\textbf {\bibinfo {volume} {17}},\ \bibinfo {pages} {e1009714}
  (\bibinfo {year} {2021})}\BibitemShut {NoStop}%
\bibitem [{\citenamefont {Mayer}\ \emph {et~al.}(2016)\citenamefont {Mayer},
  \citenamefont {Mora}, \citenamefont {Rivoire},\ and\ \citenamefont
  {Walczak}}]{Mayer2016-mw}%
  \BibitemOpen
  \bibfield  {author} {\bibinfo {author} {\bibfnamefont {A.}~\bibnamefont
  {Mayer}}, \bibinfo {author} {\bibfnamefont {T.}~\bibnamefont {Mora}},
  \bibinfo {author} {\bibfnamefont {O.}~\bibnamefont {Rivoire}},\ and\ \bibinfo
  {author} {\bibfnamefont {A.~M.}\ \bibnamefont {Walczak}},\ }\href@noop {}
  {\bibfield  {journal} {\bibinfo  {journal} {Proc. Natl. Acad. Sci. U. S. A.}\
  }\textbf {\bibinfo {volume} {113}},\ \bibinfo {pages} {8630} (\bibinfo {year}
  {2016})}\BibitemShut {NoStop}%
\bibitem [{\citenamefont {van~den Berg}\ and\ \citenamefont
  {Kiselëv}(2004)}]{Van_den_Berg2004-pn}%
  \BibitemOpen
  \bibfield  {author} {\bibinfo {author} {\bibfnamefont {H.~A.}\ \bibnamefont
  {van~den Berg}}\ and\ \bibinfo {author} {\bibfnamefont {Y.~N.}\ \bibnamefont
  {Kiselëv}},\ }\href@noop {} {\bibfield  {journal} {\bibinfo  {journal}
  {Bull. Math. Biol.}\ }\textbf {\bibinfo {volume} {66}},\ \bibinfo {pages}
  {1345} (\bibinfo {year} {2004})}\BibitemShut {NoStop}%
\bibitem [{\citenamefont {Green}\ \emph {et~al.}(2009)\citenamefont {Green},
  \citenamefont {Ferguson}, \citenamefont {Zitvogel},\ and\ \citenamefont
  {Kroemer}}]{Green2009-el}%
  \BibitemOpen
  \bibfield  {author} {\bibinfo {author} {\bibfnamefont {D.~R.}\ \bibnamefont
  {Green}}, \bibinfo {author} {\bibfnamefont {T.}~\bibnamefont {Ferguson}},
  \bibinfo {author} {\bibfnamefont {L.}~\bibnamefont {Zitvogel}},\ and\
  \bibinfo {author} {\bibfnamefont {G.}~\bibnamefont {Kroemer}},\ }\href@noop
  {} {\bibfield  {journal} {\bibinfo  {journal} {Nat. Rev. Immunol.}\ }\textbf
  {\bibinfo {volume} {9}},\ \bibinfo {pages} {353} (\bibinfo {year}
  {2009})}\BibitemShut {NoStop}%
\bibitem [{\citenamefont {Labbé}\ and\ \citenamefont
  {Saleh}(2008)}]{Labbe2008-px}%
  \BibitemOpen
  \bibfield  {author} {\bibinfo {author} {\bibfnamefont {K.}~\bibnamefont
  {Labbé}}\ and\ \bibinfo {author} {\bibfnamefont {M.}~\bibnamefont {Saleh}},\
  }\href@noop {} {\bibfield  {journal} {\bibinfo  {journal} {Cell Death
  Differ.}\ }\textbf {\bibinfo {volume} {15}},\ \bibinfo {pages} {1339}
  (\bibinfo {year} {2008})}\BibitemShut {NoStop}%
\bibitem [{\citenamefont {Butler}\ \emph {et~al.}(2013)\citenamefont {Butler},
  \citenamefont {Kardar},\ and\ \citenamefont {Chakraborty}}]{Butler2013-li}%
  \BibitemOpen
  \bibfield  {author} {\bibinfo {author} {\bibfnamefont {T.~C.}\ \bibnamefont
  {Butler}}, \bibinfo {author} {\bibfnamefont {M.}~\bibnamefont {Kardar}},\
  and\ \bibinfo {author} {\bibfnamefont {A.~K.}\ \bibnamefont {Chakraborty}},\
  }\href@noop {} {\bibfield  {journal} {\bibinfo  {journal} {Proc. Natl. Acad.
  Sci. U. S. A.}\ }\textbf {\bibinfo {volume} {110}},\ \bibinfo {pages} {11833}
  (\bibinfo {year} {2013})}\BibitemShut {NoStop}%
\bibitem [{\citenamefont {Cicchese}\ \emph {et~al.}(2018)\citenamefont
  {Cicchese}, \citenamefont {Evans}, \citenamefont {Hult}, \citenamefont
  {Joslyn}, \citenamefont {Wessler}, \citenamefont {Millar}, \citenamefont
  {Marino}, \citenamefont {Cilfone}, \citenamefont {Mattila}, \citenamefont
  {Linderman},\ and\ \citenamefont {Kirschner}}]{Cicchese2018-po}%
  \BibitemOpen
  \bibfield  {author} {\bibinfo {author} {\bibfnamefont {J.~M.}\ \bibnamefont
  {Cicchese}}, \bibinfo {author} {\bibfnamefont {S.}~\bibnamefont {Evans}},
  \bibinfo {author} {\bibfnamefont {C.}~\bibnamefont {Hult}}, \bibinfo {author}
  {\bibfnamefont {L.~R.}\ \bibnamefont {Joslyn}}, \bibinfo {author}
  {\bibfnamefont {T.}~\bibnamefont {Wessler}}, \bibinfo {author} {\bibfnamefont
  {J.~A.}\ \bibnamefont {Millar}}, \bibinfo {author} {\bibfnamefont
  {S.}~\bibnamefont {Marino}}, \bibinfo {author} {\bibfnamefont {N.~A.}\
  \bibnamefont {Cilfone}}, \bibinfo {author} {\bibfnamefont {J.~T.}\
  \bibnamefont {Mattila}}, \bibinfo {author} {\bibfnamefont {J.~J.}\
  \bibnamefont {Linderman}},\ and\ \bibinfo {author} {\bibfnamefont {D.~E.}\
  \bibnamefont {Kirschner}},\ }\href@noop {} {\bibfield  {journal} {\bibinfo
  {journal} {Immunol. Rev.}\ }\textbf {\bibinfo {volume} {285}},\ \bibinfo
  {pages} {147} (\bibinfo {year} {2018})}\BibitemShut {NoStop}%
\bibitem [{\citenamefont {Medzhitov}(2021)}]{Medzhitov2021-to}%
  \BibitemOpen
  \bibfield  {author} {\bibinfo {author} {\bibfnamefont {R.}~\bibnamefont
  {Medzhitov}},\ }\href@noop {} {\bibfield  {journal} {\bibinfo  {journal}
  {Science}\ }\textbf {\bibinfo {volume} {374}},\ \bibinfo {pages} {1070}
  (\bibinfo {year} {2021})}\BibitemShut {NoStop}%
\bibitem [{\citenamefont {François}\ and\ \citenamefont
  {Hakim}(2004)}]{Francois2004-mc}%
  \BibitemOpen
  \bibfield  {author} {\bibinfo {author} {\bibfnamefont {P.}~\bibnamefont
  {François}}\ and\ \bibinfo {author} {\bibfnamefont {V.}~\bibnamefont
  {Hakim}},\ }\href@noop {} {\bibfield  {journal} {\bibinfo  {journal} {Proc.
  Natl. Acad. Sci. U. S. A.}\ }\textbf {\bibinfo {volume} {101}},\ \bibinfo
  {pages} {580} (\bibinfo {year} {2004})}\BibitemShut {NoStop}%
\bibitem [{\citenamefont {Shinar}\ \emph {et~al.}(2007)\citenamefont {Shinar},
  \citenamefont {Milo}, \citenamefont {Martínez},\ and\ \citenamefont
  {Alon}}]{Shinar2007-mz}%
  \BibitemOpen
  \bibfield  {author} {\bibinfo {author} {\bibfnamefont {G.}~\bibnamefont
  {Shinar}}, \bibinfo {author} {\bibfnamefont {R.}~\bibnamefont {Milo}},
  \bibinfo {author} {\bibfnamefont {M.~R.}\ \bibnamefont {Martínez}},\ and\
  \bibinfo {author} {\bibfnamefont {U.}~\bibnamefont {Alon}},\ }\href@noop {}
  {\bibfield  {journal} {\bibinfo  {journal} {Proc. Natl. Acad. Sci. U. S. A.}\
  }\textbf {\bibinfo {volume} {104}},\ \bibinfo {pages} {19931} (\bibinfo
  {year} {2007})}\BibitemShut {NoStop}%
\bibitem [{\citenamefont {Lalanne}\ and\ \citenamefont
  {François}(2013)}]{Lalanne2013-rr}%
  \BibitemOpen
  \bibfield  {author} {\bibinfo {author} {\bibfnamefont {J.-B.}\ \bibnamefont
  {Lalanne}}\ and\ \bibinfo {author} {\bibfnamefont {P.}~\bibnamefont
  {François}},\ }\href@noop {} {\bibfield  {journal} {\bibinfo  {journal}
  {Phys. Rev. Lett.}\ }\textbf {\bibinfo {volume} {110}},\ \bibinfo {pages}
  {218102} (\bibinfo {year} {2013})}\BibitemShut {NoStop}%
\bibitem [{\citenamefont {Hart}\ \emph {et~al.}(2014)\citenamefont {Hart},
  \citenamefont {Reich-Zeliger}, \citenamefont {Antebi}, \citenamefont
  {Zaretsky}, \citenamefont {Mayo}, \citenamefont {Alon},\ and\ \citenamefont
  {Friedman}}]{Hart2014-ms}%
  \BibitemOpen
  \bibfield  {author} {\bibinfo {author} {\bibfnamefont {Y.}~\bibnamefont
  {Hart}}, \bibinfo {author} {\bibfnamefont {S.}~\bibnamefont {Reich-Zeliger}},
  \bibinfo {author} {\bibfnamefont {Y.~E.}\ \bibnamefont {Antebi}}, \bibinfo
  {author} {\bibfnamefont {I.}~\bibnamefont {Zaretsky}}, \bibinfo {author}
  {\bibfnamefont {A.~E.}\ \bibnamefont {Mayo}}, \bibinfo {author}
  {\bibfnamefont {U.}~\bibnamefont {Alon}},\ and\ \bibinfo {author}
  {\bibfnamefont {N.}~\bibnamefont {Friedman}},\ }\href@noop {} {\bibfield
  {journal} {\bibinfo  {journal} {Cell}\ }\textbf {\bibinfo {volume} {158}},\
  \bibinfo {pages} {1022} (\bibinfo {year} {2014})}\BibitemShut {NoStop}%
\bibitem [{\citenamefont {Alon}(2019)}]{Alon2019-ep}%
  \BibitemOpen
  \bibfield  {author} {\bibinfo {author} {\bibfnamefont {U.}~\bibnamefont
  {Alon}},\ }\href@noop {} {{\selectlanguage {english}\emph {\bibinfo {title}
  {An introduction to systems biology: Design principles of biological
  circuits}}}}\ (\bibinfo  {publisher} {Chapman and Hall/CRC},\ \bibinfo
  {address} {Second edition. | Boca Raton, Fla. : CRC Press, [2019]},\ \bibinfo
  {year} {2019})\BibitemShut {NoStop}%
\bibitem [{\citenamefont {Sokolowski}\ \emph {et~al.}(2025)\citenamefont
  {Sokolowski}, \citenamefont {Gregor}, \citenamefont {Bialek},\ and\
  \citenamefont {Tkačik}}]{Sokolowski2023-um}%
  \BibitemOpen
  \bibfield  {author} {\bibinfo {author} {\bibfnamefont {T.~R.}\ \bibnamefont
  {Sokolowski}}, \bibinfo {author} {\bibfnamefont {T.}~\bibnamefont {Gregor}},
  \bibinfo {author} {\bibfnamefont {W.}~\bibnamefont {Bialek}},\ and\ \bibinfo
  {author} {\bibfnamefont {G.}~\bibnamefont {Tkačik}},\ }\href
  {https://doi.org/10.1073/pnas.2402925121} {\bibfield  {journal} {\bibinfo
  {journal} {Proceedings of the National Academy of Sciences}\ }\textbf
  {\bibinfo {volume} {122}},\ \bibinfo {pages} {e2402925121} (\bibinfo {year}
  {2025})},\ \Eprint
  {https://arxiv.org/abs/https://www.pnas.org/doi/pdf/10.1073/pnas.2402925121}
  {https://www.pnas.org/doi/pdf/10.1073/pnas.2402925121} \BibitemShut {NoStop}%
\bibitem [{\citenamefont {Perelson}\ and\ \citenamefont
  {Oster}(1979)}]{Perelson1979-kq}%
  \BibitemOpen
  \bibfield  {author} {\bibinfo {author} {\bibfnamefont {A.~S.}\ \bibnamefont
  {Perelson}}\ and\ \bibinfo {author} {\bibfnamefont {G.~F.}\ \bibnamefont
  {Oster}},\ }\href@noop {} {\bibfield  {journal} {\bibinfo  {journal} {J.
  Theor. Biol.}\ }\textbf {\bibinfo {volume} {81}},\ \bibinfo {pages} {645}
  (\bibinfo {year} {1979})}\BibitemShut {NoStop}%
\bibitem [{\citenamefont {Mayer}\ \emph {et~al.}(2015)\citenamefont {Mayer},
  \citenamefont {Balasubramanian}, \citenamefont {Mora},\ and\ \citenamefont
  {Walczak}}]{Mayer2015-ii}%
  \BibitemOpen
  \bibfield  {author} {\bibinfo {author} {\bibfnamefont {A.}~\bibnamefont
  {Mayer}}, \bibinfo {author} {\bibfnamefont {V.}~\bibnamefont
  {Balasubramanian}}, \bibinfo {author} {\bibfnamefont {T.}~\bibnamefont
  {Mora}},\ and\ \bibinfo {author} {\bibfnamefont {A.~M.}\ \bibnamefont
  {Walczak}},\ }\href@noop {} {\bibfield  {journal} {\bibinfo  {journal} {Proc.
  Natl. Acad. Sci. U. S. A.}\ }\textbf {\bibinfo {volume} {112}},\ \bibinfo
  {pages} {5950} (\bibinfo {year} {2015})}\BibitemShut {NoStop}%
\bibitem [{\citenamefont {Bradde}\ \emph {et~al.}(2020)\citenamefont {Bradde},
  \citenamefont {Nourmohammad}, \citenamefont {Goyal},\ and\ \citenamefont
  {Balasubramanian}}]{Bradde2020-lt}%
  \BibitemOpen
  \bibfield  {author} {\bibinfo {author} {\bibfnamefont {S.}~\bibnamefont
  {Bradde}}, \bibinfo {author} {\bibfnamefont {A.}~\bibnamefont
  {Nourmohammad}}, \bibinfo {author} {\bibfnamefont {S.}~\bibnamefont
  {Goyal}},\ and\ \bibinfo {author} {\bibfnamefont {V.}~\bibnamefont
  {Balasubramanian}},\ }\href@noop {} {\bibfield  {journal} {\bibinfo
  {journal} {Proc. Natl. Acad. Sci. U. S. A.}\ }\textbf {\bibinfo {volume}
  {117}},\ \bibinfo {pages} {5144} (\bibinfo {year} {2020})}\BibitemShut
  {NoStop}%
\bibitem [{\citenamefont {Mayer}\ \emph
  {et~al.}(2019{\natexlab{a}})\citenamefont {Mayer}, \citenamefont
  {Balasubramanian}, \citenamefont {Walczak},\ and\ \citenamefont
  {Mora}}]{Mayer2019-tx}%
  \BibitemOpen
  \bibfield  {author} {\bibinfo {author} {\bibfnamefont {A.}~\bibnamefont
  {Mayer}}, \bibinfo {author} {\bibfnamefont {V.}~\bibnamefont
  {Balasubramanian}}, \bibinfo {author} {\bibfnamefont {A.~M.}\ \bibnamefont
  {Walczak}},\ and\ \bibinfo {author} {\bibfnamefont {T.}~\bibnamefont
  {Mora}},\ }\href@noop {} {\bibfield  {journal} {\bibinfo  {journal} {Proc.
  Natl. Acad. Sci. U. S. A.}\ }\textbf {\bibinfo {volume} {116}},\ \bibinfo
  {pages} {8815} (\bibinfo {year} {2019}{\natexlab{a}})}\BibitemShut {NoStop}%
\bibitem [{\citenamefont {Schnaack}\ and\ \citenamefont
  {Nourmohammad}(2021)}]{Schnaack2021-wh}%
  \BibitemOpen
  \bibfield  {author} {\bibinfo {author} {\bibfnamefont {O.~H.}\ \bibnamefont
  {Schnaack}}\ and\ \bibinfo {author} {\bibfnamefont {A.}~\bibnamefont
  {Nourmohammad}},\ }\href@noop {} {\bibfield  {journal} {\bibinfo  {journal}
  {Elife}\ }\textbf {\bibinfo {volume} {10}} (\bibinfo {year}
  {2021})}\BibitemShut {NoStop}%
\bibitem [{\citenamefont {Chardès}\ \emph {et~al.}(2021)\citenamefont
  {Chardès}, \citenamefont {Vergassola}, \citenamefont {Walczak},\ and\
  \citenamefont {Mora}}]{Chardes2021-pf}%
  \BibitemOpen
  \bibfield  {author} {\bibinfo {author} {\bibfnamefont {V.}~\bibnamefont
  {Chardès}}, \bibinfo {author} {\bibfnamefont {M.}~\bibnamefont
  {Vergassola}}, \bibinfo {author} {\bibfnamefont {A.~M.}\ \bibnamefont
  {Walczak}},\ and\ \bibinfo {author} {\bibfnamefont {T.}~\bibnamefont
  {Mora}},\ }\href@noop {} {\bibfield  {journal} {\bibinfo  {journal} {arXiv
  [q-bio.PE]}\ ,\ \bibinfo {pages} {2021.07.26.453765}} (\bibinfo {year}
  {2021})}\BibitemShut {NoStop}%
\bibitem [{\citenamefont {Schnaack}\ \emph {et~al.}(2022)\citenamefont
  {Schnaack}, \citenamefont {Peliti},\ and\ \citenamefont
  {Nourmohammad}}]{Schnaack2022-jf}%
  \BibitemOpen
  \bibfield  {author} {\bibinfo {author} {\bibfnamefont {O.~H.}\ \bibnamefont
  {Schnaack}}, \bibinfo {author} {\bibfnamefont {L.}~\bibnamefont {Peliti}},\
  and\ \bibinfo {author} {\bibfnamefont {A.}~\bibnamefont {Nourmohammad}},\
  }\href@noop {} {\bibfield  {journal} {\bibinfo  {journal} {Phys. Rev. X}\
  }\textbf {\bibinfo {volume} {12}},\ \bibinfo {pages} {021063} (\bibinfo
  {year} {2022})}\BibitemShut {NoStop}%
\bibitem [{\citenamefont {Qin}\ \emph {et~al.}(2023)\citenamefont {Qin},
  \citenamefont {Mace},\ and\ \citenamefont {Barton}}]{Qin2023-is}%
  \BibitemOpen
  \bibfield  {author} {\bibinfo {author} {\bibfnamefont {Y.}~\bibnamefont
  {Qin}}, \bibinfo {author} {\bibfnamefont {E.~M.}\ \bibnamefont {Mace}},\ and\
  \bibinfo {author} {\bibfnamefont {J.~P.}\ \bibnamefont {Barton}},\
  }\href@noop {} {\bibfield  {journal} {\bibinfo  {journal} {Proc. Natl. Acad.
  Sci. U. S. A.}\ }\textbf {\bibinfo {volume} {120}},\ \bibinfo {pages}
  {e2305859120} (\bibinfo {year} {2023})}\BibitemShut {NoStop}%
\bibitem [{\citenamefont {Milligan}\ and\ \citenamefont
  {Barrett}(2015)}]{Milligan2015-yu}%
  \BibitemOpen
  \bibfield  {author} {\bibinfo {author} {\bibfnamefont {G.~N.}\ \bibnamefont
  {Milligan}}\ and\ \bibinfo {author} {\bibfnamefont {A.~D.~T.}\ \bibnamefont
  {Barrett}},\ }\href@noop {} {{\selectlanguage {english}\emph {\bibinfo
  {title} {Vaccinology: An Essential Guide}}}},\ edited by\ \bibinfo {editor}
  {\bibfnamefont {G.~N.}\ \bibnamefont {Milligan}}\ and\ \bibinfo {editor}
  {\bibfnamefont {A.~D.~T.}\ \bibnamefont {Barrett}}\ (\bibinfo  {publisher}
  {Wiley-Blackwell},\ \bibinfo {address} {Hoboken, NJ},\ \bibinfo {year}
  {2015})\BibitemShut {NoStop}%
\bibitem [{\citenamefont {Abuin}\ \emph {et~al.}(2020)\citenamefont {Abuin},
  \citenamefont {Anderson}, \citenamefont {Ferramosca}, \citenamefont
  {Hernandez-Vargas},\ and\ \citenamefont {Gonzalez}}]{Abuin2020-tf}%
  \BibitemOpen
  \bibfield  {author} {\bibinfo {author} {\bibfnamefont {P.}~\bibnamefont
  {Abuin}}, \bibinfo {author} {\bibfnamefont {A.}~\bibnamefont {Anderson}},
  \bibinfo {author} {\bibfnamefont {A.}~\bibnamefont {Ferramosca}}, \bibinfo
  {author} {\bibfnamefont {E.~A.}\ \bibnamefont {Hernandez-Vargas}},\ and\
  \bibinfo {author} {\bibfnamefont {A.~H.}\ \bibnamefont {Gonzalez}},\
  }\href@noop {} {\bibfield  {journal} {\bibinfo  {journal} {Annu. Rev.
  Control}\ }\textbf {\bibinfo {volume} {50}},\ \bibinfo {pages} {457}
  (\bibinfo {year} {2020})}\BibitemShut {NoStop}%
\bibitem [{\citenamefont {van Stipdonk}\ \emph {et~al.}(2001)\citenamefont {van
  Stipdonk}, \citenamefont {Lemmens},\ and\ \citenamefont
  {Schoenberger}}]{Van_Stipdonk2001-nm}%
  \BibitemOpen
  \bibfield  {author} {\bibinfo {author} {\bibfnamefont {M.~J.}\ \bibnamefont
  {van Stipdonk}}, \bibinfo {author} {\bibfnamefont {E.~E.}\ \bibnamefont
  {Lemmens}},\ and\ \bibinfo {author} {\bibfnamefont {S.~P.}\ \bibnamefont
  {Schoenberger}},\ }\href@noop {} {\bibfield  {journal} {\bibinfo  {journal}
  {Nat. Immunol.}\ }\textbf {\bibinfo {volume} {2}},\ \bibinfo {pages} {423}
  (\bibinfo {year} {2001})}\BibitemShut {NoStop}%
\bibitem [{\citenamefont {François}\ \emph {et~al.}(2013)\citenamefont
  {François}, \citenamefont {Voisinne}, \citenamefont {Siggia}, \citenamefont
  {Altan-Bonnet},\ and\ \citenamefont {Vergassola}}]{Francois2013-yl}%
  \BibitemOpen
  \bibfield  {author} {\bibinfo {author} {\bibfnamefont {P.}~\bibnamefont
  {François}}, \bibinfo {author} {\bibfnamefont {G.}~\bibnamefont {Voisinne}},
  \bibinfo {author} {\bibfnamefont {E.~D.}\ \bibnamefont {Siggia}}, \bibinfo
  {author} {\bibfnamefont {G.}~\bibnamefont {Altan-Bonnet}},\ and\ \bibinfo
  {author} {\bibfnamefont {M.}~\bibnamefont {Vergassola}},\ }\href@noop {}
  {\bibfield  {journal} {\bibinfo  {journal} {Proc. Natl. Acad. Sci. U. S. A.}\
  }\textbf {\bibinfo {volume} {110}},\ \bibinfo {pages} {E888} (\bibinfo {year}
  {2013})}\BibitemShut {NoStop}%
\bibitem [{\citenamefont {Gattinoni}\ \emph {et~al.}(2011)\citenamefont
  {Gattinoni}, \citenamefont {Lugli}, \citenamefont {Ji}, \citenamefont {Pos},
  \citenamefont {Paulos}, \citenamefont {Quigley}, \citenamefont {Almeida},
  \citenamefont {Gostick}, \citenamefont {Yu}, \citenamefont {Carpenito},
  \citenamefont {Wang}, \citenamefont {Douek}, \citenamefont {Price},
  \citenamefont {June}, \citenamefont {Marincola}, \citenamefont {Roederer},\
  and\ \citenamefont {Restifo}}]{Gattinoni2011-ft}%
  \BibitemOpen
  \bibfield  {author} {\bibinfo {author} {\bibfnamefont {L.}~\bibnamefont
  {Gattinoni}}, \bibinfo {author} {\bibfnamefont {E.}~\bibnamefont {Lugli}},
  \bibinfo {author} {\bibfnamefont {Y.}~\bibnamefont {Ji}}, \bibinfo {author}
  {\bibfnamefont {Z.}~\bibnamefont {Pos}}, \bibinfo {author} {\bibfnamefont
  {C.~M.}\ \bibnamefont {Paulos}}, \bibinfo {author} {\bibfnamefont {M.~F.}\
  \bibnamefont {Quigley}}, \bibinfo {author} {\bibfnamefont {J.~R.}\
  \bibnamefont {Almeida}}, \bibinfo {author} {\bibfnamefont {E.}~\bibnamefont
  {Gostick}}, \bibinfo {author} {\bibfnamefont {Z.}~\bibnamefont {Yu}},
  \bibinfo {author} {\bibfnamefont {C.}~\bibnamefont {Carpenito}}, \bibinfo
  {author} {\bibfnamefont {E.}~\bibnamefont {Wang}}, \bibinfo {author}
  {\bibfnamefont {D.~C.}\ \bibnamefont {Douek}}, \bibinfo {author}
  {\bibfnamefont {D.~A.}\ \bibnamefont {Price}}, \bibinfo {author}
  {\bibfnamefont {C.~H.}\ \bibnamefont {June}}, \bibinfo {author}
  {\bibfnamefont {F.~M.}\ \bibnamefont {Marincola}}, \bibinfo {author}
  {\bibfnamefont {M.}~\bibnamefont {Roederer}},\ and\ \bibinfo {author}
  {\bibfnamefont {N.~P.}\ \bibnamefont {Restifo}},\ }\href@noop {} {\bibfield
  {journal} {\bibinfo  {journal} {Nat. Med.}\ }\textbf {\bibinfo {volume}
  {17}},\ \bibinfo {pages} {1290} (\bibinfo {year} {2011})}\BibitemShut
  {NoStop}%
\bibitem [{\citenamefont {Cui}\ and\ \citenamefont {Kaech}(2010)}]{Cui2010-cz}%
  \BibitemOpen
  \bibfield  {author} {\bibinfo {author} {\bibfnamefont {W.}~\bibnamefont
  {Cui}}\ and\ \bibinfo {author} {\bibfnamefont {S.~M.}\ \bibnamefont
  {Kaech}},\ }\href {https://doi.org/10.1111/j.1600-065X.2010.00926.x}
  {\bibfield  {journal} {\bibinfo  {journal} {Immunol. Rev.}\ }\textbf
  {\bibinfo {volume} {236}},\ \bibinfo {pages} {151} (\bibinfo {year}
  {2010})}\BibitemShut {NoStop}%
\bibitem [{\citenamefont {de~Ronde}\ \emph {et~al.}(2012)\citenamefont
  {de~Ronde}, \citenamefont {Rein~ten Wolde},\ and\ \citenamefont
  {Mugler}}]{De_Ronde2012-py}%
  \BibitemOpen
  \bibfield  {author} {\bibinfo {author} {\bibfnamefont {W.}~\bibnamefont
  {de~Ronde}}, \bibinfo {author} {\bibfnamefont {P.}~\bibnamefont {Rein~ten
  Wolde}},\ and\ \bibinfo {author} {\bibfnamefont {A.}~\bibnamefont {Mugler}},\
  }\href@noop {} {\bibfield  {journal} {\bibinfo  {journal} {Biophys. J.}\
  }\textbf {\bibinfo {volume} {103}},\ \bibinfo {pages} {1097} (\bibinfo {year}
  {2012})}\BibitemShut {NoStop}%
\bibitem [{\citenamefont {Walczak}\ \emph {et~al.}(2010)\citenamefont
  {Walczak}, \citenamefont {Tkacik},\ and\ \citenamefont
  {Bialek}}]{Walczak2010-gs}%
  \BibitemOpen
  \bibfield  {author} {\bibinfo {author} {\bibfnamefont {A.~M.}\ \bibnamefont
  {Walczak}}, \bibinfo {author} {\bibfnamefont {G.}~\bibnamefont {Tkacik}},\
  and\ \bibinfo {author} {\bibfnamefont {W.}~\bibnamefont {Bialek}},\
  }\href@noop {} {\bibfield  {journal} {\bibinfo  {journal} {Phys. Rev. E Stat.
  Nonlin. Soft Matter Phys.}\ }\textbf {\bibinfo {volume} {81}},\ \bibinfo
  {pages} {041905} (\bibinfo {year} {2010})}\BibitemShut {NoStop}%
\bibitem [{\citenamefont {Oyler-Yaniv}\ \emph {et~al.}(2017)\citenamefont
  {Oyler-Yaniv}, \citenamefont {Oyler-Yaniv}, \citenamefont {Whitlock},
  \citenamefont {Liu}, \citenamefont {Germain}, \citenamefont {Huse},
  \citenamefont {Altan-Bonnet},\ and\ \citenamefont
  {Krichevsky}}]{Oyler-Yaniv2017-yr}%
  \BibitemOpen
  \bibfield  {author} {\bibinfo {author} {\bibfnamefont {A.}~\bibnamefont
  {Oyler-Yaniv}}, \bibinfo {author} {\bibfnamefont {J.}~\bibnamefont
  {Oyler-Yaniv}}, \bibinfo {author} {\bibfnamefont {B.~M.}\ \bibnamefont
  {Whitlock}}, \bibinfo {author} {\bibfnamefont {Z.}~\bibnamefont {Liu}},
  \bibinfo {author} {\bibfnamefont {R.~N.}\ \bibnamefont {Germain}}, \bibinfo
  {author} {\bibfnamefont {M.}~\bibnamefont {Huse}}, \bibinfo {author}
  {\bibfnamefont {G.}~\bibnamefont {Altan-Bonnet}},\ and\ \bibinfo {author}
  {\bibfnamefont {O.}~\bibnamefont {Krichevsky}},\ }\href@noop {} {\bibfield
  {journal} {\bibinfo  {journal} {Immunity}\ }\textbf {\bibinfo {volume}
  {46}},\ \bibinfo {pages} {609} (\bibinfo {year} {2017})}\BibitemShut
  {NoStop}%
\bibitem [{\citenamefont {Achar}\ \emph {et~al.}(2022)\citenamefont {Achar},
  \citenamefont {Bourassa}, \citenamefont {Rademaker}, \citenamefont {Lee},
  \citenamefont {Kondo}, \citenamefont {Salazar-Cavazos}, \citenamefont
  {Davies}, \citenamefont {Taylor}, \citenamefont {François},\ and\
  \citenamefont {Altan-Bonnet}}]{Achar2022-fl}%
  \BibitemOpen
  \bibfield  {author} {\bibinfo {author} {\bibfnamefont {S.~R.}\ \bibnamefont
  {Achar}}, \bibinfo {author} {\bibfnamefont {F.~X.~P.}\ \bibnamefont
  {Bourassa}}, \bibinfo {author} {\bibfnamefont {T.~J.}\ \bibnamefont
  {Rademaker}}, \bibinfo {author} {\bibfnamefont {A.}~\bibnamefont {Lee}},
  \bibinfo {author} {\bibfnamefont {T.}~\bibnamefont {Kondo}}, \bibinfo
  {author} {\bibfnamefont {E.}~\bibnamefont {Salazar-Cavazos}}, \bibinfo
  {author} {\bibfnamefont {J.~S.}\ \bibnamefont {Davies}}, \bibinfo {author}
  {\bibfnamefont {N.}~\bibnamefont {Taylor}}, \bibinfo {author} {\bibfnamefont
  {P.}~\bibnamefont {François}},\ and\ \bibinfo {author} {\bibfnamefont
  {G.}~\bibnamefont {Altan-Bonnet}},\ }\href@noop {} {\bibfield  {journal}
  {\bibinfo  {journal} {Science}\ }\textbf {\bibinfo {volume} {376}},\ \bibinfo
  {pages} {880} (\bibinfo {year} {2022})}\BibitemShut {NoStop}%
\bibitem [{\citenamefont {Voisinne}\ \emph {et~al.}(2015)\citenamefont
  {Voisinne}, \citenamefont {Nixon}, \citenamefont {Melbinger}, \citenamefont
  {Gasteiger}, \citenamefont {Vergassola},\ and\ \citenamefont
  {Altan-Bonnet}}]{Voisinne2015-ed}%
  \BibitemOpen
  \bibfield  {author} {\bibinfo {author} {\bibfnamefont {G.}~\bibnamefont
  {Voisinne}}, \bibinfo {author} {\bibfnamefont {B.~G.}\ \bibnamefont {Nixon}},
  \bibinfo {author} {\bibfnamefont {A.}~\bibnamefont {Melbinger}}, \bibinfo
  {author} {\bibfnamefont {G.}~\bibnamefont {Gasteiger}}, \bibinfo {author}
  {\bibfnamefont {M.}~\bibnamefont {Vergassola}},\ and\ \bibinfo {author}
  {\bibfnamefont {G.}~\bibnamefont {Altan-Bonnet}},\ }\href@noop {} {\bibfield
  {journal} {\bibinfo  {journal} {Cell Rep.}\ }\textbf {\bibinfo {volume}
  {11}},\ \bibinfo {pages} {1208} (\bibinfo {year} {2015})}\BibitemShut
  {NoStop}%
\bibitem [{\citenamefont {Adler}\ \emph {et~al.}(2014)\citenamefont {Adler},
  \citenamefont {Mayo},\ and\ \citenamefont {Alon}}]{Adler2014-lz}%
  \BibitemOpen
  \bibfield  {author} {\bibinfo {author} {\bibfnamefont {M.}~\bibnamefont
  {Adler}}, \bibinfo {author} {\bibfnamefont {A.}~\bibnamefont {Mayo}},\ and\
  \bibinfo {author} {\bibfnamefont {U.}~\bibnamefont {Alon}},\ }\href@noop {}
  {\bibfield  {journal} {\bibinfo  {journal} {PLoS Comput. Biol.}\ }\textbf
  {\bibinfo {volume} {10}},\ \bibinfo {pages} {e1003781} (\bibinfo {year}
  {2014})}\BibitemShut {NoStop}%
\bibitem [{\citenamefont {Heinzel}\ \emph {et~al.}(2018)\citenamefont
  {Heinzel}, \citenamefont {Marchingo}, \citenamefont {Horton},\ and\
  \citenamefont {Hodgkin}}]{Heinzel2018-wr}%
  \BibitemOpen
  \bibfield  {author} {\bibinfo {author} {\bibfnamefont {S.}~\bibnamefont
  {Heinzel}}, \bibinfo {author} {\bibfnamefont {J.~M.}\ \bibnamefont
  {Marchingo}}, \bibinfo {author} {\bibfnamefont {M.~B.}\ \bibnamefont
  {Horton}},\ and\ \bibinfo {author} {\bibfnamefont {P.~D.}\ \bibnamefont
  {Hodgkin}},\ }\href@noop {} {\bibfield  {journal} {\bibinfo  {journal} {Curr.
  Opin. Immunol.}\ }\textbf {\bibinfo {volume} {51}},\ \bibinfo {pages} {32}
  (\bibinfo {year} {2018})}\BibitemShut {NoStop}%
\bibitem [{\citenamefont {Gupta}\ \emph {et~al.}(2021)\citenamefont {Gupta},
  \citenamefont {Chevée}, \citenamefont {Kirosingh}, \citenamefont {Davis},\
  and\ \citenamefont {Schneider}}]{Gupta2021-vo}%
  \BibitemOpen
  \bibfield  {author} {\bibinfo {author} {\bibfnamefont {A.~S.}\ \bibnamefont
  {Gupta}}, \bibinfo {author} {\bibfnamefont {V.}~\bibnamefont {Chevée}},
  \bibinfo {author} {\bibfnamefont {A.~S.}\ \bibnamefont {Kirosingh}}, \bibinfo
  {author} {\bibfnamefont {N.~M.}\ \bibnamefont {Davis}},\ and\ \bibinfo
  {author} {\bibfnamefont {D.~S.}\ \bibnamefont {Schneider}},\ }\href@noop {}
  {\bibfield  {journal} {\bibinfo  {journal} {bioRxiv}\ ,\ \bibinfo {pages}
  {2021.09.29.462483}} (\bibinfo {year} {2021})}\BibitemShut {NoStop}%
\bibitem [{\citenamefont {Lebel}\ \emph {et~al.}(2025)\citenamefont {Lebel},
  \citenamefont {Gupta}, \citenamefont {Chevée}, \citenamefont {Alon},\ and\
  \citenamefont {Schneider}}]{Lebel2025-ah}%
  \BibitemOpen
  \bibfield  {author} {\bibinfo {author} {\bibfnamefont {Y.}~\bibnamefont
  {Lebel}}, \bibinfo {author} {\bibfnamefont {A.~S.}\ \bibnamefont {Gupta}},
  \bibinfo {author} {\bibfnamefont {V.}~\bibnamefont {Chevée}}, \bibinfo
  {author} {\bibfnamefont {U.}~\bibnamefont {Alon}},\ and\ \bibinfo {author}
  {\bibfnamefont {D.~S.}\ \bibnamefont {Schneider}},\ }\href@noop {} {\bibfield
   {journal} {\bibinfo  {journal} {bioRxiv}\ ,\ \bibinfo {pages}
  {2025.03.04.641508}} (\bibinfo {year} {2025})}\BibitemShut {NoStop}%
\bibitem [{\citenamefont {Johnson}\ \emph {et~al.}(2011)\citenamefont
  {Johnson}, \citenamefont {Kochin}, \citenamefont {McAfee}, \citenamefont
  {Stromnes}, \citenamefont {Regoes}, \citenamefont {Ahmed}, \citenamefont
  {Blattman},\ and\ \citenamefont {Antia}}]{Johnson2011-ja}%
  \BibitemOpen
  \bibfield  {author} {\bibinfo {author} {\bibfnamefont {P.~L.~F.}\
  \bibnamefont {Johnson}}, \bibinfo {author} {\bibfnamefont {B.~F.}\
  \bibnamefont {Kochin}}, \bibinfo {author} {\bibfnamefont {M.~S.}\
  \bibnamefont {McAfee}}, \bibinfo {author} {\bibfnamefont {I.~M.}\
  \bibnamefont {Stromnes}}, \bibinfo {author} {\bibfnamefont {R.~R.}\
  \bibnamefont {Regoes}}, \bibinfo {author} {\bibfnamefont {R.}~\bibnamefont
  {Ahmed}}, \bibinfo {author} {\bibfnamefont {J.~N.}\ \bibnamefont
  {Blattman}},\ and\ \bibinfo {author} {\bibfnamefont {R.}~\bibnamefont
  {Antia}},\ }\href@noop {} {\bibfield  {journal} {\bibinfo  {journal} {J.
  Virol.}\ }\textbf {\bibinfo {volume} {85}},\ \bibinfo {pages} {5565}
  (\bibinfo {year} {2011})}\BibitemShut {NoStop}%
\bibitem [{\citenamefont {Iwasaki}\ and\ \citenamefont
  {Medzhitov}(2015)}]{Iwasaki2015-qi}%
  \BibitemOpen
  \bibfield  {author} {\bibinfo {author} {\bibfnamefont {A.}~\bibnamefont
  {Iwasaki}}\ and\ \bibinfo {author} {\bibfnamefont {R.}~\bibnamefont
  {Medzhitov}},\ }\href {https://doi.org/10.1038/ni.3123} {\bibfield  {journal}
  {\bibinfo  {journal} {Nat. Immunol.}\ }\textbf {\bibinfo {volume} {16}},\
  \bibinfo {pages} {343} (\bibinfo {year} {2015})}\BibitemShut {NoStop}%
\bibitem [{\citenamefont {Shoval}\ \emph {et~al.}(2012)\citenamefont {Shoval},
  \citenamefont {Sheftel}, \citenamefont {Shinar}, \citenamefont {Hart},
  \citenamefont {Ramote}, \citenamefont {Mayo}, \citenamefont {Dekel},
  \citenamefont {Kavanagh},\ and\ \citenamefont {Alon}}]{Shoval2012-ln}%
  \BibitemOpen
  \bibfield  {author} {\bibinfo {author} {\bibfnamefont {O.}~\bibnamefont
  {Shoval}}, \bibinfo {author} {\bibfnamefont {H.}~\bibnamefont {Sheftel}},
  \bibinfo {author} {\bibfnamefont {G.}~\bibnamefont {Shinar}}, \bibinfo
  {author} {\bibfnamefont {Y.}~\bibnamefont {Hart}}, \bibinfo {author}
  {\bibfnamefont {O.}~\bibnamefont {Ramote}}, \bibinfo {author} {\bibfnamefont
  {A.}~\bibnamefont {Mayo}}, \bibinfo {author} {\bibfnamefont {E.}~\bibnamefont
  {Dekel}}, \bibinfo {author} {\bibfnamefont {K.}~\bibnamefont {Kavanagh}},\
  and\ \bibinfo {author} {\bibfnamefont {U.}~\bibnamefont {Alon}},\ }\href@noop
  {} {\bibfield  {journal} {\bibinfo  {journal} {Science}\ }\textbf {\bibinfo
  {volume} {336}},\ \bibinfo {pages} {1157} (\bibinfo {year}
  {2012})}\BibitemShut {NoStop}%
\bibitem [{\citenamefont {Ko\c{c}illari}\ \emph {et~al.}(2018)\citenamefont
  {Ko\c{c}illari}, \citenamefont {Fariselli}, \citenamefont {Trovato},
  \citenamefont {Seno},\ and\ \citenamefont {Maritan}}]{Kocillari2018-pf}%
  \BibitemOpen
  \bibfield  {author} {\bibinfo {author} {\bibfnamefont {L.}~\bibnamefont
  {Ko\c{c}illari}}, \bibinfo {author} {\bibfnamefont {P.}~\bibnamefont
  {Fariselli}}, \bibinfo {author} {\bibfnamefont {A.}~\bibnamefont {Trovato}},
  \bibinfo {author} {\bibfnamefont {F.}~\bibnamefont {Seno}},\ and\ \bibinfo
  {author} {\bibfnamefont {A.}~\bibnamefont {Maritan}},\ }\href@noop {}
  {\bibfield  {journal} {\bibinfo  {journal} {Sci. Rep.}\ }\textbf {\bibinfo
  {volume} {8}},\ \bibinfo {pages} {9141} (\bibinfo {year} {2018})}\BibitemShut
  {NoStop}%
\bibitem [{\citenamefont {Zehn}\ \emph {et~al.}(2009)\citenamefont {Zehn},
  \citenamefont {Lee},\ and\ \citenamefont {Bevan}}]{Zehn2009-up}%
  \BibitemOpen
  \bibfield  {author} {\bibinfo {author} {\bibfnamefont {D.}~\bibnamefont
  {Zehn}}, \bibinfo {author} {\bibfnamefont {S.~Y.}\ \bibnamefont {Lee}},\ and\
  \bibinfo {author} {\bibfnamefont {M.~J.}\ \bibnamefont {Bevan}},\ }\href@noop
  {} {\bibfield  {journal} {\bibinfo  {journal} {Nature}\ }\textbf {\bibinfo
  {volume} {458}},\ \bibinfo {pages} {211} (\bibinfo {year}
  {2009})}\BibitemShut {NoStop}%
\bibitem [{\citenamefont {De~Boer}\ \emph {et~al.}(2001)\citenamefont
  {De~Boer}, \citenamefont {Oprea}, \citenamefont {Antia}, \citenamefont
  {Murali-Krishna}, \citenamefont {Ahmed},\ and\ \citenamefont
  {Perelson}}]{De_Boer2001-wl}%
  \BibitemOpen
  \bibfield  {author} {\bibinfo {author} {\bibfnamefont {R.~J.}\ \bibnamefont
  {De~Boer}}, \bibinfo {author} {\bibfnamefont {M.}~\bibnamefont {Oprea}},
  \bibinfo {author} {\bibfnamefont {R.}~\bibnamefont {Antia}}, \bibinfo
  {author} {\bibfnamefont {K.}~\bibnamefont {Murali-Krishna}}, \bibinfo
  {author} {\bibfnamefont {R.}~\bibnamefont {Ahmed}},\ and\ \bibinfo {author}
  {\bibfnamefont {A.~S.}\ \bibnamefont {Perelson}},\ }\href@noop {} {\bibfield
  {journal} {\bibinfo  {journal} {J. Virol.}\ }\textbf {\bibinfo {volume}
  {75}},\ \bibinfo {pages} {10663} (\bibinfo {year} {2001})}\BibitemShut
  {NoStop}%
\bibitem [{\citenamefont {Blattman}\ \emph {et~al.}(2003)\citenamefont
  {Blattman}, \citenamefont {Grayson}, \citenamefont {Wherry}, \citenamefont
  {Kaech}, \citenamefont {Smith},\ and\ \citenamefont
  {Ahmed}}]{Blattman2003-sc}%
  \BibitemOpen
  \bibfield  {author} {\bibinfo {author} {\bibfnamefont {J.~N.}\ \bibnamefont
  {Blattman}}, \bibinfo {author} {\bibfnamefont {J.~M.}\ \bibnamefont
  {Grayson}}, \bibinfo {author} {\bibfnamefont {E.~J.}\ \bibnamefont {Wherry}},
  \bibinfo {author} {\bibfnamefont {S.~M.}\ \bibnamefont {Kaech}}, \bibinfo
  {author} {\bibfnamefont {K.~A.}\ \bibnamefont {Smith}},\ and\ \bibinfo
  {author} {\bibfnamefont {R.}~\bibnamefont {Ahmed}},\ }\href@noop {}
  {\bibfield  {journal} {\bibinfo  {journal} {Nat. Med.}\ }\textbf {\bibinfo
  {volume} {9}},\ \bibinfo {pages} {540} (\bibinfo {year} {2003})}\BibitemShut
  {NoStop}%
\bibitem [{\citenamefont {Mayer}\ \emph
  {et~al.}(2019{\natexlab{b}})\citenamefont {Mayer}, \citenamefont {Zhang},
  \citenamefont {Perelson},\ and\ \citenamefont {Wingreen}}]{Mayer2019-pj}%
  \BibitemOpen
  \bibfield  {author} {\bibinfo {author} {\bibfnamefont {A.}~\bibnamefont
  {Mayer}}, \bibinfo {author} {\bibfnamefont {Y.}~\bibnamefont {Zhang}},
  \bibinfo {author} {\bibfnamefont {A.~S.}\ \bibnamefont {Perelson}},\ and\
  \bibinfo {author} {\bibfnamefont {N.~S.}\ \bibnamefont {Wingreen}},\
  }\href@noop {} {\bibfield  {journal} {\bibinfo  {journal} {Proc. Natl. Acad.
  Sci. U. S. A.}\ }\textbf {\bibinfo {volume} {116}},\ \bibinfo {pages} {5914}
  (\bibinfo {year} {2019}{\natexlab{b}})}\BibitemShut {NoStop}%
\bibitem [{\citenamefont {Baulu}\ \emph {et~al.}(2023)\citenamefont {Baulu},
  \citenamefont {Gardet}, \citenamefont {Chuvin},\ and\ \citenamefont
  {Depil}}]{Baulu2023-zd}%
  \BibitemOpen
  \bibfield  {author} {\bibinfo {author} {\bibfnamefont {E.}~\bibnamefont
  {Baulu}}, \bibinfo {author} {\bibfnamefont {C.}~\bibnamefont {Gardet}},
  \bibinfo {author} {\bibfnamefont {N.}~\bibnamefont {Chuvin}},\ and\ \bibinfo
  {author} {\bibfnamefont {S.}~\bibnamefont {Depil}},\ }\href@noop {}
  {\bibfield  {journal} {\bibinfo  {journal} {Sci. Adv.}\ }\textbf {\bibinfo
  {volume} {9}},\ \bibinfo {pages} {eadf3700} (\bibinfo {year}
  {2023})}\BibitemShut {NoStop}%
\bibitem [{\citenamefont {Zugasti}\ \emph {et~al.}(2025)\citenamefont
  {Zugasti}, \citenamefont {Espinosa-Aroca}, \citenamefont {Fidyt},
  \citenamefont {Mulens-Arias}, \citenamefont {Diaz-Beya}, \citenamefont
  {Juan}, \citenamefont {Urbano-Ispizua}, \citenamefont {Esteve}, \citenamefont
  {Velasco-Hernandez},\ and\ \citenamefont {Men\~{A}ndez}}]{Zugasti2025-dj}%
  \BibitemOpen
  \bibfield  {author} {\bibinfo {author} {\bibfnamefont {I.}~\bibnamefont
  {Zugasti}}, \bibinfo {author} {\bibfnamefont {L.}~\bibnamefont
  {Espinosa-Aroca}}, \bibinfo {author} {\bibfnamefont {K.}~\bibnamefont
  {Fidyt}}, \bibinfo {author} {\bibfnamefont {V.}~\bibnamefont {Mulens-Arias}},
  \bibinfo {author} {\bibfnamefont {M.}~\bibnamefont {Diaz-Beya}}, \bibinfo
  {author} {\bibfnamefont {M.}~\bibnamefont {Juan}}, \bibinfo {author}
  {\bibfnamefont {.~A.}\ \bibnamefont {Urbano-Ispizua}}, \bibinfo {author}
  {\bibfnamefont {J.}~\bibnamefont {Esteve}}, \bibinfo {author} {\bibfnamefont
  {T.}~\bibnamefont {Velasco-Hernandez}},\ and\ \bibinfo {author}
  {\bibfnamefont {P.}~\bibnamefont {Men\~{A}ndez}},\ }\href@noop {} {\bibfield
  {journal} {\bibinfo  {journal} {Signal Transduct. Target. Ther.}\ }\textbf
  {\bibinfo {volume} {10}},\ \bibinfo {pages} {210} (\bibinfo {year}
  {2025})}\BibitemShut {NoStop}%
\bibitem [{\citenamefont {Rotte}\ \emph {et~al.}(2022)\citenamefont {Rotte},
  \citenamefont {Frigault}, \citenamefont {Ansari}, \citenamefont {Gliner},
  \citenamefont {Heery},\ and\ \citenamefont {Shah}}]{Rotte2022-kq}%
  \BibitemOpen
  \bibfield  {author} {\bibinfo {author} {\bibfnamefont {A.}~\bibnamefont
  {Rotte}}, \bibinfo {author} {\bibfnamefont {M.~J.}\ \bibnamefont {Frigault}},
  \bibinfo {author} {\bibfnamefont {A.}~\bibnamefont {Ansari}}, \bibinfo
  {author} {\bibfnamefont {B.}~\bibnamefont {Gliner}}, \bibinfo {author}
  {\bibfnamefont {C.}~\bibnamefont {Heery}},\ and\ \bibinfo {author}
  {\bibfnamefont {B.}~\bibnamefont {Shah}},\ }\href@noop {} {\bibfield
  {journal} {\bibinfo  {journal} {J. Immunother. Cancer}\ }\textbf {\bibinfo
  {volume} {10}},\ \bibinfo {pages} {e005678} (\bibinfo {year}
  {2022})}\BibitemShut {NoStop}%
\bibitem [{\citenamefont {Hoffmann}\ and\ \citenamefont
  {Slansky}(2020)}]{Hoffmann2020-yg}%
  \BibitemOpen
  \bibfield  {author} {\bibinfo {author} {\bibfnamefont {M.~M.}\ \bibnamefont
  {Hoffmann}}\ and\ \bibinfo {author} {\bibfnamefont {J.~E.}\ \bibnamefont
  {Slansky}},\ }\href@noop {} {\bibfield  {journal} {\bibinfo  {journal} {Mol.
  Carcinog.}\ }\textbf {\bibinfo {volume} {59}},\ \bibinfo {pages} {862}
  (\bibinfo {year} {2020})}\BibitemShut {NoStop}%
\bibitem [{\citenamefont {Liu}\ \emph {et~al.}(2025)\citenamefont {Liu},
  \citenamefont {Greenwood}, \citenamefont {Bonzanini}, \citenamefont
  {Motmaen}, \citenamefont {Meyerberg}, \citenamefont {Dao}, \citenamefont
  {Xiang}, \citenamefont {Ault}, \citenamefont {Sharp}, \citenamefont {Wang},
  \citenamefont {Visani}, \citenamefont {Vafeados}, \citenamefont {Roullier},
  \citenamefont {Nourmohammad}, \citenamefont {Scheinberg}, \citenamefont
  {Garcia},\ and\ \citenamefont {Baker}}]{Liu2025-qk}%
  \BibitemOpen
  \bibfield  {author} {\bibinfo {author} {\bibfnamefont {B.}~\bibnamefont
  {Liu}}, \bibinfo {author} {\bibfnamefont {N.~F.}\ \bibnamefont {Greenwood}},
  \bibinfo {author} {\bibfnamefont {J.~E.}\ \bibnamefont {Bonzanini}}, \bibinfo
  {author} {\bibfnamefont {A.}~\bibnamefont {Motmaen}}, \bibinfo {author}
  {\bibfnamefont {J.}~\bibnamefont {Meyerberg}}, \bibinfo {author}
  {\bibfnamefont {T.}~\bibnamefont {Dao}}, \bibinfo {author} {\bibfnamefont
  {X.}~\bibnamefont {Xiang}}, \bibinfo {author} {\bibfnamefont
  {R.}~\bibnamefont {Ault}}, \bibinfo {author} {\bibfnamefont {J.}~\bibnamefont
  {Sharp}}, \bibinfo {author} {\bibfnamefont {C.}~\bibnamefont {Wang}},
  \bibinfo {author} {\bibfnamefont {G.~M.}\ \bibnamefont {Visani}}, \bibinfo
  {author} {\bibfnamefont {D.~K.}\ \bibnamefont {Vafeados}}, \bibinfo {author}
  {\bibfnamefont {N.}~\bibnamefont {Roullier}}, \bibinfo {author}
  {\bibfnamefont {A.}~\bibnamefont {Nourmohammad}}, \bibinfo {author}
  {\bibfnamefont {D.~A.}\ \bibnamefont {Scheinberg}}, \bibinfo {author}
  {\bibfnamefont {K.~C.}\ \bibnamefont {Garcia}},\ and\ \bibinfo {author}
  {\bibfnamefont {D.}~\bibnamefont {Baker}},\ }\href@noop {} {\bibfield
  {journal} {\bibinfo  {journal} {Science}\ }\textbf {\bibinfo {volume}
  {389}},\ \bibinfo {pages} {386} (\bibinfo {year} {2025})}\BibitemShut
  {NoStop}%
\bibitem [{\citenamefont {Kondo}\ \emph {et~al.}(2025)\citenamefont {Kondo},
  \citenamefont {Bourassa}, \citenamefont {Achar}, \citenamefont {DuSold},
  \citenamefont {Céspedes}, \citenamefont {Ando}, \citenamefont {Dwivedi},
  \citenamefont {Moraly}, \citenamefont {Chien}, \citenamefont {Majdoul},
  \citenamefont {Kenet}, \citenamefont {Wahlsten}, \citenamefont {Kvalvaag},
  \citenamefont {Jenkins}, \citenamefont {Kim}, \citenamefont {Ade},
  \citenamefont {Yu}, \citenamefont {Gaud}, \citenamefont {Davila},
  \citenamefont {Love}, \citenamefont {Yang}, \citenamefont {Dustin},
  \citenamefont {Altan-Bonnet}, \citenamefont {François},\ and\ \citenamefont
  {Taylor}}]{Kondo2025-nl}%
  \BibitemOpen
  \bibfield  {author} {\bibinfo {author} {\bibfnamefont {T.}~\bibnamefont
  {Kondo}}, \bibinfo {author} {\bibfnamefont {F.~X.~P.}\ \bibnamefont
  {Bourassa}}, \bibinfo {author} {\bibfnamefont {S.}~\bibnamefont {Achar}},
  \bibinfo {author} {\bibfnamefont {J.}~\bibnamefont {DuSold}}, \bibinfo
  {author} {\bibfnamefont {P.~F.}\ \bibnamefont {Céspedes}}, \bibinfo {author}
  {\bibfnamefont {M.}~\bibnamefont {Ando}}, \bibinfo {author} {\bibfnamefont
  {A.}~\bibnamefont {Dwivedi}}, \bibinfo {author} {\bibfnamefont
  {J.}~\bibnamefont {Moraly}}, \bibinfo {author} {\bibfnamefont
  {C.}~\bibnamefont {Chien}}, \bibinfo {author} {\bibfnamefont
  {S.}~\bibnamefont {Majdoul}}, \bibinfo {author} {\bibfnamefont {A.~L.}\
  \bibnamefont {Kenet}}, \bibinfo {author} {\bibfnamefont {M.}~\bibnamefont
  {Wahlsten}}, \bibinfo {author} {\bibfnamefont {A.}~\bibnamefont {Kvalvaag}},
  \bibinfo {author} {\bibfnamefont {E.}~\bibnamefont {Jenkins}}, \bibinfo
  {author} {\bibfnamefont {S.~P.}\ \bibnamefont {Kim}}, \bibinfo {author}
  {\bibfnamefont {C.~M.}\ \bibnamefont {Ade}}, \bibinfo {author} {\bibfnamefont
  {Z.}~\bibnamefont {Yu}}, \bibinfo {author} {\bibfnamefont {G.}~\bibnamefont
  {Gaud}}, \bibinfo {author} {\bibfnamefont {M.}~\bibnamefont {Davila}},
  \bibinfo {author} {\bibfnamefont {P.}~\bibnamefont {Love}}, \bibinfo {author}
  {\bibfnamefont {J.~C.}\ \bibnamefont {Yang}}, \bibinfo {author}
  {\bibfnamefont {M.~L.}\ \bibnamefont {Dustin}}, \bibinfo {author}
  {\bibfnamefont {G.}~\bibnamefont {Altan-Bonnet}}, \bibinfo {author}
  {\bibfnamefont {P.}~\bibnamefont {François}},\ and\ \bibinfo {author}
  {\bibfnamefont {N.}~\bibnamefont {Taylor}},\ }\href@noop {} {\bibfield
  {journal} {\bibinfo  {journal} {Cell}\ }\textbf {\bibinfo {volume} {0}}
  (\bibinfo {year} {2025})}\BibitemShut {NoStop}%
\bibitem [{\citenamefont {Visani}\ \emph {et~al.}(2025)\citenamefont {Visani},
  \citenamefont {Pun}, \citenamefont {Minervina}, \citenamefont {Bradley},
  \citenamefont {Thomas},\ and\ \citenamefont {Nourmohammad}}]{Visani2025-wx}%
  \BibitemOpen
  \bibfield  {author} {\bibinfo {author} {\bibfnamefont {G.~M.}\ \bibnamefont
  {Visani}}, \bibinfo {author} {\bibfnamefont {M.~N.}\ \bibnamefont {Pun}},
  \bibinfo {author} {\bibfnamefont {A.~A.}\ \bibnamefont {Minervina}}, \bibinfo
  {author} {\bibfnamefont {P.}~\bibnamefont {Bradley}}, \bibinfo {author}
  {\bibfnamefont {P.}~\bibnamefont {Thomas}},\ and\ \bibinfo {author}
  {\bibfnamefont {A.}~\bibnamefont {Nourmohammad}},\ }\href@noop {} {\bibfield
  {journal} {\bibinfo  {journal} {bioRxivorg}\ ,\ \bibinfo {pages}
  {2025.02.28.640903}} (\bibinfo {year} {2025})}\BibitemShut {NoStop}%
\bibitem [{\citenamefont {Escobar}\ \emph {et~al.}(2023)\citenamefont
  {Escobar}, \citenamefont {Tooley}, \citenamefont {Oliveras}, \citenamefont
  {Huang}, \citenamefont {Cheng}, \citenamefont {Bookstaver}, \citenamefont
  {Edwards}, \citenamefont {Froimchuk}, \citenamefont {Xue}, \citenamefont
  {Mangani}, \citenamefont {Krishnan}, \citenamefont {Hazel}, \citenamefont
  {Rutigliani}, \citenamefont {Jewell}, \citenamefont {Biasco},\ and\
  \citenamefont {Anderson}}]{Escobar2023-ot}%
  \BibitemOpen
  \bibfield  {author} {\bibinfo {author} {\bibfnamefont {G.}~\bibnamefont
  {Escobar}}, \bibinfo {author} {\bibfnamefont {K.}~\bibnamefont {Tooley}},
  \bibinfo {author} {\bibfnamefont {J.~P.}\ \bibnamefont {Oliveras}}, \bibinfo
  {author} {\bibfnamefont {L.}~\bibnamefont {Huang}}, \bibinfo {author}
  {\bibfnamefont {H.}~\bibnamefont {Cheng}}, \bibinfo {author} {\bibfnamefont
  {M.~L.}\ \bibnamefont {Bookstaver}}, \bibinfo {author} {\bibfnamefont
  {C.}~\bibnamefont {Edwards}}, \bibinfo {author} {\bibfnamefont
  {E.}~\bibnamefont {Froimchuk}}, \bibinfo {author} {\bibfnamefont
  {C.}~\bibnamefont {Xue}}, \bibinfo {author} {\bibfnamefont {D.}~\bibnamefont
  {Mangani}}, \bibinfo {author} {\bibfnamefont {R.~K.}\ \bibnamefont
  {Krishnan}}, \bibinfo {author} {\bibfnamefont {N.}~\bibnamefont {Hazel}},
  \bibinfo {author} {\bibfnamefont {C.}~\bibnamefont {Rutigliani}}, \bibinfo
  {author} {\bibfnamefont {C.~M.}\ \bibnamefont {Jewell}}, \bibinfo {author}
  {\bibfnamefont {L.}~\bibnamefont {Biasco}},\ and\ \bibinfo {author}
  {\bibfnamefont {A.~C.}\ \bibnamefont {Anderson}},\ }\href@noop {} {\bibfield
  {journal} {\bibinfo  {journal} {Cancer Cell}\ }\textbf {\bibinfo {volume}
  {41}},\ \bibinfo {pages} {1662} (\bibinfo {year} {2023})}\BibitemShut
  {NoStop}%
\bibitem [{\citenamefont {Muthalagu}\ \emph {et~al.}(2014)\citenamefont
  {Muthalagu}, \citenamefont {Junttila}, \citenamefont {Wiese}, \citenamefont
  {Wolf}, \citenamefont {Morton}, \citenamefont {Bauer}, \citenamefont {Evan},
  \citenamefont {Eilers},\ and\ \citenamefont {Murphy}}]{Muthalagu2014-ni}%
  \BibitemOpen
  \bibfield  {author} {\bibinfo {author} {\bibfnamefont {N.}~\bibnamefont
  {Muthalagu}}, \bibinfo {author} {\bibfnamefont {M.~R.}\ \bibnamefont
  {Junttila}}, \bibinfo {author} {\bibfnamefont {K.~E.}\ \bibnamefont {Wiese}},
  \bibinfo {author} {\bibfnamefont {E.}~\bibnamefont {Wolf}}, \bibinfo {author}
  {\bibfnamefont {J.}~\bibnamefont {Morton}}, \bibinfo {author} {\bibfnamefont
  {B.}~\bibnamefont {Bauer}}, \bibinfo {author} {\bibfnamefont {G.~I.}\
  \bibnamefont {Evan}}, \bibinfo {author} {\bibfnamefont {M.}~\bibnamefont
  {Eilers}},\ and\ \bibinfo {author} {\bibfnamefont {D.~J.}\ \bibnamefont
  {Murphy}},\ }\href@noop {} {\bibfield  {journal} {\bibinfo  {journal} {Cell
  Rep.}\ }\textbf {\bibinfo {volume} {8}},\ \bibinfo {pages} {1347} (\bibinfo
  {year} {2014})}\BibitemShut {NoStop}%
\bibitem [{\citenamefont {Tisoncik}\ \emph {et~al.}(2012)\citenamefont
  {Tisoncik}, \citenamefont {Korth}, \citenamefont {Simmons}, \citenamefont
  {Farrar}, \citenamefont {Martin},\ and\ \citenamefont
  {Katze}}]{Tisoncik2012-kx}%
  \BibitemOpen
  \bibfield  {author} {\bibinfo {author} {\bibfnamefont {J.~R.}\ \bibnamefont
  {Tisoncik}}, \bibinfo {author} {\bibfnamefont {M.~J.}\ \bibnamefont {Korth}},
  \bibinfo {author} {\bibfnamefont {C.~P.}\ \bibnamefont {Simmons}}, \bibinfo
  {author} {\bibfnamefont {J.}~\bibnamefont {Farrar}}, \bibinfo {author}
  {\bibfnamefont {T.~R.}\ \bibnamefont {Martin}},\ and\ \bibinfo {author}
  {\bibfnamefont {M.~G.}\ \bibnamefont {Katze}},\ }\href@noop {} {\bibfield
  {journal} {\bibinfo  {journal} {Microbiol. Mol. Biol. Rev.}\ }\textbf
  {\bibinfo {volume} {76}},\ \bibinfo {pages} {16} (\bibinfo {year}
  {2012})}\BibitemShut {NoStop}%
\bibitem [{\citenamefont {Yoshikawa}\ \emph {et~al.}(2022)\citenamefont
  {Yoshikawa}, \citenamefont {Wu}, \citenamefont {Inoue}, \citenamefont
  {Kasuya}, \citenamefont {Matsushita}, \citenamefont {Takahashi},
  \citenamefont {Kuroda}, \citenamefont {Hosoda}, \citenamefont {Suzuki},\ and\
  \citenamefont {Kagoya}}]{Yoshikawa2022-hd}%
  \BibitemOpen
  \bibfield  {author} {\bibinfo {author} {\bibfnamefont {T.}~\bibnamefont
  {Yoshikawa}}, \bibinfo {author} {\bibfnamefont {Z.}~\bibnamefont {Wu}},
  \bibinfo {author} {\bibfnamefont {S.}~\bibnamefont {Inoue}}, \bibinfo
  {author} {\bibfnamefont {H.}~\bibnamefont {Kasuya}}, \bibinfo {author}
  {\bibfnamefont {H.}~\bibnamefont {Matsushita}}, \bibinfo {author}
  {\bibfnamefont {Y.}~\bibnamefont {Takahashi}}, \bibinfo {author}
  {\bibfnamefont {H.}~\bibnamefont {Kuroda}}, \bibinfo {author} {\bibfnamefont
  {W.}~\bibnamefont {Hosoda}}, \bibinfo {author} {\bibfnamefont
  {S.}~\bibnamefont {Suzuki}},\ and\ \bibinfo {author} {\bibfnamefont
  {Y.}~\bibnamefont {Kagoya}},\ }\href@noop {} {\bibfield  {journal} {\bibinfo
  {journal} {Blood}\ }\textbf {\bibinfo {volume} {139}},\ \bibinfo {pages}
  {2156} (\bibinfo {year} {2022})}\BibitemShut {NoStop}%
\bibitem [{\citenamefont {Yi}\ \emph {et~al.}(2025)\citenamefont {Yi},
  \citenamefont {Cohen}, \citenamefont {Zimmerman}, \citenamefont {Dündar},
  \citenamefont {Zumbo}, \citenamefont {Eltilib}, \citenamefont {Brophy},
  \citenamefont {Arkin}, \citenamefont {Feucht}, \citenamefont {Gormally},
  \citenamefont {Hackett}, \citenamefont {Kropp}, \citenamefont {Etxeberria},
  \citenamefont {Chandran}, \citenamefont {Zhao}, \citenamefont {Cai},
  \citenamefont {Daniyan}, \citenamefont {Park}, \citenamefont {Lareau},
  \citenamefont {Hsu}, \citenamefont {Sadelain}, \citenamefont {Betel},\ and\
  \citenamefont {Klebanoff}}]{Yi2025-bb}%
  \BibitemOpen
  \bibfield  {author} {\bibinfo {author} {\bibfnamefont {F.}~\bibnamefont
  {Yi}}, \bibinfo {author} {\bibfnamefont {T.}~\bibnamefont {Cohen}}, \bibinfo
  {author} {\bibfnamefont {N.}~\bibnamefont {Zimmerman}}, \bibinfo {author}
  {\bibfnamefont {F.}~\bibnamefont {Dündar}}, \bibinfo {author} {\bibfnamefont
  {P.}~\bibnamefont {Zumbo}}, \bibinfo {author} {\bibfnamefont
  {R.}~\bibnamefont {Eltilib}}, \bibinfo {author} {\bibfnamefont {E.~J.}\
  \bibnamefont {Brophy}}, \bibinfo {author} {\bibfnamefont {H.}~\bibnamefont
  {Arkin}}, \bibinfo {author} {\bibfnamefont {J.}~\bibnamefont {Feucht}},
  \bibinfo {author} {\bibfnamefont {M.~V.}\ \bibnamefont {Gormally}}, \bibinfo
  {author} {\bibfnamefont {C.~S.}\ \bibnamefont {Hackett}}, \bibinfo {author}
  {\bibfnamefont {K.~N.}\ \bibnamefont {Kropp}}, \bibinfo {author}
  {\bibfnamefont {I.}~\bibnamefont {Etxeberria}}, \bibinfo {author}
  {\bibfnamefont {S.~S.}\ \bibnamefont {Chandran}}, \bibinfo {author}
  {\bibfnamefont {Z.}~\bibnamefont {Zhao}}, \bibinfo {author} {\bibfnamefont
  {W.}~\bibnamefont {Cai}}, \bibinfo {author} {\bibfnamefont {A.~F.}\
  \bibnamefont {Daniyan}}, \bibinfo {author} {\bibfnamefont {J.~H.}\
  \bibnamefont {Park}}, \bibinfo {author} {\bibfnamefont {C.~A.}\ \bibnamefont
  {Lareau}}, \bibinfo {author} {\bibfnamefont {K.~C.}\ \bibnamefont {Hsu}},
  \bibinfo {author} {\bibfnamefont {M.}~\bibnamefont {Sadelain}}, \bibinfo
  {author} {\bibfnamefont {D.}~\bibnamefont {Betel}},\ and\ \bibinfo {author}
  {\bibfnamefont {C.~A.}\ \bibnamefont {Klebanoff}},\ }\href@noop {} {\bibfield
   {journal} {\bibinfo  {journal} {Nat. Cancer}\ ,\ \bibinfo {pages} {1}}
  (\bibinfo {year} {2025})}\BibitemShut {NoStop}%
\bibitem [{\citenamefont {Knudsen}\ \emph {et~al.}(2025)\citenamefont
  {Knudsen}, \citenamefont {Escobar}, \citenamefont {Korell}, \citenamefont
  {Kienka}, \citenamefont {Nobrega}, \citenamefont {Anderson}, \citenamefont
  {Cheng}, \citenamefont {Zschummel}, \citenamefont {Armstrong}, \citenamefont
  {Bouffard}, \citenamefont {Kann}, \citenamefont {Goncalves}, \citenamefont
  {Pope}, \citenamefont {Pezeshki}, \citenamefont {Rojas}, \citenamefont
  {Suermondt}, \citenamefont {Phillips}, \citenamefont {Berger}, \citenamefont
  {Park}, \citenamefont {Salas-Benito}, \citenamefont {Darnell}, \citenamefont
  {Birocchi}, \citenamefont {Leick}, \citenamefont {Larson}, \citenamefont
  {Doench}, \citenamefont {Sen}, \citenamefont {Yates}, \citenamefont
  {Manguso},\ and\ \citenamefont {Maus}}]{Knudsen2025-xb}%
  \BibitemOpen
  \bibfield  {author} {\bibinfo {author} {\bibfnamefont {N.~H.}\ \bibnamefont
  {Knudsen}}, \bibinfo {author} {\bibfnamefont {G.}~\bibnamefont {Escobar}},
  \bibinfo {author} {\bibfnamefont {F.}~\bibnamefont {Korell}}, \bibinfo
  {author} {\bibfnamefont {T.}~\bibnamefont {Kienka}}, \bibinfo {author}
  {\bibfnamefont {C.}~\bibnamefont {Nobrega}}, \bibinfo {author} {\bibfnamefont
  {S.}~\bibnamefont {Anderson}}, \bibinfo {author} {\bibfnamefont {A.~Y.}\
  \bibnamefont {Cheng}}, \bibinfo {author} {\bibfnamefont {M.}~\bibnamefont
  {Zschummel}}, \bibinfo {author} {\bibfnamefont {A.}~\bibnamefont
  {Armstrong}}, \bibinfo {author} {\bibfnamefont {A.}~\bibnamefont {Bouffard}},
  \bibinfo {author} {\bibfnamefont {M.~C.}\ \bibnamefont {Kann}}, \bibinfo
  {author} {\bibfnamefont {S.}~\bibnamefont {Goncalves}}, \bibinfo {author}
  {\bibfnamefont {H.~W.}\ \bibnamefont {Pope}}, \bibinfo {author}
  {\bibfnamefont {M.}~\bibnamefont {Pezeshki}}, \bibinfo {author}
  {\bibfnamefont {A.}~\bibnamefont {Rojas}}, \bibinfo {author} {\bibfnamefont
  {J.~S. M.~T.}\ \bibnamefont {Suermondt}}, \bibinfo {author} {\bibfnamefont
  {M.}~\bibnamefont {Phillips}}, \bibinfo {author} {\bibfnamefont {T.~R.}\
  \bibnamefont {Berger}}, \bibinfo {author} {\bibfnamefont {S.}~\bibnamefont
  {Park}}, \bibinfo {author} {\bibfnamefont {D.}~\bibnamefont {Salas-Benito}},
  \bibinfo {author} {\bibfnamefont {E.~P.}\ \bibnamefont {Darnell}}, \bibinfo
  {author} {\bibfnamefont {F.}~\bibnamefont {Birocchi}}, \bibinfo {author}
  {\bibfnamefont {M.~B.}\ \bibnamefont {Leick}}, \bibinfo {author}
  {\bibfnamefont {R.~C.}\ \bibnamefont {Larson}}, \bibinfo {author}
  {\bibfnamefont {J.~G.}\ \bibnamefont {Doench}}, \bibinfo {author}
  {\bibfnamefont {D.}~\bibnamefont {Sen}}, \bibinfo {author} {\bibfnamefont
  {K.~B.}\ \bibnamefont {Yates}}, \bibinfo {author} {\bibfnamefont {R.~T.}\
  \bibnamefont {Manguso}},\ and\ \bibinfo {author} {\bibfnamefont {M.~V.}\
  \bibnamefont {Maus}},\ }\href@noop {} {\bibfield  {journal} {\bibinfo
  {journal} {Nature}\ ,\ \bibinfo {pages} {1}} (\bibinfo {year}
  {2025})}\BibitemShut {NoStop}%
\bibitem [{\citenamefont {Datlinger}\ \emph {et~al.}(2025)\citenamefont
  {Datlinger}, \citenamefont {Pankevich}, \citenamefont {Arnold}, \citenamefont
  {Pranckevicius}, \citenamefont {Lin}, \citenamefont {Romanovskaia},
  \citenamefont {Schaefer}, \citenamefont {Piras}, \citenamefont {Orts},
  \citenamefont {Nemc}, \citenamefont {Biesaga}, \citenamefont {Chan},
  \citenamefont {Neuwirth}, \citenamefont {Artemov}, \citenamefont {Li},
  \citenamefont {Ladstätter}, \citenamefont {Krausgruber},\ and\ \citenamefont
  {Bock}}]{Datlinger2025-oi}%
  \BibitemOpen
  \bibfield  {author} {\bibinfo {author} {\bibfnamefont {P.}~\bibnamefont
  {Datlinger}}, \bibinfo {author} {\bibfnamefont {E.~V.}\ \bibnamefont
  {Pankevich}}, \bibinfo {author} {\bibfnamefont {C.~D.}\ \bibnamefont
  {Arnold}}, \bibinfo {author} {\bibfnamefont {N.}~\bibnamefont
  {Pranckevicius}}, \bibinfo {author} {\bibfnamefont {J.}~\bibnamefont {Lin}},
  \bibinfo {author} {\bibfnamefont {D.}~\bibnamefont {Romanovskaia}}, \bibinfo
  {author} {\bibfnamefont {M.}~\bibnamefont {Schaefer}}, \bibinfo {author}
  {\bibfnamefont {F.}~\bibnamefont {Piras}}, \bibinfo {author} {\bibfnamefont
  {A.-C.}\ \bibnamefont {Orts}}, \bibinfo {author} {\bibfnamefont
  {A.}~\bibnamefont {Nemc}}, \bibinfo {author} {\bibfnamefont {P.~N.}\
  \bibnamefont {Biesaga}}, \bibinfo {author} {\bibfnamefont {M.}~\bibnamefont
  {Chan}}, \bibinfo {author} {\bibfnamefont {T.}~\bibnamefont {Neuwirth}},
  \bibinfo {author} {\bibfnamefont {A.~V.}\ \bibnamefont {Artemov}}, \bibinfo
  {author} {\bibfnamefont {W.}~\bibnamefont {Li}}, \bibinfo {author}
  {\bibfnamefont {S.}~\bibnamefont {Ladstätter}}, \bibinfo {author}
  {\bibfnamefont {T.}~\bibnamefont {Krausgruber}},\ and\ \bibinfo {author}
  {\bibfnamefont {C.}~\bibnamefont {Bock}},\ }\href@noop {} {\bibfield
  {journal} {\bibinfo  {journal} {Nature}\ ,\ \bibinfo {pages} {1}} (\bibinfo
  {year} {2025})}\BibitemShut {NoStop}%
\bibitem [{\citenamefont {Ang}\ \emph {et~al.}(2005)\citenamefont {Ang},
  \citenamefont {Chong},\ and\ \citenamefont {Li}}]{Ang2005-qv}%
  \BibitemOpen
  \bibfield  {author} {\bibinfo {author} {\bibfnamefont {K.~H.}\ \bibnamefont
  {Ang}}, \bibinfo {author} {\bibfnamefont {G.}~\bibnamefont {Chong}},\ and\
  \bibinfo {author} {\bibfnamefont {Y.}~\bibnamefont {Li}},\ }\href@noop {}
  {\bibfield  {journal} {\bibinfo  {journal} {IEEE Trans. Control Syst.
  Technol.}\ }\textbf {\bibinfo {volume} {13}},\ \bibinfo {pages} {559}
  (\bibinfo {year} {2005})}\BibitemShut {NoStop}%
\bibitem [{\citenamefont {Kaech}\ and\ \citenamefont
  {Cui}(2012)}]{Kaech2012-ye}%
  \BibitemOpen
  \bibfield  {author} {\bibinfo {author} {\bibfnamefont {S.~M.}\ \bibnamefont
  {Kaech}}\ and\ \bibinfo {author} {\bibfnamefont {W.}~\bibnamefont {Cui}},\
  }\href@noop {} {\bibfield  {journal} {\bibinfo  {journal} {Nat. Rev.
  Immunol.}\ }\textbf {\bibinfo {volume} {12}},\ \bibinfo {pages} {749}
  (\bibinfo {year} {2012})}\BibitemShut {NoStop}%
\bibitem [{\citenamefont {Bresser}\ \emph {et~al.}(2022)\citenamefont
  {Bresser}, \citenamefont {Kok}, \citenamefont {Swain}, \citenamefont {King},
  \citenamefont {Jacobs}, \citenamefont {Weber}, \citenamefont {Perié},
  \citenamefont {Duffy}, \citenamefont {de~Boer}, \citenamefont {Scheeren},\
  and\ \citenamefont {Schumacher}}]{Bresser2022-iw}%
  \BibitemOpen
  \bibfield  {author} {\bibinfo {author} {\bibfnamefont {K.}~\bibnamefont
  {Bresser}}, \bibinfo {author} {\bibfnamefont {L.}~\bibnamefont {Kok}},
  \bibinfo {author} {\bibfnamefont {A.~C.}\ \bibnamefont {Swain}}, \bibinfo
  {author} {\bibfnamefont {L.~A.}\ \bibnamefont {King}}, \bibinfo {author}
  {\bibfnamefont {L.}~\bibnamefont {Jacobs}}, \bibinfo {author} {\bibfnamefont
  {T.~S.}\ \bibnamefont {Weber}}, \bibinfo {author} {\bibfnamefont
  {L.}~\bibnamefont {Perié}}, \bibinfo {author} {\bibfnamefont {K.~R.}\
  \bibnamefont {Duffy}}, \bibinfo {author} {\bibfnamefont {R.~J.}\ \bibnamefont
  {de~Boer}}, \bibinfo {author} {\bibfnamefont {F.~A.}\ \bibnamefont
  {Scheeren}},\ and\ \bibinfo {author} {\bibfnamefont {T.~N.}\ \bibnamefont
  {Schumacher}},\ }\href@noop {} {\bibfield  {journal} {\bibinfo  {journal}
  {Nat. Immunol.}\ }\textbf {\bibinfo {volume} {23}},\ \bibinfo {pages} {791}
  (\bibinfo {year} {2022})}\BibitemShut {NoStop}%
\bibitem [{\citenamefont {Grassmann}\ \emph {et~al.}(2020)\citenamefont
  {Grassmann}, \citenamefont {Mihatsch}, \citenamefont {Mir}, \citenamefont
  {Kazeroonian}, \citenamefont {Rahimi}, \citenamefont {Flommersfeld},
  \citenamefont {Schober}, \citenamefont {Hensel}, \citenamefont {Leube},
  \citenamefont {Pachmayr}, \citenamefont {Kretschmer}, \citenamefont {Zhang},
  \citenamefont {Jolly}, \citenamefont {Chaudhry}, \citenamefont {Schiemann},
  \citenamefont {Cicin-Sain}, \citenamefont {Höfer}, \citenamefont {Busch},
  \citenamefont {Flossdorf},\ and\ \citenamefont
  {Buchholz}}]{Grassmann2020-ek}%
  \BibitemOpen
  \bibfield  {author} {\bibinfo {author} {\bibfnamefont {S.}~\bibnamefont
  {Grassmann}}, \bibinfo {author} {\bibfnamefont {L.}~\bibnamefont {Mihatsch}},
  \bibinfo {author} {\bibfnamefont {J.}~\bibnamefont {Mir}}, \bibinfo {author}
  {\bibfnamefont {A.}~\bibnamefont {Kazeroonian}}, \bibinfo {author}
  {\bibfnamefont {R.}~\bibnamefont {Rahimi}}, \bibinfo {author} {\bibfnamefont
  {S.}~\bibnamefont {Flommersfeld}}, \bibinfo {author} {\bibfnamefont
  {K.}~\bibnamefont {Schober}}, \bibinfo {author} {\bibfnamefont
  {I.}~\bibnamefont {Hensel}}, \bibinfo {author} {\bibfnamefont
  {J.}~\bibnamefont {Leube}}, \bibinfo {author} {\bibfnamefont {L.~O.}\
  \bibnamefont {Pachmayr}}, \bibinfo {author} {\bibfnamefont {L.}~\bibnamefont
  {Kretschmer}}, \bibinfo {author} {\bibfnamefont {Q.}~\bibnamefont {Zhang}},
  \bibinfo {author} {\bibfnamefont {A.}~\bibnamefont {Jolly}}, \bibinfo
  {author} {\bibfnamefont {M.~Z.}\ \bibnamefont {Chaudhry}}, \bibinfo {author}
  {\bibfnamefont {M.}~\bibnamefont {Schiemann}}, \bibinfo {author}
  {\bibfnamefont {L.}~\bibnamefont {Cicin-Sain}}, \bibinfo {author}
  {\bibfnamefont {T.}~\bibnamefont {Höfer}}, \bibinfo {author} {\bibfnamefont
  {D.~H.}\ \bibnamefont {Busch}}, \bibinfo {author} {\bibfnamefont
  {M.}~\bibnamefont {Flossdorf}},\ and\ \bibinfo {author} {\bibfnamefont
  {V.~R.}\ \bibnamefont {Buchholz}},\ }\href@noop {} {\bibfield  {journal}
  {\bibinfo  {journal} {Nat. Immunol.}\ }\textbf {\bibinfo {volume} {21}},\
  \bibinfo {pages} {1563} (\bibinfo {year} {2020})}\BibitemShut {NoStop}%
\bibitem [{\citenamefont {Xia}\ \emph {et~al.}(2024)\citenamefont {Xia},
  \citenamefont {Lu}, \citenamefont {Tobin}, \citenamefont {Luo}, \citenamefont
  {Moeller}, \citenamefont {Shon}, \citenamefont {Du}, \citenamefont {Linton},
  \citenamefont {Sui}, \citenamefont {Horns},\ and\ \citenamefont
  {Elowitz}}]{Xia2024-fr}%
  \BibitemOpen
  \bibfield  {author} {\bibinfo {author} {\bibfnamefont {S.}~\bibnamefont
  {Xia}}, \bibinfo {author} {\bibfnamefont {A.~C.}\ \bibnamefont {Lu}},
  \bibinfo {author} {\bibfnamefont {V.}~\bibnamefont {Tobin}}, \bibinfo
  {author} {\bibfnamefont {K.}~\bibnamefont {Luo}}, \bibinfo {author}
  {\bibfnamefont {L.}~\bibnamefont {Moeller}}, \bibinfo {author} {\bibfnamefont
  {D.~J.}\ \bibnamefont {Shon}}, \bibinfo {author} {\bibfnamefont
  {R.}~\bibnamefont {Du}}, \bibinfo {author} {\bibfnamefont {J.~M.}\
  \bibnamefont {Linton}}, \bibinfo {author} {\bibfnamefont {M.}~\bibnamefont
  {Sui}}, \bibinfo {author} {\bibfnamefont {F.}~\bibnamefont {Horns}},\ and\
  \bibinfo {author} {\bibfnamefont {M.~B.}\ \bibnamefont {Elowitz}},\
  }\href@noop {} {\bibfield  {journal} {\bibinfo  {journal} {Cell}\ }\textbf
  {\bibinfo {volume} {187}},\ \bibinfo {pages} {2785} (\bibinfo {year}
  {2024})}\BibitemShut {NoStop}%
\bibitem [{\citenamefont {Li}\ \emph {et~al.}(2022)\citenamefont {Li},
  \citenamefont {Israni}, \citenamefont {Gagnon}, \citenamefont {Gan},
  \citenamefont {Raymond}, \citenamefont {Sander}, \citenamefont {Roybal},
  \citenamefont {Joung}, \citenamefont {Wong},\ and\ \citenamefont
  {Khalil}}]{Li2022-rf}%
  \BibitemOpen
  \bibfield  {author} {\bibinfo {author} {\bibfnamefont {H.-S.}\ \bibnamefont
  {Li}}, \bibinfo {author} {\bibfnamefont {D.~V.}\ \bibnamefont {Israni}},
  \bibinfo {author} {\bibfnamefont {K.~A.}\ \bibnamefont {Gagnon}}, \bibinfo
  {author} {\bibfnamefont {K.~A.}\ \bibnamefont {Gan}}, \bibinfo {author}
  {\bibfnamefont {M.~H.}\ \bibnamefont {Raymond}}, \bibinfo {author}
  {\bibfnamefont {J.~D.}\ \bibnamefont {Sander}}, \bibinfo {author}
  {\bibfnamefont {K.~T.}\ \bibnamefont {Roybal}}, \bibinfo {author}
  {\bibfnamefont {J.~K.}\ \bibnamefont {Joung}}, \bibinfo {author}
  {\bibfnamefont {W.~W.}\ \bibnamefont {Wong}},\ and\ \bibinfo {author}
  {\bibfnamefont {A.~S.}\ \bibnamefont {Khalil}},\ }\href@noop {} {\bibfield
  {journal} {\bibinfo  {journal} {Science}\ }\textbf {\bibinfo {volume}
  {378}},\ \bibinfo {pages} {1227} (\bibinfo {year} {2022})}\BibitemShut
  {NoStop}%
\bibitem [{\citenamefont {Allen}\ \emph {et~al.}(2022)\citenamefont {Allen},
  \citenamefont {Frankel}, \citenamefont {Reddy}, \citenamefont {Bhargava},
  \citenamefont {Yoshida}, \citenamefont {Stark}, \citenamefont {Purl},
  \citenamefont {Lee}, \citenamefont {Yee}, \citenamefont {Yu}, \citenamefont
  {Li}, \citenamefont {Garcia}, \citenamefont {El-Samad}, \citenamefont
  {Roybal}, \citenamefont {Spitzer},\ and\ \citenamefont {Lim}}]{Allen2022-lo}%
  \BibitemOpen
  \bibfield  {author} {\bibinfo {author} {\bibfnamefont {G.~M.}\ \bibnamefont
  {Allen}}, \bibinfo {author} {\bibfnamefont {N.~W.}\ \bibnamefont {Frankel}},
  \bibinfo {author} {\bibfnamefont {N.~R.}\ \bibnamefont {Reddy}}, \bibinfo
  {author} {\bibfnamefont {H.~K.}\ \bibnamefont {Bhargava}}, \bibinfo {author}
  {\bibfnamefont {M.~A.}\ \bibnamefont {Yoshida}}, \bibinfo {author}
  {\bibfnamefont {S.~R.}\ \bibnamefont {Stark}}, \bibinfo {author}
  {\bibfnamefont {M.}~\bibnamefont {Purl}}, \bibinfo {author} {\bibfnamefont
  {J.}~\bibnamefont {Lee}}, \bibinfo {author} {\bibfnamefont {J.~L.}\
  \bibnamefont {Yee}}, \bibinfo {author} {\bibfnamefont {W.}~\bibnamefont
  {Yu}}, \bibinfo {author} {\bibfnamefont {A.~W.}\ \bibnamefont {Li}}, \bibinfo
  {author} {\bibfnamefont {K.~C.}\ \bibnamefont {Garcia}}, \bibinfo {author}
  {\bibfnamefont {H.}~\bibnamefont {El-Samad}}, \bibinfo {author}
  {\bibfnamefont {K.~T.}\ \bibnamefont {Roybal}}, \bibinfo {author}
  {\bibfnamefont {M.~H.}\ \bibnamefont {Spitzer}},\ and\ \bibinfo {author}
  {\bibfnamefont {W.~A.}\ \bibnamefont {Lim}},\ }\href@noop {} {\bibfield
  {journal} {\bibinfo  {journal} {Science}\ }\textbf {\bibinfo {volume}
  {378}},\ \bibinfo {pages} {eaba1624} (\bibinfo {year} {2022})}\BibitemShut
  {NoStop}%
\bibitem [{\citenamefont {Lässig}\ \emph {et~al.}(2023)\citenamefont
  {Lässig}, \citenamefont {Mustonen},\ and\ \citenamefont
  {Nourmohammad}}]{Lassig2023-ky}%
  \BibitemOpen
  \bibfield  {author} {\bibinfo {author} {\bibfnamefont {M.}~\bibnamefont
  {Lässig}}, \bibinfo {author} {\bibfnamefont {V.}~\bibnamefont {Mustonen}},\
  and\ \bibinfo {author} {\bibfnamefont {A.}~\bibnamefont {Nourmohammad}},\
  }\href@noop {} {\bibfield  {journal} {\bibinfo  {journal} {Nat. Rev. Genet.}\
  }\textbf {\bibinfo {volume} {24}},\ \bibinfo {pages} {851} (\bibinfo {year}
  {2023})}\BibitemShut {NoStop}%
\bibitem [{\citenamefont {van Dorp}\ \emph {et~al.}(2025)\citenamefont {van
  Dorp}, \citenamefont {Gray}, \citenamefont {Paik}, \citenamefont {Farber},\
  and\ \citenamefont {Yates}}]{van-Dorp2025-my}%
  \BibitemOpen
  \bibfield  {author} {\bibinfo {author} {\bibfnamefont {C.~H.}\ \bibnamefont
  {van Dorp}}, \bibinfo {author} {\bibfnamefont {J.~I.}\ \bibnamefont {Gray}},
  \bibinfo {author} {\bibfnamefont {D.~H.}\ \bibnamefont {Paik}}, \bibinfo
  {author} {\bibfnamefont {D.~L.}\ \bibnamefont {Farber}},\ and\ \bibinfo
  {author} {\bibfnamefont {A.~J.}\ \bibnamefont {Yates}},\ }\href@noop {}
  {\bibfield  {journal} {\bibinfo  {journal} {PLoS Comput. Biol.}\ }\textbf
  {\bibinfo {volume} {21}},\ \bibinfo {pages} {e1013242} (\bibinfo {year}
  {2025})}\BibitemShut {NoStop}%
\bibitem [{\citenamefont {Jerison}\ \emph {et~al.}(2025)\citenamefont
  {Jerison}, \citenamefont {Romeo},\ and\ \citenamefont
  {Quake}}]{Jerison2025-bl}%
  \BibitemOpen
  \bibfield  {author} {\bibinfo {author} {\bibfnamefont {E.~R.}\ \bibnamefont
  {Jerison}}, \bibinfo {author} {\bibfnamefont {N.}~\bibnamefont {Romeo}},\
  and\ \bibinfo {author} {\bibfnamefont {S.~R.}\ \bibnamefont {Quake}},\
  }\href@noop {} {\bibfield  {journal} {\bibinfo  {journal} {bioRxiv}\ ,\
  \bibinfo {pages} {2025.01.28.635318}} (\bibinfo {year} {2025})}\BibitemShut
  {NoStop}%
\bibitem [{\citenamefont {Straub}\ \emph {et~al.}(2023)\citenamefont {Straub},
  \citenamefont {Grassmann}, \citenamefont {Jarosch}, \citenamefont {Richter},
  \citenamefont {Hilgendorf}, \citenamefont {Hammel}, \citenamefont {Wagner},
  \citenamefont {Buchholz}, \citenamefont {Schober},\ and\ \citenamefont
  {Busch}}]{Straub2023-oa}%
  \BibitemOpen
  \bibfield  {author} {\bibinfo {author} {\bibfnamefont {A.}~\bibnamefont
  {Straub}}, \bibinfo {author} {\bibfnamefont {S.}~\bibnamefont {Grassmann}},
  \bibinfo {author} {\bibfnamefont {S.}~\bibnamefont {Jarosch}}, \bibinfo
  {author} {\bibfnamefont {L.}~\bibnamefont {Richter}}, \bibinfo {author}
  {\bibfnamefont {P.}~\bibnamefont {Hilgendorf}}, \bibinfo {author}
  {\bibfnamefont {M.}~\bibnamefont {Hammel}}, \bibinfo {author} {\bibfnamefont
  {K.~I.}\ \bibnamefont {Wagner}}, \bibinfo {author} {\bibfnamefont {V.~R.}\
  \bibnamefont {Buchholz}}, \bibinfo {author} {\bibfnamefont {K.}~\bibnamefont
  {Schober}},\ and\ \bibinfo {author} {\bibfnamefont {D.~H.}\ \bibnamefont
  {Busch}},\ }\href@noop {} {\bibfield  {journal} {\bibinfo  {journal}
  {Immunity}\ } (\bibinfo {year} {2023})}\BibitemShut {NoStop}%
\bibitem [{\citenamefont {Chao}\ \emph {et~al.}(2004)\citenamefont {Chao},
  \citenamefont {Davenport}, \citenamefont {Forrest},\ and\ \citenamefont
  {Perelson}}]{Chao2004-jq}%
  \BibitemOpen
  \bibfield  {author} {\bibinfo {author} {\bibfnamefont {D.~L.}\ \bibnamefont
  {Chao}}, \bibinfo {author} {\bibfnamefont {M.~P.}\ \bibnamefont {Davenport}},
  \bibinfo {author} {\bibfnamefont {S.}~\bibnamefont {Forrest}},\ and\ \bibinfo
  {author} {\bibfnamefont {A.~S.}\ \bibnamefont {Perelson}},\ }\href@noop {}
  {\bibfield  {journal} {\bibinfo  {journal} {J. Theor. Biol.}\ }\textbf
  {\bibinfo {volume} {228}},\ \bibinfo {pages} {227} (\bibinfo {year}
  {2004})}\BibitemShut {NoStop}%
\bibitem [{\citenamefont {Ganusov}\ \emph {et~al.}(2011)\citenamefont
  {Ganusov}, \citenamefont {Barber},\ and\ \citenamefont
  {De~Boer}}]{Ganusov2011-bl}%
  \BibitemOpen
  \bibfield  {author} {\bibinfo {author} {\bibfnamefont {V.~V.}\ \bibnamefont
  {Ganusov}}, \bibinfo {author} {\bibfnamefont {D.~L.}\ \bibnamefont
  {Barber}},\ and\ \bibinfo {author} {\bibfnamefont {R.~J.}\ \bibnamefont
  {De~Boer}},\ }\href@noop {} {\bibfield  {journal} {\bibinfo  {journal} {PLoS
  One}\ }\textbf {\bibinfo {volume} {6}},\ \bibinfo {pages} {e15959} (\bibinfo
  {year} {2011})}\BibitemShut {NoStop}%
\bibitem [{\citenamefont {Halle}\ \emph {et~al.}(2016)\citenamefont {Halle},
  \citenamefont {Keyser}, \citenamefont {Stahl}, \citenamefont {Busche},
  \citenamefont {Marquardt}, \citenamefont {Zheng}, \citenamefont {Galla},
  \citenamefont {Heissmeyer}, \citenamefont {Heller}, \citenamefont {Boelter},
  \citenamefont {Wagner}, \citenamefont {Bischoff}, \citenamefont {Martens},
  \citenamefont {Braun}, \citenamefont {Werth}, \citenamefont {Uvarovskii},
  \citenamefont {Kempf}, \citenamefont {Meyer-Hermann}, \citenamefont {Arens},
  \citenamefont {Kremer}, \citenamefont {Sutter}, \citenamefont {Messerle},\
  and\ \citenamefont {Förster}}]{Halle2016-kw}%
  \BibitemOpen
  \bibfield  {author} {\bibinfo {author} {\bibfnamefont {S.}~\bibnamefont
  {Halle}}, \bibinfo {author} {\bibfnamefont {K.~A.}\ \bibnamefont {Keyser}},
  \bibinfo {author} {\bibfnamefont {F.~R.}\ \bibnamefont {Stahl}}, \bibinfo
  {author} {\bibfnamefont {A.}~\bibnamefont {Busche}}, \bibinfo {author}
  {\bibfnamefont {A.}~\bibnamefont {Marquardt}}, \bibinfo {author}
  {\bibfnamefont {X.}~\bibnamefont {Zheng}}, \bibinfo {author} {\bibfnamefont
  {M.}~\bibnamefont {Galla}}, \bibinfo {author} {\bibfnamefont
  {V.}~\bibnamefont {Heissmeyer}}, \bibinfo {author} {\bibfnamefont
  {K.}~\bibnamefont {Heller}}, \bibinfo {author} {\bibfnamefont
  {J.}~\bibnamefont {Boelter}}, \bibinfo {author} {\bibfnamefont
  {K.}~\bibnamefont {Wagner}}, \bibinfo {author} {\bibfnamefont
  {Y.}~\bibnamefont {Bischoff}}, \bibinfo {author} {\bibfnamefont
  {R.}~\bibnamefont {Martens}}, \bibinfo {author} {\bibfnamefont
  {A.}~\bibnamefont {Braun}}, \bibinfo {author} {\bibfnamefont
  {K.}~\bibnamefont {Werth}}, \bibinfo {author} {\bibfnamefont
  {A.}~\bibnamefont {Uvarovskii}}, \bibinfo {author} {\bibfnamefont
  {H.}~\bibnamefont {Kempf}}, \bibinfo {author} {\bibfnamefont
  {M.}~\bibnamefont {Meyer-Hermann}}, \bibinfo {author} {\bibfnamefont
  {R.}~\bibnamefont {Arens}}, \bibinfo {author} {\bibfnamefont
  {M.}~\bibnamefont {Kremer}}, \bibinfo {author} {\bibfnamefont
  {G.}~\bibnamefont {Sutter}}, \bibinfo {author} {\bibfnamefont
  {M.}~\bibnamefont {Messerle}},\ and\ \bibinfo {author} {\bibfnamefont
  {R.}~\bibnamefont {Förster}},\ }\href@noop {} {\bibfield  {journal}
  {\bibinfo  {journal} {Immunity}\ }\textbf {\bibinfo {volume} {44}},\ \bibinfo
  {pages} {233} (\bibinfo {year} {2016})}\BibitemShut {NoStop}%
\end{thebibliography}

\end{document}


\noindent{\bf \Large Supplementary Information\\}
\vspace{-1ex}

\noindent {\bf \large Design principles of the cytotoxic CD8+ T-cell response}\\

\vspace{-2ex}
\noindent Obinna A. Ukogu, Zachary Montague, Gr\'egoire Altan-Bonnet \& Armita Nourmohammad

\tableofcontents

\clearpage

\section{Model of T-cell response to infections}

\subsection*{Competitive binding of T-cells to healthy and infected cells}

\label{section: binding}
T-cells are cross-reactive and can bind to a broad range of antigens. A cognate T-cell for a pathogen-derived antigen can be cross-reactive to antigens derived from self-proteins. Consequently, when analyzing T-cell recognition, it is essential to account for the potential competition between binding to {cells presenting} cognate non-self antigens and to self-antigens.

To model this effect in a simple way, we consider the following coupled reactions:
\begin{align}
\begin{split}
  &E^{(u)} + I^{(u)} \underset{k_I^d}{\stackrel{k_I^a}{\rightleftharpoons}} C_I\\
  &E^{(u)} + S^{(u)}  \underset{k_S^d}{\stackrel{k_S^a}{\rightleftharpoons}} C_S
\end{split}
\end{align}
where $E^{(u)}$ denotes the number of unbound effector T-cells, $I^{(u)}$ and $S^{(u)}$ represent the numbers of unbound {infected and susceptible cells}, respectively, and $C_I$ and $C_S$ correspond to the bound complexes of effector T-cells with {those cells}. Note that we take the number of infected and susceptible cells as a proxy for the amount non-self and self antigens presented, respectively. The parameters $k^a_{(\cdot)}$ and $k^d_{(\cdot)}$ are the rate constants for the association and dissociation of complexes $C_{(\cdot)}$, where the placeholder $(\cdot)$ can be substituted with either $I$ (infected) or $S$ ({susceptible} healthy). The dynamics of these complexes over time follows, 

\EQA
\nonumber \frac{\d C_I }{\d t} &=& k_I^a E^{(u)} I^{(u)} - k_I^d C_I\\
\nonumber \frac{\d C_S }{\d t} &=& k_S^a E^{(u)} S^{(u)} - k_S^d C_S.
\EEA

We assume a separation of timescales: binding and unbinding events occur much more rapidly than T-cell proliferation and effector killing of bound cells. The bound complexes can be treated as being in quasi-equilibrium at all times, i.e., $\d C_{(\cdot)}/\d t= 0$, which yields,
\EQ
E^{(u)} I^{(u)} = K_I C_I, \qquad\qquad E^{(u)} S^{(u)} = K_S C_S,
\label{eq:equib_conc}
\EE
where $K_I = {k_I^d}/{k_I^a}$ and $K_S = {k_S^d}/{k_S^a}$ are the dissociation constants of the effector cells from {infected and healthy cells}, respectively.

Expressing eq.~\ref{eq:equib_conc} in terms of the total concentrations of effectors $E = E^{(u)} + C_I + C_S$, {infected cells} $I= I^{(u)} + C_I $, and {healthy cells} $S= S^{(u)} + C_S $, we arrive at,
\EQA
\nonumber 0 &=&(E - C_I -C_S) ( I - C_I) - K_I C_I \\
\nonumber 0 &=& (E - C_I -C_S) (S- C_S) - K_S C_S
\EEA

Assuming that the number of bound complexes is small relative to that of unbound 

 cells, we can neglect the second order terms $\mathcal{O}(C_I^2),\mathcal{O}(C_S^2),\mathcal{O}(C_I C_S)$, and solve for these abundances in terms of the total number of effectors, and the infected and healthy cells,
\EQA
C_I = \frac{E I}{E+ I + K_I + \frac{E+K_I}{E+ K_S} S },\qquad C_S = \frac{E S}{E+ S + K_S + \frac{E+K_S}{E+ K_I} I }
\label{clearance-term}.
\EEA

The correction terms in the denominators ($\frac{E+K_I}{E+ K_S} S$, and $\frac{E+K_S}{E+ K_I} I$) capture the competition between binding to self- and non-self-derived antigens. In the limit $K_S \gg K_I$ (i.e., when T-cells bind more strongly to infected cells than healthy ones), the antagonistic effect vanishes and $C_I$ approaches the form for response in the presence of a single-antigen only, consistent with ref.~\cite{Chao2004-jq, Mayer2019-pj}. However, as $K_S \to K_I$ (self and non-self antigens become comparable in affinity), the fraction of effector cells bound to infection-derived antigens decreases, reflecting the impact of self/non-self cross-reactivity and antagonism on recognition. Prior work has explored the impact of antagonistic effects on T-cell response in much more detail ~\cite{Altan-Bonnet2005,Francois2013-yl}. Here, we only adopt a high-level perspective to capture how such effects shape responses to low-immunogenicity infections and cancer.

The relative magnitude of infection- to self-derived antigenicities (i.e, the ratio of the respective recognition thresholds) determine the {\em immunogenicity} of an infection, which we define as, 
\begin{equation}
\label{eq:kappa}
\text{immunogenicity:}~~~~\kappa = K_I^{-1}K_S.
\end{equation}
This dimensionless measure reflects how readily cognate T-cells recognize antigens presented by infected cells as ``non-self'' relative to those presented by uninfected cells (self).


\subsection*{Dynamics of growing (acute) infections} In the absence of infection, the population of susceptible cells ($S$) are homeostatically maintained by slow birth-death processes at a steady state, $S(t=0) = S_\text{max}$. Upon infection, there are initially a small number of infected cells---$I(t=0) = I_0$---but susceptible cells gradually become infected as they come into contact with other infected cells at rate $b_I$. The infected state of cells and their death, at rate $d_I$, produce chemical signals, namely the antigen signal $\sigma_\Ag(t)$ and the infection signal $\sigma_\infc(t)$, which are detected and transmitted by the innate immune system. The innate response then triggers na\"ive T-cells in {the} adaptive immune response ($N$) to differentiate into effector cells ($E$). Differentiated effector cells kill infected and susceptible cells, producing chemical response signals ($\sigma_\res(t)$) that feedback on the response. The dynamics of susceptible and infected cells follow,
\begin{align}
\label{infection-dynamics}
  \begin{split}
    \dot{S} &= - b_I I S -d_{E}C_{S},
    \\
    \dot{I} &= b_I S I- d_I I -d_{E}C_I,
  \end{split}
\end{align}
where $d_{E}$ is the maximal clearance rate of infected/susceptible cells by effector T-cells, and $C_{S } = \frac{E S}{E+ S + K_S + \frac{E+K_S}{E+ K_I} I }$ and $C_I = \frac{E I}{E+ I + K_I + \frac{E+K_I}{E+ K_S} S }$ represent the number of susceptible or infected cells bound to an effector cell at time $t$, respectively; see section~\ref{section: binding} for {a} derivation of these quantities. This framework closely reflects viral or intracellular bacterial infections, but also captures features of extracellular pathogens that damage or invade cells indirectly (e.g., via toxins); we do not distinguish these mechanisms here.

The extent of pathogen reproduction during the lifetime of an infected cell is determined by the product of the infection rate and the characteristic survival time of the cell. Specifically, over the infected-cell lifetime $\tau_\text{death} =d_I^{-1}$, an infected cell produces on average $b_I S \cdot \tau_\text{death} = b_I S/d_I$ new infections. In epidemiology, this term is often referred to as the \emph{effective reproduction number}, whose maximum value---attained when $S=S_\text{max}$---is the \emph{basic reproduction number}, ${b_IS_\text{max}}/{d_I}$~\cite{Milligan2015-yu, Abuin2020-tf}. This motivates defining the
 \emph{intra-host basic reproduction number},
\begin{equation}
\label{eq:R_0}
R_0 = \frac{b_I S_{max}}{d_I},
\end{equation}
{which measures} the expected number of secondary infected cells generated by a single infected cell in an otherwise fully susceptible cell population. Note that $R_0$ scales with $S_\text{max}$, the total number of cells in the modeled tissue. We simulate these dynamics over a time horizon $T>0$ days, chosen to match the typical timescale of acute immune responses (approximately $2-3$ weeks). A window that is too short underestimates cumulative harm, whereas a longer window increases computational cost. Figs.~\ref{paramsensitivityFigure1},~\ref{paramsensitivityFigure2}  demonstrate how total harm and effector cell fold expansion depend on the choice of $T$, over a wide range from 1 to 100 days. In our main simulations, we set $T=30$ days to  ensure that most responses generated by our model are fully
captured~\cite{Zehn2009-up,Badovinac2007-ii,Lessler2009-aw}. For related approaches see refs.~\cite{Chao2004-jq, Deenick2003-pn, Cheon2021-os}.

\subsection*{Dynamics of slow-growing tumors}
In Fig.~\mainfigsix, we show the efficacy of different T-cell response designs for tumor clearance. In this context, we model a tumor as a slowly growing cell population---maximum rate $b_I$---which grows by out-competing healthy tissue cells for limited space/resources, resulting in a logistic growth of tumor cells \cite{Yin2019-nk}. We assume that at the time of treatment, the tumor is already at detectable size, $I_0 = 10\%\, S_\text{max}$, and that tumor cells die at negligible rates during our simulation window, $d_I = 0$. The resulting dynamics are

\begin{align}
\label{cancer-dynamics}
  \begin{split}
    \dot{S} &= - b_I \left(1 -\frac{I}{S_\text{max}}\right) I -d_{E}C_{S},
    \\
    \dot{I} &= b_I \left(1 -\frac{I}{S_\text{max}}\right) I-d_{E}C_I.
  \end{split}
\end{align}

\subsection*{Sampling pathogen space}
{To model diverse {immune challenges}, we sample a broad range of immunogenicities ($\kappa$), and basic reproduction numbers ($R_0$). We take $K_{S} = S_{max} \geq K_I$ so that effector cells typically bind more strongly to infected cells than healthy cells. We note that $K_I$ and $K_S$ are effective parameters reflecting the dynamics of antigen detection in the complex tissue environment. The killer T-cell response is delayed with respect to pathogen exposure---typically around a few days \cite{De_Boer2001-wl, Lessler2009-aw}---so there are limits on its effectiveness against fast-replicating pathogens. We sample a range of reproduction numbers ($R_0$) to include non-replicating {immune challenges} $R_0 = 1$ and fast replicating {immune challenges} which deplete the reservoir of susceptible cells. Quantitatively, we can consider when the infected cell population peaks in the absence of an immune response $T_{I_\text{max}}$ by setting $\d {I}/\d t = 0$ and $E=0$ in eq. \ref{infection-dynamics} to find:
\[
S(T_{I_\text{max}}) = \frac{S_\text{max}}{R_0}.
\]
In our simulations, we choose the maximum $R_0$ such that $S(T_{I_\text{max}}) \ll S_\text{max}$, representing a pathogen that quickly infects most susceptible cells. Here, we chose $(K_I, R_0) \in [10^{-3}S_\text{max},S_\text{max}]\times [1, 5]${, discretized into a $9\times9$ grid,} to reflect the aforementioned considerations; see Table \ref{table:parameters} for details on simulation parameter ranges.

\subsection*{Harm during  T-cell response to pathogens}
\label{section: harm signals}

We express harm (damage) to a host in terms of cell deaths. This depends on the details of the programmed immune response, which we parametrize by $\vec\theta$ {(see below for details)}. The rate of harm at a given time point can be expressed as 
\begin{align}
\begin{split}
  h(t) &=
  h_{\infc}(t,\vec \theta) + \underbrace{ h_{E,I}(t,\vec \theta) + h_{E,S}(t,\vec\theta)}_{=h_\res(t)},\label{eq:harm}
  \end{split}
\end{align}
where $h_{E,I}(t) = d_E C_I(t)$ and $h_{E,S}(t) = d_E C_s(t)$  denote the loss of infected and healthy cells, respectively, caused by the effector T-cell response, amounting to the total damage by the response $h_\res(t)$. Here, $C_{I}(t)$ and $C_{S}(t)$ are the expected number of infected and susceptible cells bound to effectors at time $t$ (eq.~\ref{clearance-term}). {We consider two scenarios:}\\
\begin{itemize}
\item{\bf Harm associated with growing (acute) infections.} The infection harm $h_\infc(t)$ captures pathogen-induced harm, i.e., cells directly killed by the infection itself. Its functional form depends on the biological context. For an acute, proliferating infection (eq.~\ref{infection-dynamics}), it follows:
\EQ
h_\infc (t,\vec \theta) = d_I I(t,\vec\theta)
\label{eq:harm_inf}\EE
reflecting cell loss due to the death of infected cells. The {\em total cell death} (or inflicted harm) then follows
\begin{align}
\label{eq: clear_tox}
  \begin{split}
 H(\vec\theta) =
 \underbrace{\int _0^{T}\left[h_\infc\left(t, \vec\theta\right) + h_{E,I}\left(t, \vec\theta\right)\right]{\rm d}t}_{H_{\rm inf.}(\vec\theta)} + \underbrace{\int _0^{T} h_{E,S}\left(t, \vec\theta\right){\rm d}t }_{H_{\rm toxic.}(\vec\theta)},
  \end{split}
\end{align}
where $H_\text{inf.}(\vec\theta)$ aggregates infection-related harm (cells killed by the pathogen and infected cells eliminated by effectors), $H_\text{toxic.}(\vec\theta)$ captures immunopathology (healthy bystander cells killed by effectors), and $T > 0$ is the simulation window. This single, interpretable metric balances clearance benefits against collateral damage, enabling principled comparison across designs and infections. \\

\item{\bf Harm associated with slow-growing tumors.} For {logistically} growing cancer tumors (eq.~\ref{cancer-dynamics}), the harm from cancer takes the form
\EQ h_\text{cancer}(t)  = b_I \left(1 -\frac{I(t)}{S_\text{max}}\right) I(t)\label{eq:harm_cancer},
\EE
reflecting net healthy cell loss due to tumor growth (eq.~\ref{cancer-dynamics}). In this case, we define the total harm as
\begin{align}
\label{eq:cancer_harm_total}
  \begin{split}
 H(\vec\theta) =
 \underbrace{\int _0^{T}h_{cancer}\left(t, \vec\theta\right){\rm d}t}_{H_{\rm cancer}(\vec\theta)} + \underbrace{\int _0^{T} h_{E,S}\left(t, \vec\theta\right){\rm d}t }_{H_{\rm toxic.}(\vec\theta)}.
  \end{split}
\end{align}
Here, $H_{\rm cancer}(\vec\theta)$ quantifies the loss of healthy cells as the tumor expands and replaces normal tissue, while $H_{\rm toxic.}(\vec \theta)$ captures collateral healthy-cell death caused by the ensuing immune response. Note that in the case of acute infections (eq.~\ref{eq: clear_tox}), loss of healthy cells is represented by $H_{\rm inf.}$, the sum of cells killed directly by the pathogen and infected cells eliminated by effectors. In cancer, by contrast, a slowly growing tumor replaces healthy cells with malignant cells. Accordingly, harm is tied to the ``birth'' of cancer cells captured by the time-integral of the birth term $h_\text{cancer}(t)$ in eq.~\ref{eq:harm_cancer}.\\
\end{itemize}

\subsection*{Signal cues for T-cell response} 
\label{section: signal dynamics}
We assume that T-cells can sense three types of signals during infection:
\begin{itemize}
  \item {\bf Antigen signal} $\sigma_\Ag(t)$, which is proportional to the number of infected cells, and expressed in units of the {antigen recognition threshold} $K_I$,
  \EQ\text{antigen signal:}~~\sigma_{\Ag}(t) ={K_I}^{-1} {I}(t) \leq K_I^{-1}S_\text{max}
  \EE
  \item {\bf Infection-induced harm signal} $\sigma_\infc (t)$, reflecting cumulative damage caused directly by the pathogen, and expressed in units of the {harm recognition threshold} $K_H$,
  \EQ
  \text{infection harm signal:}~~\sigma_{\infc}(t) = K_H^{-1}\int_0^t e^{-(t-s) d_H}\,h_{\infc}(s)\,{\rm d}s \leq K_H^{-1}S_\text{max}.
  \EE
  \item {\bf Response-induced harm signal} $\sigma_\res (t)$, reflecting cumulative damage caused by the effector response, and expressed in units of the 
  {harm recognition threshold} $K_H$,
  \EQ
  \text{response harm signal:}~~\sigma_{\res}(t) = K_H^{-1}\int_0^t e^{-(t-s) d_H}\,h_{\res}(s)\,{\rm d}s \leq K_H^{-1}S_\text{max}.
  \EE
\end{itemize}
The harm-derived signals are time-integrated, reflecting the accumulation of cytokines during an immune response, and are exponentially discounted at rate $d_H$, reflecting the natural decay of cytokines in the absence of ongoing stimulation. Note that $K_I$ plays a dual role: it represents both the recognition threshold of effectors binding to and killing infected cells (see section~\ref{section: binding}) and the recognition threshold for binding to and stimulating T-cells. 

Finally, although we do not explicitly model innate or regulatory T-cell dynamics, these populations act primarily as sources and sinks of pro- and anti-inflammatory mediators. Their effects are therefore captured in part by our representation of antigen- and response-induced harm signal production and decay. With this interpretation, the design parameters and the latent biological factors that set their values implicitly encode the contributions of these cell populations, albeit at a coarse-grained level. Moreover, in our framework, highly immunogenic antigens (with large $\kappa$) are those that elicit strong CD8$^+$ T-cell responses, whereas antigens with low immunogenicity have more subtle effects. This, in part, reflects the suppressive role of Tregs in limiting effector responses against self-antigens, which in our definition correspond to low-immunogenicity targets. Future extensions could more explicitly incorporate the time-varying impact of Tregs and the innate immune system on the dynamics of the immune response, similar to the approach in ref.~\cite{Voisinne2015-ed}.

\subsection*{Lineage-based model of CD$8^+$ T-cell response}
\label{section: lineage-model}

To characterize the immune response to a {pathogenic challenge}, we model the large-population dynamics of susceptible and infected cells in a host as deterministic processes, coupled to a stochastic, time-inhomogeneous birth-death process for the CD$8^+$ T-cell response. 

The T-cell repertoire typically contains many precursors, and responses to a given pathogen are often polyclonal. An interesting problem would be to assess how interactions between different T-cell clones impact the response in an individual, which we defer to future work. 
In our simulations, we model a response mediated by a single clone of na\"ive T-cells (i.e., cells sharing the same TCR). Upon activation, each na\"ive cell can seed a lineage whose progeny differentiate into effector cells that act against the source of the immune challenge (e.g., infection or tumor) or memory cells which do not participate in resolving the ongoing immune challenge.

In what follows, we consider an initial population of $N_\text{lin.}$ monoclonal na\"ive T-cells, and track the lineages of their progeny in response to {an immune challenge}. Each cell's decisions---to engage an antigen-presenting cell (APC), become activated, divide, differentiate, or die---are governed by time-dependent antigen and infection harm signals (eqs.~\ref{eq:rate_regulation}). We assume these signals are lineage-independent: cells in the same state but belonging to different lineages share identical transition rates. Inter-lineage variability therefore arises solely from stochastic differences in transition timing, consistent with experimental observations~\cite{Marchingo2016,Plambeck2022,Cheon2021-os}.\\

\paragraph{Recruitment of na\"ive T-cells to interact with APCs ($N \to [\text{APC} \cdot N]$).} 
Na\"ive T-cells recirculate through blood and lymph and may also reside in tissues~\cite{Henrickson2008-cr}. Upon infection, APCs become activated and acquire antigen as infected cells die~\cite{Uhl2009-hg}. This process is coupled to the production of an infection harm signal $\sigma_\infc$, which serves as a proxy for the ``danger'' associated with tissue damage caused by the {immune challenge}. As a result, na\"ive T-cells can be recruited to the sites of infection and interact with APCs to initiate an immune response. We model the recruitment rate of na\"ive T-cells to bind to the APCs as
\begin{align}
\label{eq:binding}
\begin{split}
r_{N \to \text{APC} \cdot N}(t) \equiv r_{\text{bind}}(t) &= \frac{r_\text{bind}^{\text{max}}}{1 + \exp{[-\log(1 + \sigma_\infc(t))]}},
\end{split}
\end{align}
where $r_\text{bind}^{\text{max}} \approx 1~\text{[day]}^{-1}$ is the upper bound on the recruitment rate of T-cells to lymph nodes for sustained APC binding~\cite{Bousso2003-ir,Belz2007,Ozga2016-ss} (see Table~\ref{table:parameters} for the biological range). The denominator modulates this rate according to the strength of the infection harm signal activating the APCs. We assume that recruitment of all the na\"ive cells to APCs is mediated by the same infection harm signal, and that these cells do not compete for APCs or infection harm signal.\\

\paragraph{Activation of na\"ive T-cells ($[\text{APC} \cdot N] \to N^*$).} When a na\"ive T-cell engages an APC to form a complex $[\text{APC} \cdot N]$, the two cells may remain bound for several hours, leading to activation of the bound T-cell (``priming phase''). During this time, the immunological synapse formed between the T-cell and APC initiates a cascade of intracellular signaling events in the na\"ive T-cell, enabling it to surpass a sharp activation threshold~\cite{Au-Yeung2014-io}. The strength of antigenic stimulation, together with co-stimulatory signals, modulates the activation rate of APC-bound na\"ive T-cells $r_{N\to N^*} (t)$. 

Experiments have shown that na\"ive T-cells can remain stably bound for 12-20 hours before the first division~\cite{Plambeck2022} to become activated $[\text{APC} \cdot N] \to N^*$. Following activation, cells commit to a rapid proliferative program~\cite{Van_Heijst2009-tj,Van_Stipdonk2003-ui}, resulting in $k=2-3$ burst-like divisions at a rate $r_{N^* \to 2N^*}$ \cite{Plambeck2022} (Fig. \ref{differentiationschematic}). Once initiated, these early doublings proceed largely independently of continued antigen or cytokine input \cite{Van_Stipdonk2001,Plambeck2022}, and set an upper bound $N^*_\text{max} = 2^k$ for the number of activated cells generated prior to differentiation in a lineage; we use $N^*_\text{max} = 2^2$ in our simulations (Table~\ref{table:parameters}).\\

\paragraph{Differentiation of activated T-cells to effector phenotype ($N^* \to E$).} During an {immune challenge}, CD$8^+$ T-cells follow a pathway from long-lived na\"ive/memory cells to short-lived effector cells. This fate is partially ``reversible'': effectors can reacquire memory-like traits (e.g., longevity), and become so-called effector memory~\cite{Kaech2001-kp, Youngblood2017-my,Abadie2023-md}.
We consider differentiation from na\"ive to effector state as the primary transition for activated cells, with un-transitioned cells retaining the relative longevity and slow proliferation of na\"ive cells. Consistent with prior work~\cite{Kaech2001-kp,Buchholz2013, Plambeck2022,Abadie2023-md}, we take the burst-like early proliferation phase as the window during which effector commitment occurs (Fig.~\ref{differentiationschematic}).

Formally, for a given activated cell $N^*$, we model effector commitment as the first arrival of a time-inhomogeneous Poisson process with instantaneous rate, $r_{N^* \to E}(t)$. Given the stochastic time interval $\tau_\text{burst}$ to complete the initial burst-like divisions, the probability that an activated cell differentiates into the effector  state is computed as the probability that the first arrival time is less than $\tau_\text{burst}$,
\begin{align}
    p_{N^* \to E} = 1 - \exp\left[-\int_0^{\tau_\text{burst}} r_{N^* \to E}(s){\rm d}s\right].
\end{align}
Given {$N^*_\text{max}$} activated cells available at the end of the burst, the number that differentiate into effectors is drawn independently as $e_\text{differentiate}\sim \text{Binom}({N^*_\text{max}},p_{N^* \to E})$, with the remaining ${N^*_\text{max}}- e_\text{differentiate}$ cells adopting a memory fate. This fate decision is executed independently across lineages (Algorithms~\ref{alg:cd8}-\ref{alg:cd8-stoch}; Fig.~\ref{differentiationschematic}).\\

\paragraph{Division and death of effector T-cells ($E\to 2E$, and $E\to\emptyset$).}
Experiments have shown that T-cell lineages do not divide indefinitely~\cite{Heinzel2017-qp}. By tracking proliferation markers in cell lineages, it has been estimated that a na\"ive precursor undergoes at most $K_\text{max} \approx 10 - 19$ division rounds before terminal arrest~\cite{Buchholz2013,Marchingo2016, Badovinac2007-ii, Zhang2011-dv}, with differentiation into effector possible along the way~\cite{Buchholz2013,Marchingo2016,Heinzel2017-qp}.} This sets an upper bound on the number of effector cells in a lineage (if all cells turn into effector), which is $E_\text{max} = 2^{K_\text{max}}$, which we set to $2^{15}$ in our simulations (Table~\ref{table:parameters}). A comparable bound for memory cells is less certain, as establishing it would require years of longitudinal tracking.

Because effector commitment is stochastic, the maximum additional effector output beyond the initial committed pool varies by lineage. At the end of the early burst, the lineage has completed $\log_2 N^*_\text{max}$ semi-synchronous divisions. Each newly committed effector cell can therefore undergo at most $K_\text{max} - \log_2 N^*_\text{max} =\log_2 E_\text{max}- \log_2 N^*_\text{max}$ additional rounds. Thus, starting from $e_\text{differentiate}$ effectors at the end of the burst phase, proliferation halts once the effector count reaches, $e_\text{stop prolif.} = e_\text{differentiate} \times \left (\frac{E_\text{max}}{N^*_\text{max}}\right)$; see Algorithms~\ref{alg:cd8}-\ref{alg:cd8-stoch}. Each differentiated effector cell undergoes division and death events at rates $r_{E \to E}(t)$ and $r_{E \to\emptyset}(t)$, respectively, which are modulated by the time-varying antigen and harm signals (see below). In lieu of tracking the division history/tree of cells in the lineage, as effector cells divide, we keep a running count of the total number of effector division events in that lineage, $B_\text{proliferation},$ and set the division rate to zero once $B_\text{proliferation} \geq e_\text{stop prolif.}$ (Algorithms~\ref{alg:cd8}-\ref{alg:cd8-stoch}).\\

In summary, the dynamics for a single  T-cell responding {to an immune challenge}, through activation by engaging APC's, differentiation, expansion, and contraction are summarized as follows,
\begin{align}
  \begin{split}
  \text{exposure}&
  \begin{cases}
   S + I &\overset{b_I}{\longrightarrow} I + I \to \sigma_\Ag\\
   I &\overset{d_I}{\longrightarrow} \emptyset \to \sigma_\infc\\
  \end{cases}\\
  \text{clearance}&
  \begin{cases}
   I/S + E &\overset{d_E}{\longrightarrow} E + \emptyset \to E + \sigma_\res\\
   \end{cases}\\
    \text{activation}&
  \begin{cases}
   \text{APC} + N &\overset{r_\text{bind}(\vec\sigma)}{\longrightarrow} \text{APC}\cdot N\\
  \text{APC} \cdot N &\overset{r_{N \to N^*}(\vec\sigma)}{\longrightarrow} N^* + \text{APC}\\
  N^* &\overset{r_{N^* \to 2N^*}}{\longrightarrow} 2N^* \overset{r_{N^* \to 2N^*}}{\longrightarrow} 4N^* \\
  \end{cases}\\
  \text{differentiation}&
  \begin{cases}
   N^{*} &\overset{r_{N^* \to E}(\vec\sigma)}{\longrightarrow} E
   \end{cases}\\
   \text{expansion}&
  \begin{cases}
   E &\overset{r_{E \to 2E}(\vec\sigma)}{\longrightarrow} E + E\\
   \end{cases}\\
   \text{contraction}&
  \begin{cases}
   E &\overset{r_{E \to \emptyset}(\vec\sigma)}{\longrightarrow} \emptyset
   \end{cases}\\
  \end{split}
\end{align}
Note that the transition rates are dependent on the three time-dependent signals $\vec\sigma = (\sigma_\Ag, \sigma_\infc,\sigma_\res)$; see below for details. The programmatic details of these reactions, when keeping track of different T-cell lineages, are provided in Algorithms~\ref{alg:cd8}-\ref{alg:cd8-stoch}. \\

\paragraph{Regulation of T-cell transition rates.} 
Activated T-cells undergo differentiation, proliferation, and eventual contraction in response to antigen and and other signaling molecules like cytokines~\cite{Hiam-Galvez2021-nj, Reiner2014, Liao2013}. These processes are shaped by both extrinsic noise (e.g., variability in cytokine levels and spatial heterogeneity of APCs) and intrinsic noise (e.g., stochastic gene expression and chromatin remodeling)~\cite{Guillemin2021}. We model the effector CD$8^+$ T-cell response as a sequence of four canonical modules:

\begin{enumerate}
  \item {\bf Activation:} priming of na\"ive T-cells, $N \to N^*$,
  \item {\bf Differentiation:} commitment of the activated T-cells to effector phenotype, $N^* \to E$, 
  \item {\bf Expansion:} division of effector cells, $E \to 2E$,
  \item {\bf Contraction:} programmed loss of effector cells, $E \to \emptyset$.
\end{enumerate}

We model the complex genetic regulation of each transition $i$ as an inhomogeneous Poisson process with a transition rate $r_i(t)$, modulated by the signals that T-cells receive at a given time, 
\begin{align}
\label{eq:rate_regulation}
  r_i(t) =& r^\text{max}\ g_i\left(\sigma_\Ag,\sigma_\infc, \sigma_\res;\vec{\psi}_i, \ell_i\right),
\end{align}
where the maximal rate $r^\text{max}$ is bounded by the cell-cycle time ($\sim 5-6$ hrs~\cite{Van_Stipdonk2001,Jenkins2008-if,Kretschmer2020-ad}), and $g$ is a monotonic {\em response function} chosen from the modified Monod-Wyman-Changeux (Hill-type) family~\cite{Walczak2010-gs, De_Ronde2012-py}:
\begin{align}
\begin{split}
&g_i(\sigma_\Ag,\sigma_\infc, \sigma_\res;\vec{\psi}_i, \ell_i) = \frac{1}{1 + \exp\left[-L_i(\sigma_\Ag,\sigma_\infc, \sigma_\res;\vec{\psi}_i, \ell_i)\right]}, \\
&\text{with} \\
&L_i(\sigma_\Ag,\sigma_\infc, \sigma_\res;\vec{\psi}_i, \ell_i)= \ell_{i} + \sum_{\alpha \in\{\Ag,\infc,\res\}}\psi^\alpha_i \log\left[1 + {\sigma_\alpha}\right].
\end{split}
\label{eq:monod}
\end{align}

The parameter $\ell_i$ sets the baseline propensity of transition $i$: in the absence of external signals ($\vec\sigma=0$), the regulatory factor evaluates to $g_{0,i}= (1+\exp(-\ell_{i}))^{-1}$. As $\ell_{i}$ increases, this baseline transition rate approaches its ceiling $r^\text{max}$. The weights $\{\psi_i^{Ag}, \psi_i^\infc, \psi_i^\res\}$ encode the sensitivity and polarity (positive / negative = up /down-regulation) of response to the antigen ($\sigma_\Ag$) and harm-induced signals ($\sigma_\infc$, and $\sigma_\res$). Each logarithmic term $\log[1 + \sigma_\alpha]$ becomes significant only once its corresponding cue $\sigma_\alpha \sim 1$. This formulation abstracts away many layers of biochemical detail, while implementing the fold-change response to stimuli observed in many biological circuits~\cite{Hart2014-ms, Oyler-Yaniv2017-yr, Achar2022-fl, Voisinne2015-ed, Adler2014-lz}. 

The transition rates governing each of the four modules are parameterized by a response function (eq.~\ref{eq:monod}), whose parameters we collect into a {\em design vector} $\vec \theta$ (up to $4\times 4$ degrees of freedom). These functions are expressive enough to include many three-signal logic circuits \cite{De_Ronde2012-py}. Introducing explicit non-linear interactions between signals would broaden the expressivity of the model but at the cost of greater complexity and limited interpretability, an extension that we leave for future work.

\section{Numerical implementation of stochastic T-cell response}
\label{section: simulations}
Parameters governing our model and their biological ranges are given in Table \ref{table:parameters}, and the pseudo-code describing our numerical approach is provided in Algorithms~\ref{alg:cd8}-\ref{alg:cd8-stoch}. Simulating our lineage-based model of T-cell response with a fully agent-based Gillespie algorithm with time-dependent rates is computationally challenging given the large number of agents ($\sim 10^7$) that must be tracked and updated \cite{Gillespie2007-ex}. We employ two approximations. First, we model the dynamics of the susceptible and infected cells, and harm signals using a piecewise, deterministic-stochastic system of ODEs, implemented numerically with first-order difference equations, and specified in Algorithms~\ref{alg:cd8}-\ref{alg:cd8-stoch}.
Second, we model the cellular immune response as a time-inhomogeneous Markov process implemented with a tau-leaping scheme~\cite{Cao2006}.

As long as the simulation time-step $\Delta t$ is chosen carefully, this approach provides a good compromise between computational speed and accuracy. In our case, it is sufficient to choose $\Delta t$ such that transition rates do not change much during this small time interval, i.e, $r(t + \Delta t) \approx r(t)$. Thus, we seek to bound
\begin{align}
  \frac{\d r}{\d t} = r^\text{max} \sum_{\sigma \in \{\Ag, \infc,\res\}}\frac{\d g}{\d\sigma}\frac{\d\sigma}{\d t},
\end{align}
where
\begin{align}
\begin{split}
  \frac{\d g}{\d\sigma} = \frac{\psi_\sigma}{1 + \sigma}\frac{1}{1 + e^{-L(\sigma)}}\frac{e^{-L(\sigma)}}{1 + e^{-L(\sigma)}}\quad\Longrightarrow\quad
  \left|\frac{\d g}{\d\sigma}\right|\leq \frac{|\psi_\sigma|}{1 + \sigma}\frac{1}{2^2}.
\end{split}
\end{align}
The upper bound is observed by noting that the product of the last two terms is equivalent to $x(1-x)$ for $x \in (0,1)$. Next, using equations from Section \ref{section: signal dynamics}, we compute
\begin{align}
\begin{split}
  \frac{\d\sigma_\Ag}{\d t} &= K_I^{-1}\frac{\d I}{\d t} = \sigma_\Ag \times\left(b_I S - d_I - d_E {C_I}/{I}\right),\\
  \frac{\d\sigma_\infc}{\d t} &= K_H^{-1} h_\infc - d_H \sigma_\infc = K_H^{-1} d_I I - d_H\sigma_\infc,\\
  \frac{\d\sigma_\res}{\d t} &= K_H^{-1}h_\res - d_H\sigma_\res = K_H^{-1}d_E C_S - d_H\sigma_\res,
\end{split}
\end{align}
from which we conclude
\begin{align}
\begin{split}
\left|\frac{\d \sigma_\Ag}{\d t}\right | &\leq \sigma_\Ag \times(b_I S+d_I + d_EC_I/I),\\
\left |\frac{\d\sigma_\infc}{\d t}\right| &\leq K_H^{-1}d_I I(t) + d_H\sigma_\infc \sim 2 d_H\sigma_\infc,\\
\left |\frac{\d\sigma_\res}{\d t}\right | &\leq K_H^{-1}d_E C_S(t) + d_H\sigma_\res \sim 2 d_H\sigma_\res
\end{split}
\end{align}
where we rely on the homeostatic forms of the dynamical equations for $\sigma_\infc$ and $\sigma_\res$ to argue that typically, $K_H^{-1}d_I I(t) \sim d_H\sigma_\infc(t)$ and $K_H^{-1}d_E C_S(t) \sim d_H\sigma_\res(t)$. Thus, we see that typically,
\begin{align}
\begin{split}
\left|\frac{\d r}{\d t}\right| &\leq r^\text{max}\frac{1}{4}\left( \frac{\sigma_\Ag }{1 + \sigma_\Ag}|\psi_\Ag|(b_I S+d_I + d_EC_I/I) + \frac{\sigma_\infc }{1 + \sigma_\infc}2d_H|\psi_\infc| + \frac{\sigma_\res }{1 + \sigma_\res}2d_H|\psi_\res|\right)\\
&\leq r^\text{max}\frac{1}{4}\left( |\psi_\Ag|(b_I S+d_I + d_EC_I/I) + 2d_H|\psi_\infc| + 2d_H|\psi_\res|\right). 
\end{split}
\label{eq:drdt_abs}
\end{align}
With parameters chosen according to Table \ref{table:parameters}, each term in eq.~\ref{eq:drdt_abs} is at most on the order of $\mathcal{O}(10)$. Thus, taking $\Delta t=\mathcal{O}(10^{-2})$ guarantees that transition rates {typically} change only minimally over each time interval, 
\EQ
|r( t +\Delta t) - r(t)| \approx \left|\frac{\d r}{\d t}\right|\Delta t < \mathcal{O}(10^{-1}).
\EE

\clearpage{}

\setcounter{algorithm}{0}
\renewcommand{\thealgorithm}{S\arabic{algorithm}}

\begin{algorithm}[H]
\caption{{\bf Piecewise, deterministic-stochastic simulation of CD8$^+$ T-cell response to an immune challenge.} Simulating the trajectories of susceptible and infected cells, and T-cells.}
\label{alg:cd8}

\begin{algorithmic}[1]
\State \textbf{Input:} {$\vec\psi,\vec\ell, b_I, K_I, K_S, K_H, r^{\max}, r^\text{max}_\text{bind}, r_{N^* \to 2N^*}, d_I, d_E, N_\text{lin.}, N^*_{\max}, E_{\max}, \Delta t, T$}

 \State \textbf{Initialize:} $S \gets S_{\max} - I_0,\;
 I \gets I_0,\; \vec N \gets \vec 1,\; \vec\sigma \gets \vec 0,\; \vec E \gets \vec 0,\; \vec e_{\text{stop prolif.}}\gets 0; \vec B_{\text{differentiate}} \gets 0,\; \vec B_{\text{proliferate}} \gets 0$
 \Statex

 \For{$t = 0,\dots, T-1$ \textbf{ step } $\Delta t$}
 \Statex
  \Statex \Comment{\textit{(i) deterministic update of immune challenge dynamics, using} \textsc{Infection} (Algorithm~\ref{alg:cd8-det})}\linebreak
  \State $(\vec r, h_\infc, h_{E,I}, h_{E,S}) \gets \Call{Infection}{S, I, \vec E, \vec\sigma;\ I_0, N_\text{lin.}, \vec\psi, \vec\ell, b_I, K_I, K_S, K_H, r^{\max}, r^\text{max}_\text{bind}, r_{N^* \to 2N^*},d_I, d_E, \Delta t}$

\Statex \Comment{---- shorthand notation: $\vec r = \{r_\text{bind}, r_{N^* \to 2N^*},r_{N \to N^*}, r_{N^* \to E}, r_{E \to 2E}, r_{E \to \emptyset}\}$----}
  \State
  \State \Comment{\textit{(ii) stochastic T-cell response for all lineages $l = 1,\dots,N_\text{lin.}$, using } \textsc{Response} (Algorithm~\ref{alg:cd8-stoch})}\linebreak
  \For{$l = 1,\dots, N_\text{lin.}$}
  \State
   \State $(N^{(l)}, [\mathrm{APC}\!\cdot\!N]^{(l)}, N^{*,(l)}, E^{(l)}, M^{(l)}, e_{\text{stop prolif.}}^{(l)}, B_{\text{differentiate}}^{(l)}, B_{\text{proliferate}}^{(l)})$
   \State \hspace{10.9em} $\gets \Call{Response}{N, N^*, [APC\cdot N], E, M, \vec r ,N^*_{\max}, E_{\max},e_\text{stop prolif.}, B_\text{differentiate}, B_\text{proliferate};~\Delta t}$  
  \EndFor
  \State

  \State \Comment{\textit{(iii) record trajectories of observables at time $t$}}
  \State Record $S, I, \vec N, \vec N^{*}, \vec E, \vec M, h_\infc, h_{E,I}, h_{E,S}$ into $[\,\cdot\,]_t$
 \EndFor
 \State
 \State \Return{$([S]_t,[I]_t,[\vec N]_t,[\vec N^{*}]_t,[\vec E]_t,[\vec M]_t,[h_\infc]_t,[h_{E,I}]_t,[h_{E,S}]_t)$}
\end{algorithmic}
\end{algorithm}

\begin{algorithm}[H]
\caption{{\bf Deterministic update of immune challenge dynamics.} Infection variables ($S, I$), harms ($ h_{\text{inf}},\ h_{E,I},\ h_{E,S}$), signals ($\sigma_\Ag, \sigma_\infc,\sigma_\res $), and transition rates ($r_\text{bind}, r_{N^* \to 2N^*}, \ r_{N \to N^*},\ r_{N^* \to E},\ r_{E \to 2E},\ r_{E \to \emptyset}$) are computed.}

\label{alg:cd8-det}
\begin{algorithmic}[1]
\Statex
\Function{Infection}{$S, I, \vec E, \vec\sigma;\ I_0, N_\text{lin.}, \vec\psi, \vec\ell,\ b_I, K_I, K_S, K_H,\ r^{\max}, r^\text{max}_\text{bind},r_{N^* \to 2N^*}, \ d_I, d_E, d_H,\ \Delta t$}
\Statex
 \Statex \Comment{\textit{(i) compute the number of T-cell bound infected and susceptible cells $C_I$ and $C_S$}}
 \State
 \[
 \begin{aligned}
  C_I \gets \frac{I \sum_{l = 1}^{N_\text{lin.}} E^{(l)}}{\sum_{l = 1}^{N_\text{lin.}} E^{(l)} + I + K_I
   + \dfrac{\sum_{l = 1}^{N_\text{lin.}} E^{(l)} + K_I}{\sum_{l = 1}^{N_\text{lin.}} E^{(l)} + K_S}\, S}\quad,\quad 
  C_S \gets \frac{S \sum_{l = 1}^{N_\text{lin.}} E^{(l)}}{\sum_{l = 1}^{N_\text{lin.}} E^{(l)} + S + K_S
   + \dfrac{\sum_{l = 1}^{N_\text{lin.}} E^{(l)} + K_S}{\sum_{l = 1}^{N_\text{lin.}} E^{(l)} + K_I}\, I}.
 \end{aligned}
 \]
 \Statex

 \Statex \Comment{\textit{(ii) compute the infection harm $h_{\text{inf}}$, and response harms $h_{E,I}$, $h_{E,S}$}}
 \State
 \[
 h_{\text{inf}} \gets d_I I\,\Delta t, \qquad
 h_{E,I} \gets d_E C_I\,\Delta t, \qquad
 h_{E,S} \gets d_E C_S\,\Delta t
 \]
 \Statex

 \Statex \Comment{\textit{(iii) update the number of susceptible and infected cells}}
 \State\[{S \gets S + (- \mathbf{1}_{I \geq I(0)}b_I I S - d_E C_{S})\Delta t}\]
\[{I \gets I + (\mathbf{1}_{I \geq I(0)}b_I I S - d_I I - d_E C_I)\Delta t}\]
\Statex

 \Statex \Comment{\textit{(iv) update the three signals $\vec \sigma$}}
\State \[\sigma_{\Ag} \gets K_I^{-1} I \]
\[\sigma_{\infc} \gets \sigma_{\infc} + (K_H^{-1}h_{\infc} - d_H\sigma_{\infc})\Delta t\]
\[\sigma_{\res} \gets \sigma_{\res} + (K_H^{-1}h_{E,I} + K_H^{-1}h_{E,S} - d_H\sigma_{\res})\Delta t\]
\Statex

 \Statex \Comment{\textit{(v) update the recruitment rate of na\"ive cells to bind to APC's $r_\text{bind}$ and the transition rates $r_x$ for all the processes $ x \in \{N\to N^*, N^*\to E, E\to 2E , E\to \emptyset\} \}$}}
 \Statex

\State\[ r_\text{bind} \gets \frac{r^\text{max}_\text{bind}}{1+ \exp[-\log(1+\sigma_\infc)]}\]

\For{$x \in \{N\to N^*, N^*\to E, E\to 2E, E\to \emptyset\}$}
  \begin{align*}
  L_x \gets \ell_x + \sum_{\alpha:\{\Ag,\infc,\res\}} \psi_\alpha \log (1+ \sigma_\alpha),\quad g_x \gets (1+\exp[-L_x])^{-1}, \quad r_{x} \gets r^\text{max} g_x
  \end{align*}
\EndFor
\Statex

 \State \Return $\vec r =\{r_\text{bind},r_{N^* \to 2N^*}, \ r_{N \to N^*},\ r_{N^* \to E},\ r_{E \to 2E},\ r_{E \to \emptyset}\},\ h_{\text{inf}},\ h_{E,I},\ h_{E,S},\ S,\ I$

\EndFunction

\end{algorithmic}
\end{algorithm}

\begin{algorithm}[H]
\caption{{\bf Stochastic response of a T-cell lineage.} Stochastic recruitment, activation, proliferation, differentiation, and death of CD$8^+$ T-cells are simulated for each lineage (lineage superscripts dropped). The number of cells from different states ($N, [\text{APC}\cdot N], N^*, E, M$), and the variables determining the lineage's fate ($ e_\text{stop prolif.}, B_\text{differentiate},B_\text{proliferate}$) are evaluated.}
\label{alg:cd8-stoch}
\begin{algorithmic}[1]
\Statex
\Function{Response}{$N, N^*, [APC\cdot N], E, M, \vec r, N^*_{\max}, E_{\max},e_\text{stop prolif.},B_\text{differentiate}, B_\text{proliferate};~\Delta t$}
\Statex \Comment{---- shorthand notation: $\vec r = \{r_\text{bind}, r_{N \to N^*}, r_{N^* \to 2N^*}, r_{N^* \to E}, r_{E \to 2E}, r_{E \to \emptyset}\}$----}
\Statex
 \State \textbf{Initialize:} ($n_\text{bound}, n_\text{activated},n^*_\text{proliferate}, ~n^*_\text{differentiate}, ~e_\text{differentiate}, ~m_\text{differentiate}, ~e_\text{contract} )\gets 0$
 \Statex
\Statex\Comment{\textit{(i) recruitment of na\"ive cells to interact with APC for activation}}
 \If{$N > 0$}
  \State $n_\text{bound} \gets \text{Binom}(N,\, r_{\text{bind}} \Delta t)$ \Comment{---- binomial sampling of na\"ive cells to bind to APC, rate $r_\text{bind}$----}
   \Statex
  \State $n_\text{activated} \gets \text{Binom}([\text{APC}\cdot N],\, r_{N\to N^*}\Delta t)$ \Comment{---- binomial sampling to activate APC-bound na\"ives, rate $r_{N\to N^*}$----}
   \Statex
 \EndIf
 \Statex
\Statex\Comment{\textit{(ii) burst-like proliferation of activated na\"ives $N^*$, differentiation to effectors $E$, or memory $M$}}
 \Statex
 \If{$0<N^{*} < N^*_{\max}$}

  \State $n_\text{proliferate}^* \gets \text{Binom}(N^*,\, r_{N^* \to 2N^*}\Delta t)$ \Comment{---- binomial sampling to replicate $N^*$, rate $r_{N^* \to 2N^*}$ (initial burst)----}
   \Statex
  \State $B_\text{differentiate} \gets B_\text{differentiate} + r_{N^*\to E}\Delta t$ \Comment{---- updating likelihood of $N^*$ differentiating to $E$ by time $t$ ----}
   \Statex
  
 \ElsIf{$N^{*} \geq N^*_{\max}$}

  \State $n_\text{differentiate}^* \gets N^{*}$ \Comment{---- the number of activated na\"ives that can differentiate----}
  \Statex
  \State $e_\text{differentiate} \gets \text{Binom}\left(N^{*},\, 1 - {\rm e}^{-B_\text{differentiate}}\right)$ \Comment{---- binomial sampling to differentiate $N^*$ to $E$ during burst----}
  \Statex
  \State $m_\text{differentiate} \gets N^{*} - e_\text{differentiate}$ \Comment{---- number of activated cells differentiating to memory----}
  \Statex
  \State $e_\text{stop prolif.} \gets \min(e_\text{differentiate}, N^*_\text{max})\times \frac{E_\text{max}}{ N^*_\text{max}}$ \Comment{---- maximum number of allowed effectors in the lineage----}
 \EndIf

 \Statex
\Statex\Comment{\textit{(iii) contraction and proliferation of effectors with rates $r_{E\to \emptyset}$, and $r_{E\to 2E}$, respectively}}

 \Statex
 \State $e_\text{contract} \gets \text{Binom}(E,\, r_{E\to \emptyset}\Delta t)$ \Comment{---- binomial sampling to remove $E$----}
  \Statex

 \If{$B_\text{proliferate} < e_\text{stop prolif.}$}

  \State $e_{\text{proliferate}} \gets \text{Binom}(E,\, \,r_{E \to 2E} \Delta t)$ \Comment{---- binomial sampling to replicate $E$----}
     \Statex
  \State $B_\text{proliferate} \gets B_\text{proliferate} + e_{\text{proliferate}}$ \Comment{---- running count of the number of effector cell divisions----}
 \EndIf
\Statex
 \Statex \Comment{\textit{(iv) update the populations of na\"ive $N$, na\"ive bound to APC $[\text{APC}\cdot N]$, activated $N^*$, effector $E$, and memory $M$ cells for a given T-cell lineage responding to infection:}}
 \Statex
 \State $N \gets N- n_\text{bound}$
 \State $[\text{APC}\cdot N] \gets [\text{APC}\cdot N] + n_\text{bound} - n_\text{activated}$
 \State $N^* \gets N^* + n_\text{activated} + n_\text{proliferate}^* - n_\text{differentiate}^*$
 \State $E \gets E + e_\text{proliferate} - e_\text{contract} + e_\text{differentiate}$
 \State $M  \gets M + m_\text{differentiate}$

\Statex
 \State \Return $N, [\text{APC}\cdot N], N^*, E, M, e_\text{stop prolif.}, B_\text{differentiate},B_\text{proliferate}$
 \EndFunction
\end{algorithmic}
\end{algorithm}

\clearpage

\section{Optimization of CD$8^+$ T-cell  response to pathogens}

\subsection*{Design architecture of T-cell programs for immune response} 

Infection clearance results from coordination of the effector trajectory $E(t)$ in response to an infection $I(t)$. Physically, this corresponds to controlling/regulating the rates, $\vec r(t)$ of cell state transitions, divisions and deaths, potentially varying $4\times 4= 16$ module-specific sensitivity and baseline parameters, 
\begin{align}
\begin{split}
\text{activation:}&~~ \psi_{N \to N^*}^{\sigma_\Ag},~
\psi_{N \to N^*}^{\sigma_\infc},~
\psi_{N \to N^*}^{\sigma_\res},~
\ell_{N \to N^*} \in \mathbb{R}, \\
\text{differentiation:}&~~ \psi_{N^* \to E}^{\sigma_\Ag},~
\psi_{N^* \to E}^{\sigma_\infc},~
\psi_{N^* \to E}^{\sigma_\res},~
\ell_{N^* \to E} \in \mathbb{R},\\
\text{proliferation:}&~~\psi_{E \to 2E}^{\sigma_\Ag},~
\psi_{E \to 2E}^{\sigma_\infc},~
\psi_{E \to 2E}^{\sigma_\res},~
\ell_{E \to 2E} \in \mathbb{R},\\
\text{contraction:}&~~ \psi_{E \to \emptyset}^{\sigma_\Ag},~
\psi_{E \to \emptyset}^{\sigma_\infc},~
\psi_{E \to \emptyset}^{\sigma_\res},~
\ell_{E \to \emptyset} \in \mathbb{R},
\end{split}
\end{align}
to generate effector response trajectories. 

Accounting for shared regulatory programs and controlling the combinatorial complexity of our model parameter space, we impose two simplifications. First, activation and proliferation are both MYC-driven~\cite{Heinzel2018-wr}; we therefore constrain the modulatory factors, $g_{N\to N^*}$ and $g_{E\to 2E}$, to share parameter values. Note, however, that their realized rates will differ because signals vary in time, as cells transition between naive and effector states. Second, we make the assumption that extracellular cues converge on coherent effector programs, and thus enforce a common set of signal sensitivities across modules. With the sign convention that positive sensitivities promote effector production, for each signal $\sigma$, we set 
\[
\psi_\sigma \equiv \psi^\sigma_{N\to N^*} = \psi^\sigma_{E\to 2E} = \psi^\sigma_{N^*\to E} = -\psi^\sigma_{E\to\emptyset}.
\]
Each module $i$ nonetheless retains its own baseline parameter $\ell_i$, or equivalently the baseline rate $g_{0;i} = (1+\exp(-\ell))^{-1}$. With this simplification we reduce the number of degrees of freedom in our model from $16$ to six (three sensitivity and three baseline rates),

\begin{align}
\begin{split}
\text{activation:}&~~ \psi_{N \to N^*}^{\sigma_\Ag} = \psi_{\sigma_\Ag},~
\psi_{N \to N^*}^{\sigma_\infc} = \psi_{\sigma_\infc},~
\psi_{N \to N^*}^{\sigma_\res} = \psi_{\sigma_\res},~
\ell_{N \to N^*} \in \mathbb{R}, \\
\text{differentiation:}&~~ \psi_{N^* \to E}^{\sigma_\Ag} = \psi_{\sigma_\Ag},~
\psi_{N^* \to E}^{\sigma_\infc} = \psi_{\sigma_\infc},~
\psi_{N^* \to E}^{\sigma_\res} = \psi_{\sigma_\res},~
\ell_{N^* \to E} \in \mathbb{R},\\
\text{proliferation:}&~~\psi_{E \to 2E}^{\sigma_\Ag} = \psi_{\sigma_\Ag},~
\psi_{E \to 2E}^{\sigma_\infc} = \psi_{\sigma_\infc},~
\psi_{E \to 2E}^{\sigma_\res} = \psi_{\sigma_\res},~
\ell_{E \to 2E} = \ell_{N \to N^*}\in \mathbb{R},\\
\text{contraction:}&~~ \psi_{E \to \emptyset}^{\sigma_\Ag} = -\psi_{\sigma_\Ag},~
\psi_{E \to \emptyset}^{\sigma_\infc} = -\psi_{\sigma_\infc},~
\psi_{E \to \emptyset}^{\sigma_\res} = -\psi_{\sigma_\res},~
\ell_{E \to \emptyset} \in \mathbb{R}.
\end{split}
\end{align}
Note the relationship, $\psi_{E \to \emptyset}^\sigma = - \psi_\sigma$ reflects coordinated signal-regulation of effector-promoting and effector-suppressing transitions. 

The parameters of these constrained response functions define a {\em response design} $\vec \theta$ as
\begin{equation}
\label{eq:theta}
\vec \theta = (\underbrace{\psi_{\Ag},\psi_{\infc},\psi_{\res}}_\text{signal feedback}, \underbrace{\ell_{N\to N^*} = \ell_{E\to 2E}, \ell_{N^*\to E},\ell_{E\to \emptyset}}_\text{module-specific baselines}).
\end{equation}
For each transition $i$, we choose the baseline response parameter, $\ell_i$, such that $|\ell_{i}| \leq \ell^{\text{max}}$ with
\begin{align*}
r_i(\ell^{\text{max}}) \sim r^\text{max}, ~r_i(-\ell^{\text{max}}) \sim T^{-1},
\end{align*}
where $T$ is the duration of the simulation. This range is set to be wide enough that at the lower baseline rate ($-\ell^{\text{max}}$) transitions are seldom observed during our simulation time window, whereas at the upper baseline rate ($\ell^{\text{max}}$) they occur near the maximum rate. Here, we take $\ell^\text{max} = 3.0$, so $r_i(\ell^{\text{max}} = 3) \approx 0.95\times r^\text{max}$. For the signal-sensitivity parameter $\psi_\sigma$, we require that under weak stimulation ($\sigma=1$) the signal term in the response function (eq.~\ref{eq:monod}) is below the threshold of response, set by the maximal baseline term, i.e., ${\psi_\sigma^\text{max}\cdot \log(1 + 1) < \ell^{\text{max}}}$, and choose $\psi_\sigma^\text{max} = 3.0$.

\subsection*{Optimizing the performance of response designs}

\paragraph{Optimal design for an immune challenge.} We define high performing designs $\vec \theta$ as those that tend to minimize the harm from an immune challenge. For a challenge characterized by immunogenicity $\kappa$ and basic reproduction number $R_0$, we define the optimal design as,
\begin{equation}
 \vec \theta^{\rm min}(\kappa, R_0) = \underset{\vec\theta}{\arg\,{\rm min}}~ H(\vec \theta).
 \label{eq: single optimization}
\end{equation}
 where the harm $H(\vec \theta)$ is computed by eq.~\ref{eq: clear_tox} for growing (acute) infections, and by eq.~\ref{eq:cancer_harm_total} for slow-growing cancer tumors.

To do this optimization for a fixed pathogenic threat $(\kappa, R_0)$, we perform a uniform grid search over $\vec\theta~\in~[-\psi^\text{max},\psi^\text{max}]^3\times[-\ell^\text{max}, \ell^\text{max}]^3$. We use a Cartesian grid with $m=13$ equally spaced points per coordinate, yielding a total of $13^6 \approx 4.8 \times 10^6$ candidate designs. For each $\vec \theta$, we simulate the immune response dynamics to compute the associated harm and identify the design that minimizes harm $\vec \theta^\text{min}(\kappa, R_0)$.\\

\noindent {\bf Multi-objective design optimization across pathogen archetypes.} To find designs that can perform well across pathogen archetypes, we pose a principled multi-objective problem for the immune program, i.e., choose a design $\vec \theta$ that minimizes the weighted sum of harms across pathogen archetypes:
\begin{equation}
 \vec \theta'(\{\lambda_i\}) = \underset{\vec\theta}{\arg\min} \sum_{i: \text{archetypes}} \lambda_i H^{(i)} (\vec \theta),
 \label{eq: pareto optimization}
\end{equation}
where $\lambda _i\in [0, 1]$ with the constraint $\sum_i \lambda_i=1$. The weights $\{\lambda_i\}$ determine the relative importance of each archetype in determining the optimal design $\vec \theta'$. The solution to this family of problems forms a simplex over which $\lambda_i's$ vary, called the \emph{Pareto front}; see ref.~\cite{Boyd2004-qp} for an extensive discussion on multi-objective optimization, and refs.~\cite{Kocillari2018-pf, Shoval2012-ln} for successful application of these methods in different biological systems.

Similar to the case of optimization for a single immune challenge, we estimate the Pareto front by sampling designs on a grid and identifying the design that minimizes the weighted harm $\sum_{i: \text{archetypes}} \lambda_i H^{(i)} (\vec \theta)$ for a given instance of $\{\lambda_i\}$.\\

\paragraph{Finding the elbow of the two-dimensional Pareto curve.} Even though optimization can be done simultaneously over all three archetypes, we examine trade-offs between pairs of archetypes $A$ and $B$, generating one-dimensional Pareto fronts, with single independent weights $\lambda = \lambda_A = 1-\lambda_B$.
{The curvature of the Pareto front quantifies the lowest achievable increase in harm from one archetype that we must trade to decrease harm from the other archetype by tuning design parameters. Given a convex Pareto front for archetypes $A$ and $B$, in some regions of the curve (values of $\lambda$) the benefit of decreasing harm from one archetype may be larger than the cost. A common heuristic for multi-objective optimization in engineering problems is to identify the point along the curve, $( H[\vec\theta(\lambda^*)]^{(A)},H[\vec\theta(\lambda^*)]^{(B)})$, called the \emph{elbow} or \emph{knee}, at which no additional benefit can be obtained by tuning parameters \cite{Satopaa2011-el}. To estimate the elbow of the Pareto front, we first interpolate the discretely sampled Pareto points to obtain a continuous curve. This procedure is sensitive to the density of sampled points near the Pareto front and requires fine-tuning. For plots involving archetype I (Fig. \mainfigfour, \ref{otherparetosFigure-ArcI}), we found that subsetting to points where $H(\vec\theta)^{(A)} < 10^{-2}S_\text{max}$, and linearly interpolating this subset of performed best. Finally, we estimated the elbow point $(\lambda^*)$ along the interpolated curve using the Kneedle algorithm \cite{Satopaa2011-el, arvai_2023_8127224}.\\

\paragraph{Identifying the ensemble of near-optimal designs.} Since we are sampling over a finite grid and our simulations are stochastic, we must contend with the likelihood that discretely sampled points will not coincide with the true minimizer in the continuous space of designs and that stochastic effects will add noise to our measurements of harm. To capture this uncertainty, we identify a \emph{near-optimal ensemble} of designs defined as the set of designs $\theta$ such that $H(\vec\theta) \leq H(\vec\theta') + \epsilon$, where $\vec\theta'$ is the sampled minimizer and $0 <\epsilon$ is a threshold. We typically choose $\epsilon= 10^{-2} S_\text{max}$, and compute the mean and variance of designs in this ensemble, as shown in Fig. \mainfigthree D.

In the case of multi-objective optimization, first we estimate the Pareto front for a pair of archetypes $\{A,B\}$. For each point on the Pareto front, $\vec \theta'(\lambda)$ (equation \ref{eq: pareto optimization}), we identify an ensemble of near-Pareto designs
\begin{align}
\Theta(\lambda) = \{\vec\theta| H(\vec\theta)^{(A)} \leq (1 + \alpha) H[\vec\theta'(\lambda)]^{(A)}, H(\vec\theta)^{(B)} \leq (1 + \alpha) H[\vec\theta'(\lambda)]^{(B)} \},
\end{align}
with $\alpha = 5\times10^{-2}$.

\subsection*{Clustering of pathogen archetypes}
For each infection scenario $(\kappa, R_0)$, we identify the ensemble of sampled designs $\vec\theta$ such that $H(\vec\theta) \leq H(\vec\theta^{\mathrm{min}}) + \epsilon$, where $\vec\theta^{\mathrm{min}}$ is the harm-minimizing design for that scenario and $\epsilon = 10^{-2} S_\text{max}$. We call this ensemble $\Theta(\kappa,R_0)$, and compute its mean, $\langle\vec\theta\rangle_{\Theta(\kappa,R_0)}$ (Fig.~\ref{clusteringFigure}). Next, we pool the mean vectors of all infection scenarios and normalize their distribution so each design parameter has zero mean and unit variance. Finally, we use a spectral clustering algorithm implemented in Python \cite{scikit-learn} to cluster scaled mean design parameters to identify distinct pathogen archetypes. We selected the cluster number $n_\text{cluster} = 3$ that maximized the mean silhouette coefficient (implemented in Python \cite{scikit-learn}), 
\[
\sum_{(\kappa,R_0)}[b(\kappa,R_0) - a(\kappa,R_0)]/ \max[a(\kappa,R_0),b(\kappa,R_0)],
\] 
where $a(\kappa,R_0)$ is the mean distance between $\langle\vec\theta\rangle_{\Theta(\kappa,R_0)}$ and every other point in its assigned cluster, and $b(\kappa,R_0)$ is the mean distance between $\langle\vec\theta\rangle_{\Theta(\kappa,R_0)}$ and every other point in its nearest, neighboring cluster. The mean silhouette coefficient measures cohesiveness of clusters and how distinct they are from each other.


\clearpage{}
\setcounter{figure}{0}
\renewcommand{\thefigure}{S\arabic{figure}}

\begin{figure*}[t!]
\centering
\includegraphics[width=0.75\textwidth]{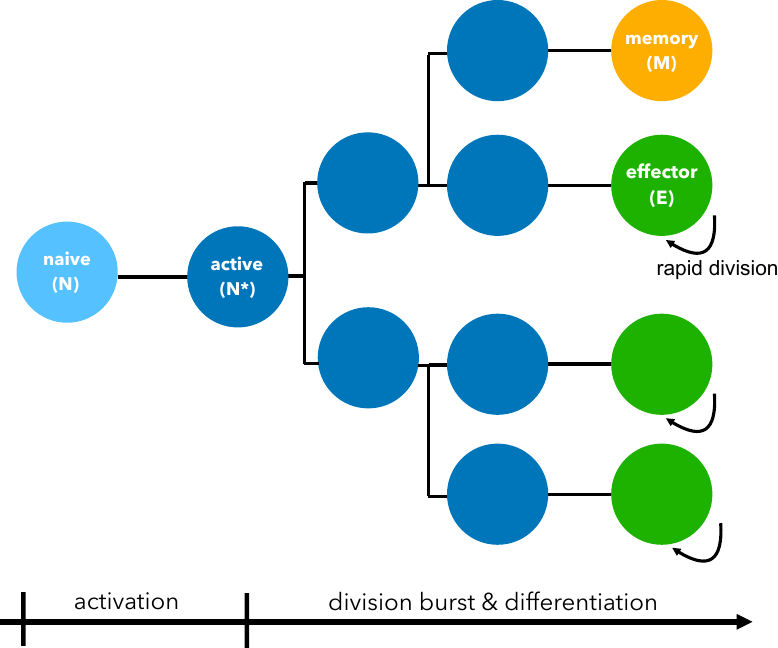}
\caption{{\bf Schematic of modeled early T-cell division bursts and differentiation.
\label{differentiationschematic}
}
}
\end{figure*}

\clearpage{}

\begin{figure*}[t!]
\centering
\includegraphics[width=0.8\textwidth]{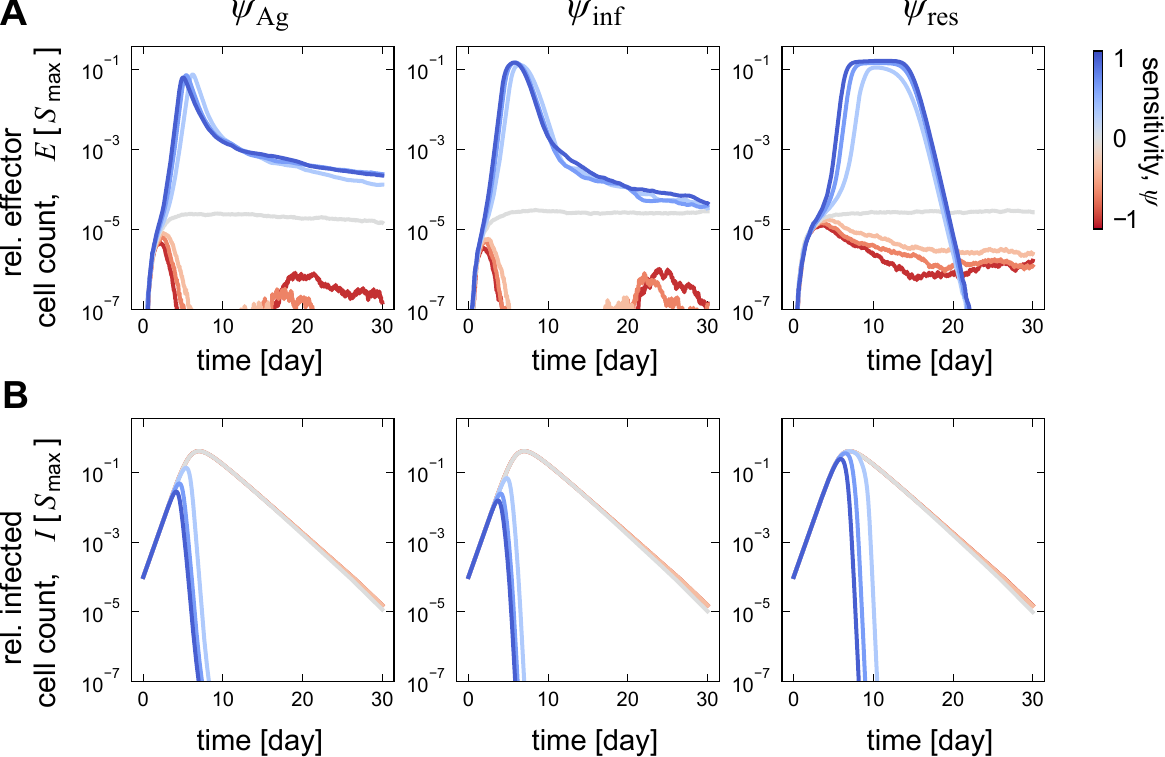}
\caption{{\bf Design parameters shape T-cell response and infection clearance.} 
{\bf (A, B)} Time course of the relative population size of (A) effector cells $E$ and (B) infected cells is shown for designs when varying (color) a single sensitivity parameter $\psi_\sigma$ (indicated at the top of each column), while keeping the rest fixed at: moderate activation / proliferation $g_{0,N\to N^*} = g_{0,E\to 2E} = 0.5 $; moderate differentiation $g_{0, N^*\to E} = 0.5$; moderate contraction $g_{0, E\to \emptyset} = 0.5$; and zero sensitivity to the other signals $\psi_{\Ag} = \psi_{\infc}= \psi_{\res} = 0 $. All population sizes are measured in units of the maximum number of susceptible cells $S_\text{max}$ and sensitivities in units of $\psi_\text{max}$. Other simulation parameters as in Fig.~\mainfigtwo.
\label{DesignResponseClearance}}
\end{figure*}

\clearpage{}

\begin{figure*}[t!]
\centering
\includegraphics[width=0.8\textwidth]{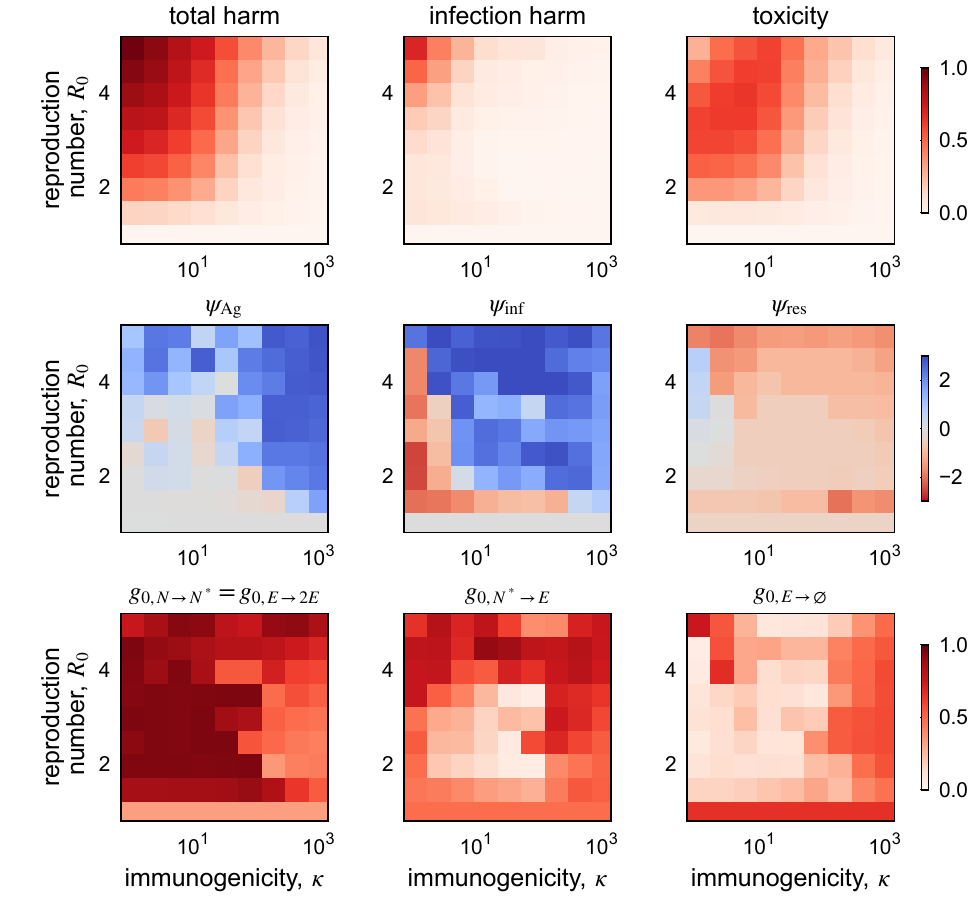}
\caption{{\bf Near-optimal performance and designs for each immune challenge.} Heat maps showing the near-optimal outcomes and design values that achieve them for each infection scenario $(\kappa,R_0)$. The first row displays the average total harm, infection harm and toxicity incurred by the near-optimal designs for each infection scenario. The last two rows display the average near-optimal design values for each scenario, $\langle\vec\theta\rangle_{\Theta(\kappa,R_0)}$. The data in the last two rows is used for clustering infection scenarios in Fig.~\mainfigthree B.
\label{clusteringFigure}
}
\end{figure*}

\clearpage{}
\begin{figure*}[t!]
\centering
\includegraphics[width=0.8\textwidth]{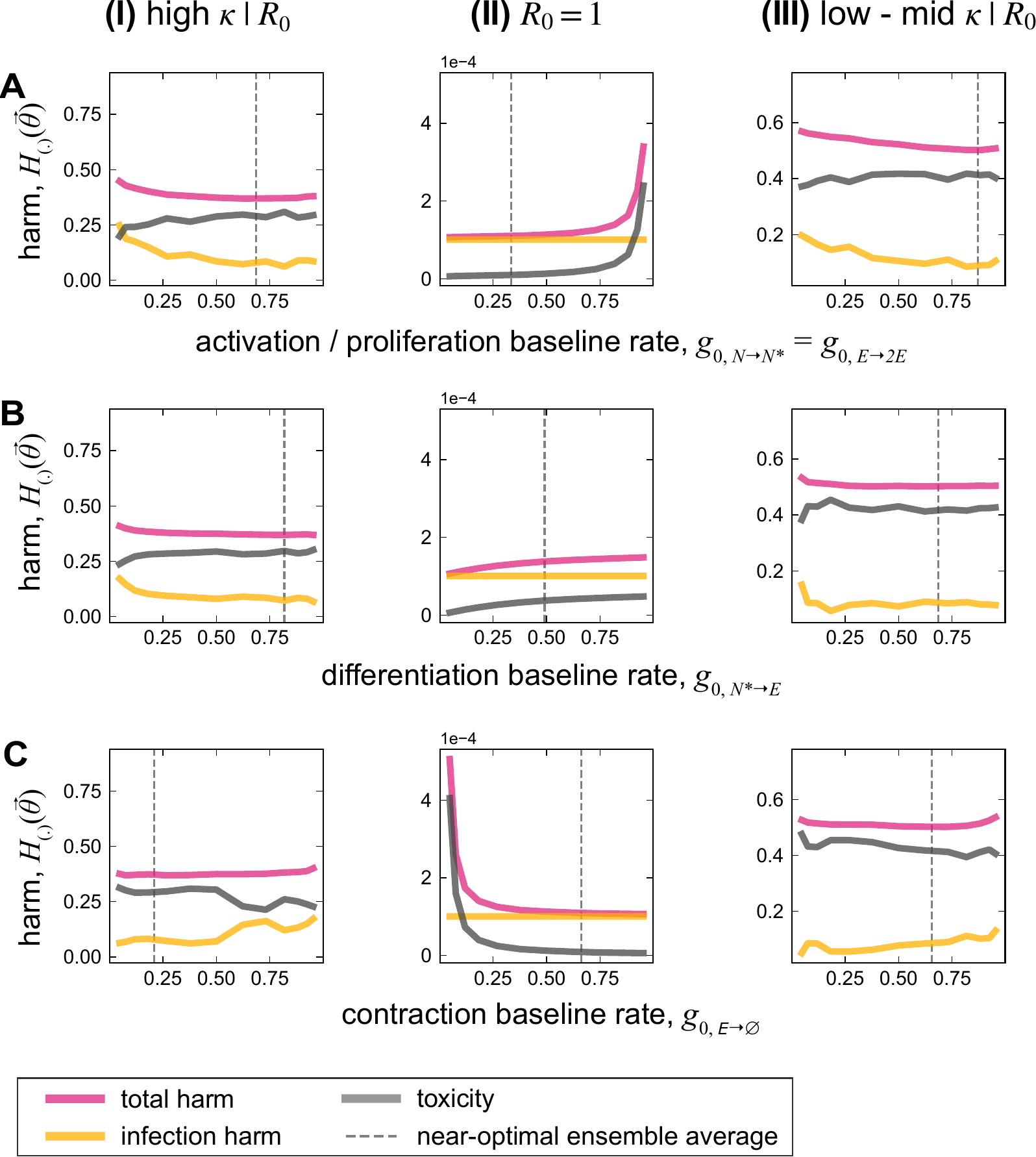}
\caption{ {\bf Dependence of harm on baseline rates in different archetypes.}
Total harm, infection harm, and toxicity (colors) are shown as a function of baseline rates for {\bf(A)} activation / proliferation $g_{0,N\to N^*} =g_{0,E\to 2E}$, {\bf (B)} differentiation $g_{0,N^*\to E}$ and {\bf (C)} contraction $g_{0,E\to \emptyset}$, in different archetypes as indicated above each column. In each panel, the value of the design parameter on the horizontal axis is fixed and total harm is minimized by varying the other design parameters. The dashed line in each panel indicates the near-optimal ensemble average values shown in bar charts in Fig.~\mainfigthree D. Simulation parameters as in Fig.~\mainfigthree.
\label{designablationFigure-baselines}}
\end{figure*}

\clearpage{}
\begin{figure*}[t!]
\centering
\includegraphics[width=0.8\textwidth]{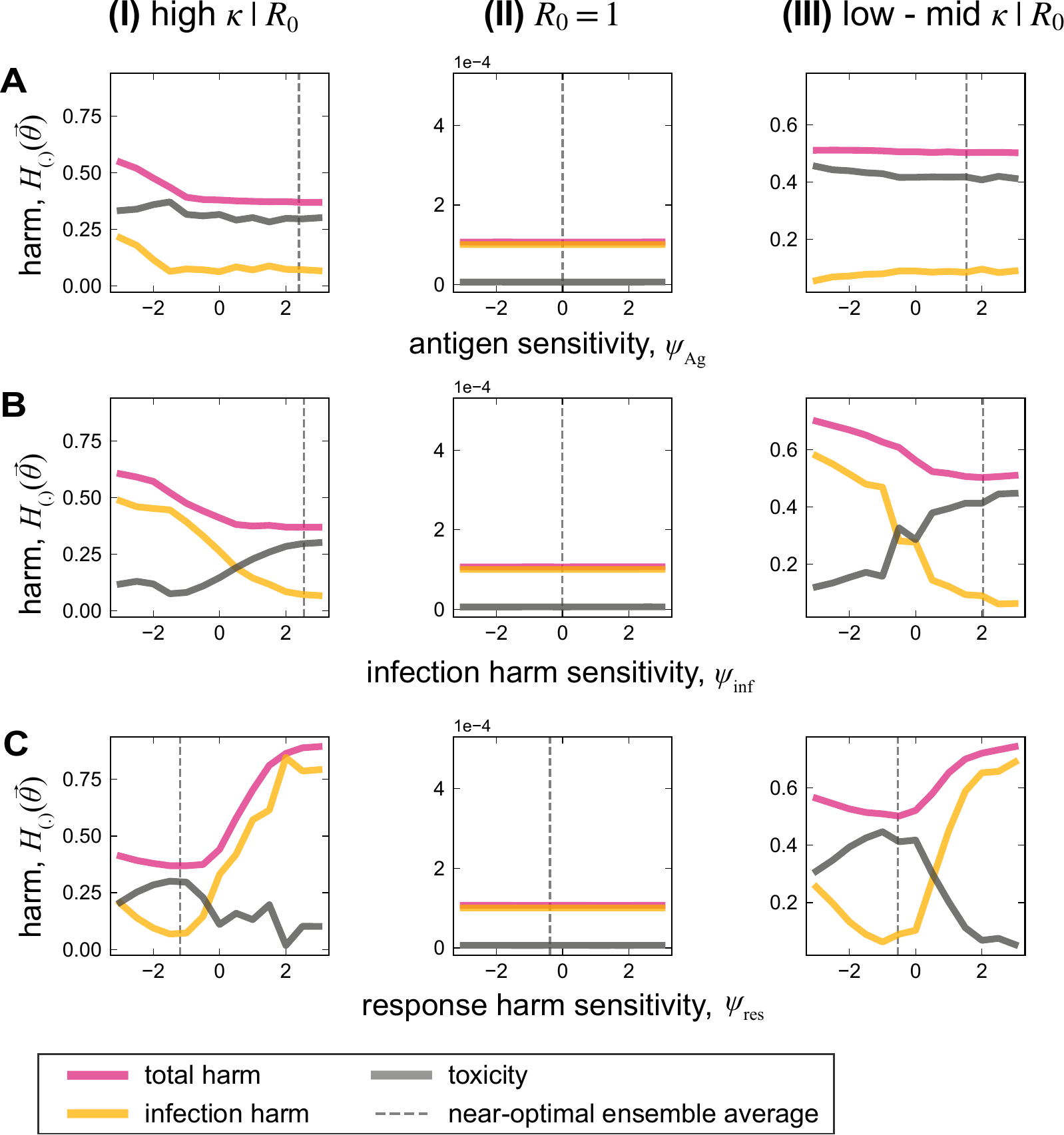}
\caption{ {\bf Dependence of harm on signal sensitivities in different archetypes.}
Similar to Fig.~\ref{designablationFigure-baselines},~but when varying the sensitivities to {\bf(A)} the antigen signal $\psi_\Ag$, {\bf (B)} the infection harm signal $\psi_\infc$, and {\bf (C)} the response harm signal $\psi_\res$.
\label{designablationFigure-psis}}
\end{figure*}

\clearpage{}
\begin{figure*}[t!]
\centering
\includegraphics[width=\textwidth]{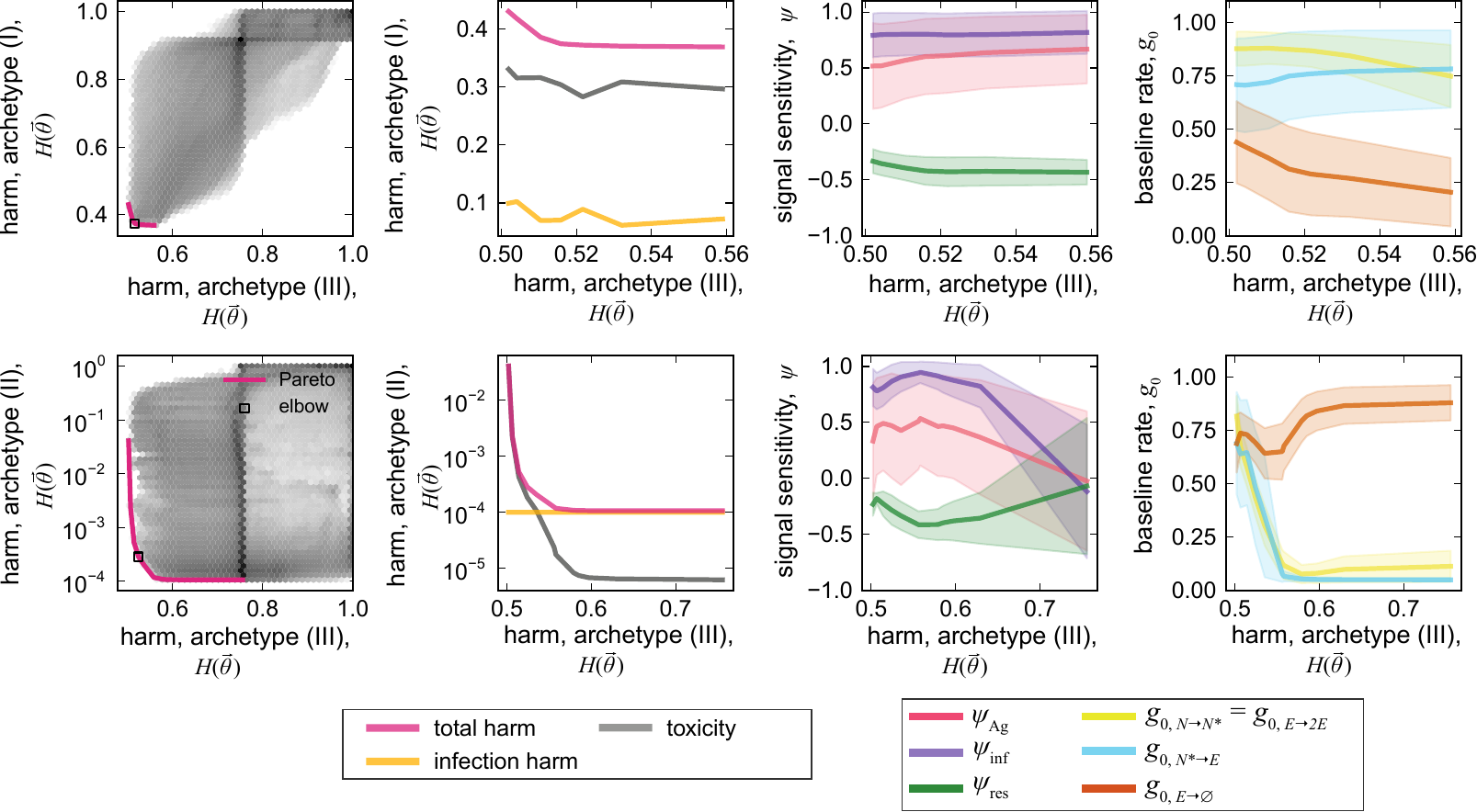}
\caption{{\bf Pareto-optimal design balancing harm from autoimmunity and other pathogen archetypes.}
Similar to Fig.~\mainfigfour, but showing the Pareto optimal designs resulting from joint optimization of {pathogen archetypes I and III}, and pathogen archetypes II and III (rows); joint optimization for archetypes I and II is shown in Fig.~\mainfigfour. Simulation parameters as in Fig.~\mainfigfour.
}\label{otherparetosFigure-ArcI} 
\end{figure*}

\clearpage{}
\begin{figure*}[t!]
\centering
\includegraphics[width=0.55\textwidth]{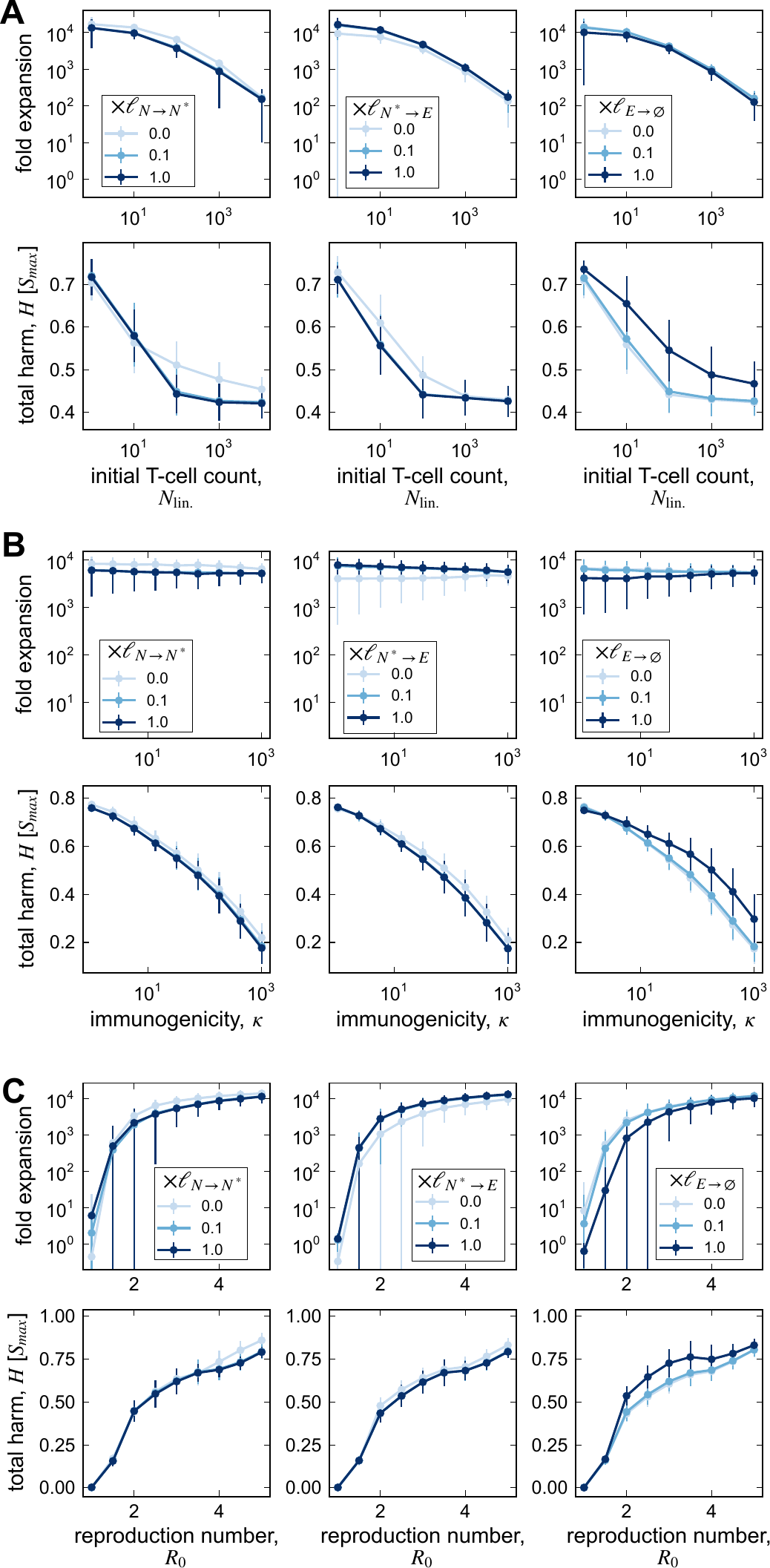}
\caption{{\bf Dependence of T-cell macro-dynamics on baseline rates.} T-cell fold expansion (top) and total harm (bottom) are shown as functions of {\bf (A)} initial T-cell number $N_\text{lin.}$, {\bf (B)} immunogenicity $\kappa$, {\bf (C)} and the intra-host basic reproduction of the infection $R_0$. In each panel, outcomes are shown for the ``elbow'' Pareto-optimal design from Fig.~\mainfigfour B while varying a single baseline rate (colors)---activation/proliferation baseline rate {$g_{0,N\to N^*} =g_{0,E\to 2E}=(1+\exp(-\ell_{N\to N^*}))^{-1} $ (left), differentiation baseline rate $g_{0, N^*\to E}=(1+\exp(-\ell_{N^*\to E}))^{-1}$ (center), or contraction baseline rate $g_{0, E \to \emptyset}=(1+\exp(-\ell_{E\to \emptyset}))^{-1}$ (right)}---with all other parameters held fixed. Trend lines and standard deviation error bars are obtained by averaging over sampled immune challenges (in Fig.~\mainfigthree B) and designs in the near-elbow ensemble. Simulation parameters as in Fig.~\mainfigfive.
}
\label{perturbexpansionFigure-baselines}
\end{figure*}

\clearpage{}
\begin{figure*}[t!]
\centering
\includegraphics[width=0.55
\textwidth]{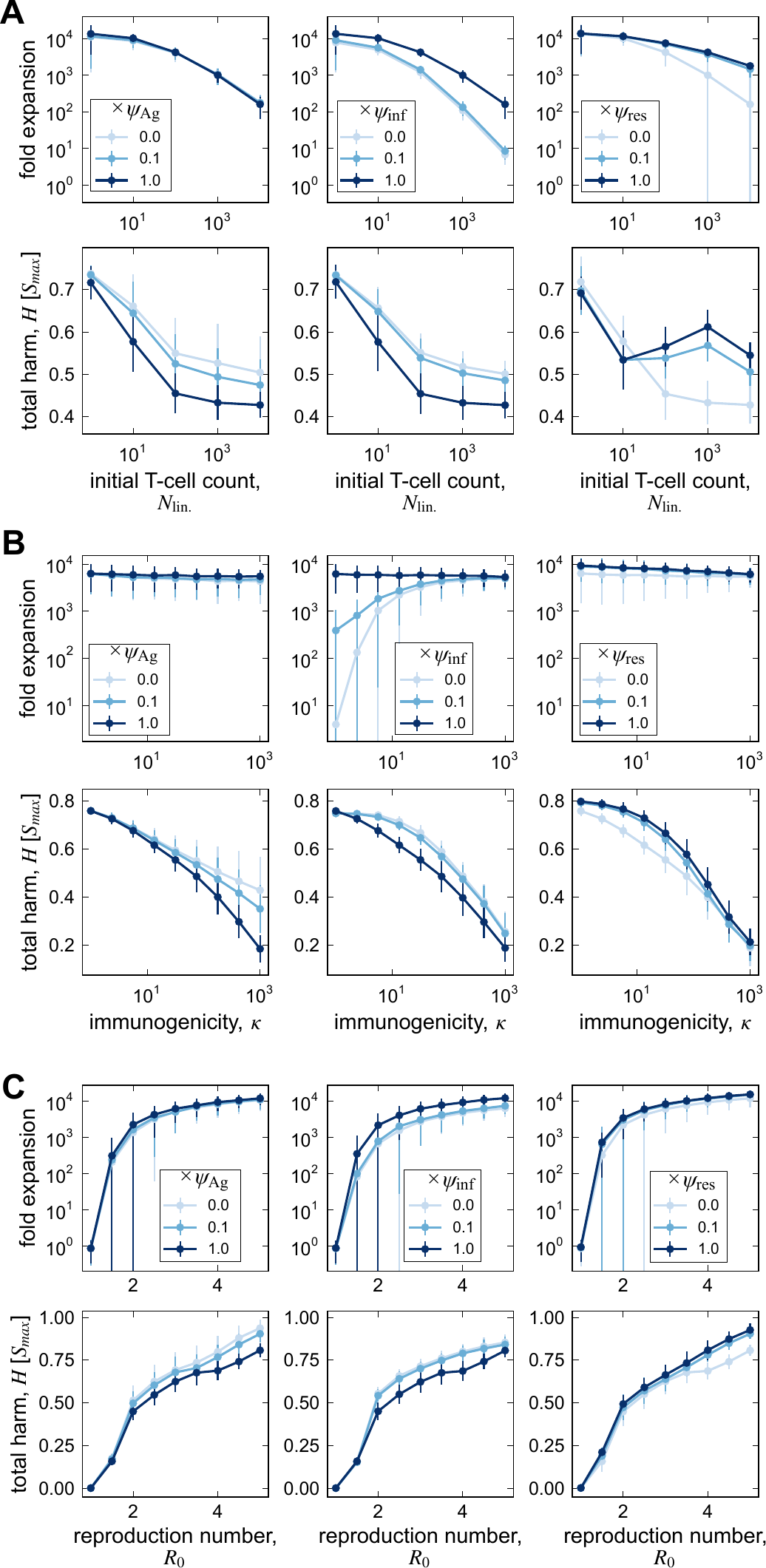}
\caption{{\bf Dependence of T-cell macro-dynamics on signal sensitivities.} T-cell fold expansion (top) and total harm (bottom) are shown as functions of {\bf (A)} initial T-cell number $N_\text{lin.}$, {\bf (B)} immunogenicity $\kappa$, {\bf (C)} and the intra-host basic reproduction of the infection $R_0$. In each panel, outcomes are shown for the ``elbow'' Pareto-optimal program from Fig.~4B while varying a single sensitivity parameter (colors)---antigen sensitivity $\psi_{\Ag}$ (left), infection-harm sensitivity $\psi_{\infc}$ (center), or response-harm sensitivity $\psi_{\res}$ (right)---with all other parameters held fixed. Trend lines and standard deviation error bars are obtained by averaging over sampled immune challenges (in Fig.~\mainfigthree B) and designs in the near-elbow ensemble. Simulation parameters as in Fig.~\mainfigfive.}
\label{perturbexpansionFigure-psis}
\end{figure*}

\clearpage{}
\begin{figure*}[t!]
\centering
\includegraphics[width=\textwidth]{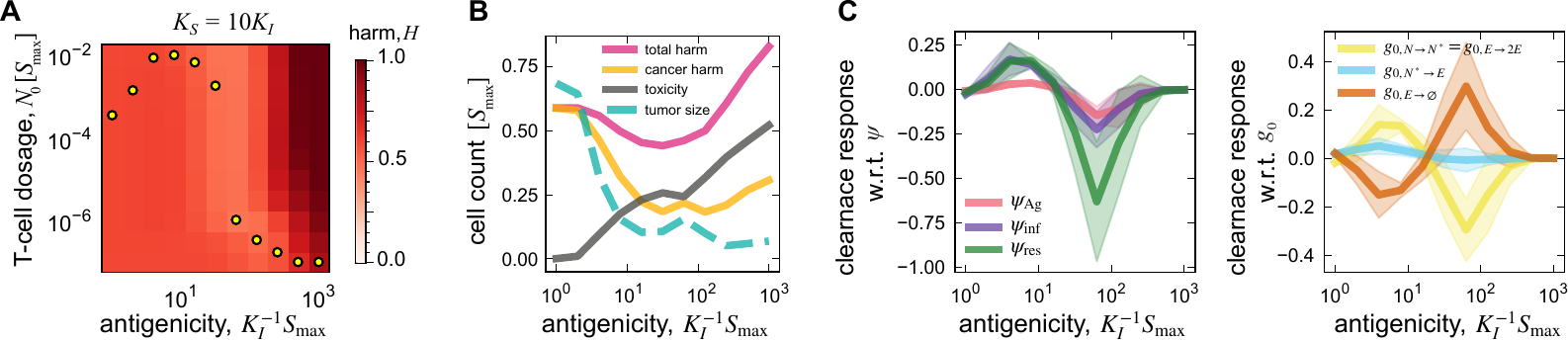}
\caption{{\bf Impact of cross-reactivity on engineered immune responses for cancer immunotherapy.} 
As in Fig.~\mainfigsix, but incorporating cross-reactivity of engineered TCRs with self and tumor antigens. We model cross-reactivity by assuming tumor and self reactivities are proportional, fixing the tumor immunogenicity $\kappa = K_I^{-1}K_S = 10 $ while varying the antigenicity of the tumor defined as $K_I^{-1} S_\text{max}$ (proxy for inverse EC$_{50}$, where larger values indicate higher antigenicity). {\bf (A)} With responses averaged over the tuned near-elbow ensemble $\tilde{\Theta}(\lambda^*)$, the heatmap shows harm as a function of T-cell dosage and tumor antigenicity. {\bf (B)} Total harm, cancer harm, immune  toxicity,  and the final tumor size following the immune response are shown as functions of tumor antigenicity at the optimal T-cell dosage (yellow circles in A). {\bf (C)} Clearance response, defined as sensitivity of protection (negative harm) to each design parameter---evaluated and averaged over $\theta \in \tilde{\Theta}(\lambda^*)$---is shown for different levels of tumor antigenicity for signal sensitivity parameters $\psi$'s (left), and for baseline parameters $g_0$'s (right). For this analysis, we assume that T-cells are administered at high dosage ($N_\text{lin.} = 10^{-2} S_\text{max}$).
 Simulation parameters as in Fig. \mainfigsix.
 }
\label{varyKSFigure}
\end{figure*}

\clearpage{}
\begin{figure*}[t!]
\centering
\includegraphics[width=\textwidth]{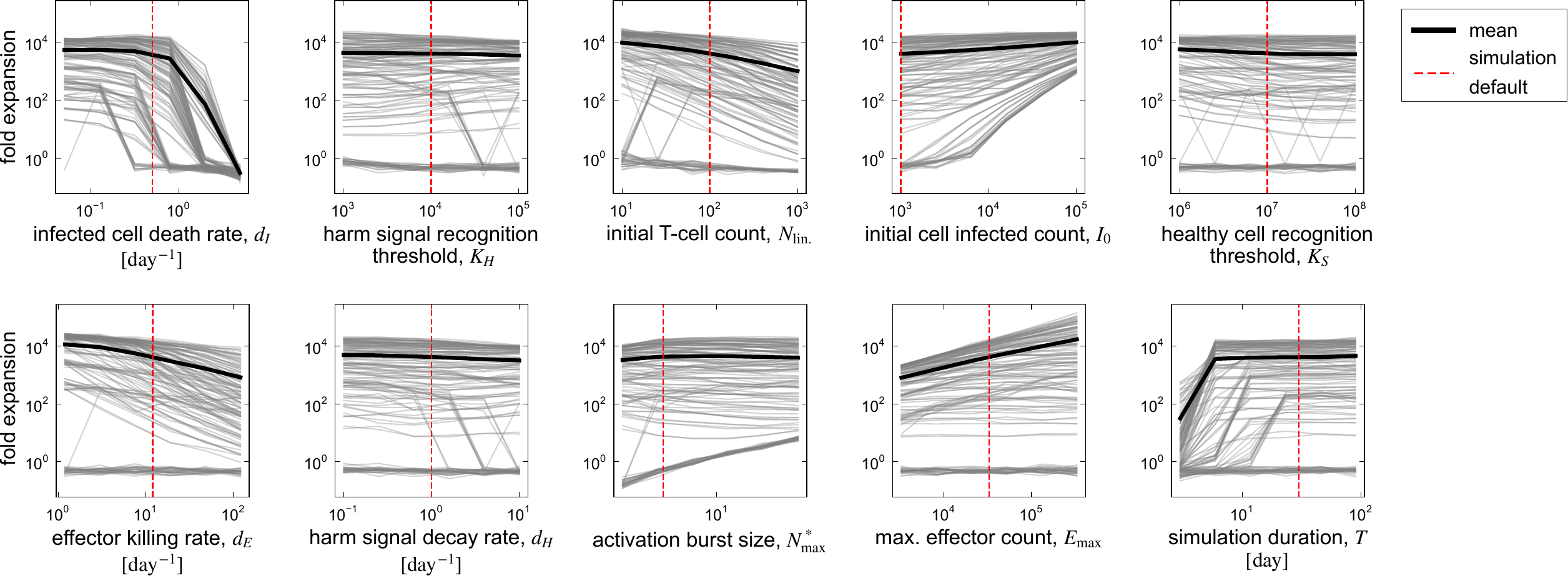}
\caption{{\bf Parameter sensitivity analysis for total harm.} Total harm, evaluated and averaged over the near-elbow ensemble $\theta \in \tilde{\Theta}(\lambda^*)$, is shown for different levels of the indicated parameter value. Gray lines represent trend lines for each infection condition $(\kappa, R_0)$ in Fig. \mainfigthree B, and the thick black line is the average trend over all infections. The red dotted line indicates the default parameter value used in simulations (SI Table~\ref{table:parameters}). Except when explicitly varied, simulation parameters are as in Fig.~\mainfigthree. 
\label{paramsensitivityFigure1}}

\end{figure*}

\clearpage{}
\begin{figure*}[t!]
\centering
\includegraphics[width=\textwidth]{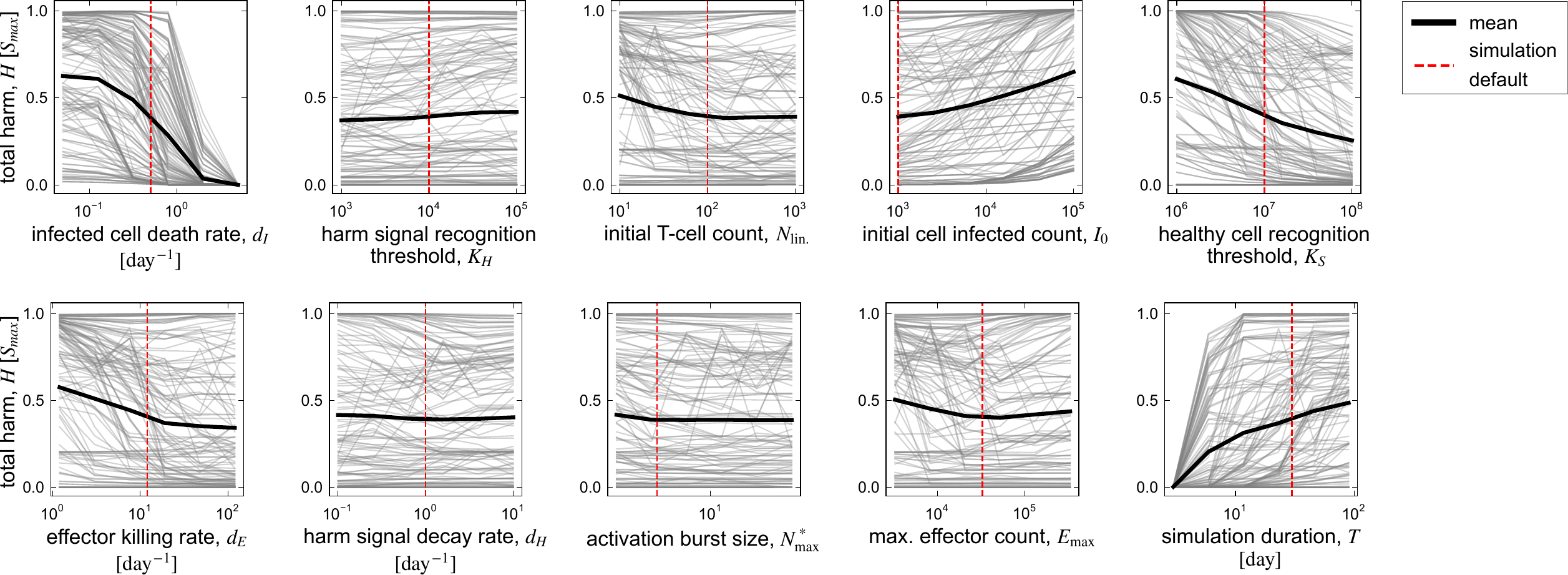}
\caption{{\bf Parameter sensitivity analysis for effector fold expansion.} Similar to Fig.~\ref{paramsensitivityFigure1} but showing the trend for effector fold expansion.
\label{paramsensitivityFigure2}
}
\end{figure*}

\clearpage{}
\begin{figure*}[t!]
\centering
\includegraphics[width=0.5\textwidth]{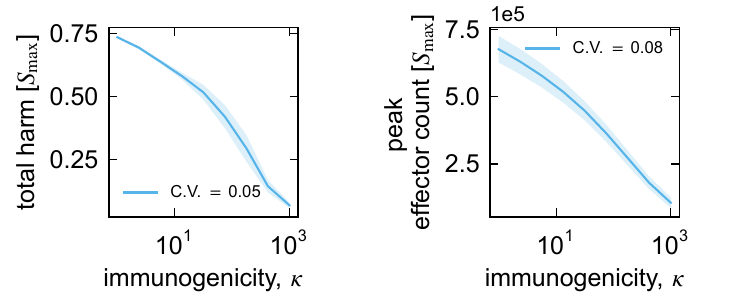}
\caption{{\bf Sensitivity of harm magnitude and T-cell expansion to stochastic cell-fate decisions.} {\bf (A)} Total harm, and {\bf (B)} effector-cell expansion,  averaged over the near-elbow design ensemble $\theta \in \Theta(\lambda^*)$, and over immune challenges spanning the range of reproduction numbers $R_0$ in Fig.~\mainfigthree, are shown as functions of immunogenicity $\kappa$. Insets report the coefficients of variation (C.V.), averaged across immune challenges spanning different $\kappa$'s. Except when explicitly varied, simulation parameters are as in Fig.~\mainfigthree.
}\label{noisesensitivityFigure1}
\end{figure*}

\clearpage{}

\setcounter{table}{0}
\renewcommand{\thetable}{S\arabic{table}}

\begin{table}[ht!]
	\centering
    \caption{Model parameters for simulations. The * symbol indicates parameters that were varied as part of the analysis.}
	\begin{adjustbox}{width=\textwidth}
	\begin{tabular}{|l|l|l|}
	\hline
	Parameter & Value & Reference(s)\\
	\hline
     initial susceptible cell count ($S_0$)  & $10^7$ & \\
     initial infected cell count ($I_0$)  &$10^3$ & \\
     infected cell death rate ($d_{I}$)  & $0.5~\text{day}^{-1}$ & \cite{Ikeda2014-cp, Ke2017-sp}\\
     susceptible cell infection rate ($b_I$)*  & $1-5 \times 10^{-7} ~\text{infected cell}^{-1}\text{day}^{-1}$ & \cite{Ikeda2014-cp}\\
     maximum cancer growth rate ($b_C$) & $0.1~\text{day}^{-1}$ & \\
     effector cell killing rate ($d_{E}$)  & $12~\text{day}^{-1}$ & \cite{Barchet2000-xr}\\
     infected cell recognition threshold ($K_{I}$)* & $10^{4}-10^{7}$ cells & \cite{Chao2004-jq}\\
     healthy cell recognition threshold ($K_{S}$)* & $10^{7}$ cells & \\
     harm signal recognition threshold ($K_{H}$) & $10^4$ cells & \\
     max. binding rate of APCs to Naive cells ($r_\text{bind}^\text{max}$) & $1~\text{day}^{-1}$ & \cite{De_Boer2001-wl,Belz2007, Mayer2019-pj}\\
     decay rate of harm signals ($d_H$) & $1.0 ~\text{day}^{-1}$ & \cite{Yiu2012-ej}\\
     initial naive cognate T-cell count ($N_\text{lin.}$)* & $10^{2}$ & \cite{Straub2023-oa}\\
     activation burst size ($N^*_\text{max}$) & $2^2$ & \cite{Plambeck2022}\\
     max. effector count per lineage ($E_\text{max}$) & $2^{15}$ & \cite{Buchholz2013,Marchingo2016, Badovinac2007-ii, Zhang2011-dv}\\
     max. rate of T-cell state transition ($r^\text{max}$) & $4~\text{day}^{-1}$& \cite{Van_Stipdonk2001-nm,Jenkins2008-if,Kretschmer2020-ad,Plambeck2022}\\
     rate of activated naive cell division ($r_{N^* \to 2N^*}$) & $\frac{24}{8.8}~\text{day}^{-1}$& \cite{Plambeck2022}\\
     signal sensitivity ($\psi_\sigma$)* & $\in [-3,3]$ & \\
     Baseline parameter ($\ell_i$)* & $\in [-3,3]$ & \\
     simulation duration ($T$) & $30~\text{days}$ & \cite{Zehn2009-up,Badovinac2007-ii, Lessler2009-aw}\\
     time step ($\Delta t$) & $10^{-2}~\text{days}$ & \\
    \hline
  \end{tabular}
  \end{adjustbox}
  \label{table:parameters}
\end{table}

\clearpage{}
